%% file: n2hp_mapping.tex
\documentclass[structabstract]{aa}  
\usepackage{graphicx}
\usepackage{txfonts}
\usepackage{natbib}
\usepackage{amsmath}
\usepackage{amssymb} 
\usepackage{import}
\usepackage{xspace}

\usepackage{multirow}
\usepackage{longtable,lscape}
\usepackage{ltxtable,booktabs}
\usepackage{array}

\usepackage{pifont}
\usepackage{color}

\newcommand{\cmark}{\ding{51}}
\newcommand{\xmark}{\ding{55}}

\begin{document}

\newcommand\micron[0]{\,$\mu$m\xspace}
\newcommand\msun[0]{\,M$_{\odot}$\xspace}
\newcommand\e[1]{\,$\times$\,10$^{#1}$}
\newcommand{\hdzco}{H$^{13}$CO$^{+}$\xspace}
\newcommand{\hco}{HCO$^{+}$\xspace}
\newcommand{\dzco}{$^{13}$CO\xspace}
\newcommand\hii{H{\sc ii}\xspace}
\newcommand{\hi}{\hbox{H{\sc i}}\xspace}
\newcommand{\htwo}{\hbox{H$_{2}$}\xspace}
\newcommand{\nhp}{N$_2$H$^{+}$\xspace}

\title{Kinematic structure of massive
  star-forming regions - I. Accretion along filaments}

\author{J. Tackenberg\inst{1},
  H. Beuther\inst{1},
  Th. Henning\inst{1},
  H. Linz\inst{1},
  T. Sakai\inst{2},
  S. E. Ragan\inst{1},
  O. Krause\inst{1},
  M. Nielbock\inst{1},
  M. Hennemann\inst{1,3},
  J. Pitann\inst{1},
  A. Schmiedeke\inst{1,4}
}

\authorrunning{J. Tackenberg et al., 2013}
\titlerunning{Flows along massive star-forming regions}

\institute{Max-Planck-Institut f\"ur Astronomie (MPIA), K\"onigstuhl 17, 69117 Heidelberg, Germany\\
  \email{last-name@mpia.de}
  \and
  Graduate School of Informatics and Engineering, The University of 
  Electro-Communications, Chofu, Tokyo 182-8585, Japan
  \and
  AIM Paris-Saclay, CEA/DSM/IRFU -- CNRS/INSU -- Universit\'e 
  Paris Diderot, CEA Saclay, 91191 Gif-sur-Yvette cedex, France
  \and
  Universit\"at zu K\"oln, Z\"ulpicher Str. 77, 50937, K\"oln, Germany
}

\date{Received September 15, 1996; accepted March 16, 1997}


\abstract
{The mid- and far-infrared view on high-mass star formation, in
  particular with the results from the {{\it Herschel}} space observatory, has shed
  light on many aspects of massive star formation. However, these continuum studies lack kinematic information.}
{We study the kinematics of the molecular gas in high-mass
  star-forming regions.}
{We complemented the PACS and SPIRE far-infrared data of 16 high-mass star-forming
  regions from the {{\it Herschel}} key project EPoS with \nhp molecular line data from
  the MOPRA and Nobeyama 45m telescope. Using the full \nhp hyperfine
  structure, we produced column density, velocity, and linewidth
  maps. These were correlated with PACS 70\micron images and PACS point
  sources. In addition, we searched for velocity gradients.
}
{For several regions, the data suggest that the linewidth on the scale of clumps is dominated
  by outflows or unresolved velocity gradients. IRDC\,18454 and
  G11.11 show two velocity components along several lines of sight. 
We find that all regions with a diameter larger than 1\,pc show either velocity
  gradients or fragment into independent structures with distinct velocities. The velocity profiles
  of three regions with a smooth gradient are consistent with gas flows along the
  filament, suggesting accretion flows onto the densest
  regions.} 
{We show that the kinematics of several regions have a significant and
      complex velocity structure. For three filaments, we suggest that gas
  flows toward the more massive clumps are present. 
}

\maketitle
%

\section{Introduction}
Despite their rarity, high-mass stars are important for all fields of
astronomy. Within the Milky Way they shape and regulate the formation
of clusters, influence the chemistry of the interstellar medium, and may even have affected the formation of the solar system
\citep{Gritschneder2012}. On
larger scales, emission from high-mass stars dominates the emission detected from
external galaxies. In addition, massive stars are the origin of heavy elements on all
scales. Nevertheless, high-mass star formation is far from being understood
\citep{Beuther2007, Zinnecker2007}. 

Sensitive infrared (IR) and (sub-)\,mm Galactic plane surveys together with
results from the {\it Herschel\footnote{{\it Herschel} is an ESA space observatory with science instruments provided by European-led Principal Investigator consortia and with important participation from NASA.}} space observatory \citep{Pilbratt2010}
have shed new light on the
cradles of massive stars/clusters and their early
formation. \citet{Perault1996} and \citet{Egan1998} discovered extinction
patches in the bright mid-IR (MIR) background using the ISO
\citep{Kessler1996} and MSX \citep{Egan2003} satellites. These extinction patches are similar to the dark
patches reported by \citet{Barnard1919}, which are today known to be connected to
low-mass star formation. Soon after this, \citet{Carey1998} established that the
so-called infrared dark clouds (IRDCs) are the
precursors of high-mass star formation. Today, the Spitzer observatory Galactic plane surveys GLIMPSE at
3.6\micron, 4.5\micron, 5.8\micron, and 8\micron \citep{Benjamin2003} and
MIPSGAL at 24\micron \citep{Carey2009} allow the systematic search for IRDCs
with unprecedented sensitivity \citep[e.g.][]{Peretto2009}. 

While Spitzer improved our MIR view of the Galaxy, the {\it Herschel}
satellite allows observations of the far-IR (FIR). With the PACS
\citep{Poglitsch2010} and SPIRE \citep{Griffin2010}
photometer, high sensitivity and spatial resolution observations between
70\micron and 500\micron are possible. 
From correlating data at MIR through submillimeter and
millimeter wavelengths, the picture emerged 
that most star-forming regions are filamentary \citep{Andre2010,
  Menshchikov2010, Molinari2010, Schneider2010, Hill2011, Hennemann2012, Peretto2012}. 

In numerical studies, the formation of dense cores and clumps is explained by
two scenarios. On the one hand, molecular clouds fragment in a self-similar
cascade down to the typical size of dense, quasi-static cores supported by
turbulence. These will then form single or multiple gravitationally bound
objects \citep{McKee2003, Zinnecker2007}. On the other hand, in the dynamical theory
molecular clouds are formed from large-scale flows of atomic gas as transient
objects \citep{MacLow2004,
  Klessen2005, Heitsch2008, Clark2012}. Within these transient structures,
supersonic turbulence compresses some fraction of the gas to
filaments, clumps, and dense cores. If gravity dominates, the cores
collapse. In contrast to the quasi-static cores, these cores
constantly grow in mass. When, by chance, some cores accrete mass faster
than others due to their higher initial gravitational potential, this
is called competitive accretion \citep{Bonnell2004}. Also dynamical, but
  of reversed reasoning, in the fragmentation-induced starvation
  scenario by
  \citet{Peters2010}, individual massive dense cores build from the large-scale
  flows, and an accompanying cluster of smaller cores drag away material from
  the main core and diminish its mass accretion.

The Earliest Phases of Star formation (EPoS, PI O. Krause) is a
Guaranteed Time {\it Herschel} Key Program for investigating 14 low-mass and 45 high-mass
star-forming regions. %
The low-mass observations have been summarized in \citet{Launhardt2013}, and
the high-mass part has been described in \citet{Ragan2012}. 
The high-mass part of the project provides an excellent target list for studying
the kinematics in star-forming regions. This is the ultimate goal of this paper, using \nhp
molecular line data.

\section{Observations and analysis}

\subsection{EPoS - A {\it Herschel} key project}
All 45 high-mass EPoS sources were observed with the {\it Herschel} satellite
\citep{Pilbratt2010} at 70\micron, 100\micron, 160\micron, 250\micron,
350\micron, and 500\micron with a spatial resolution of 5.6\arcsec,
6.8\arcsec, 11.3\arcsec, 18.1\arcsec, 25.2\arcsec, and 36.6\arcsec,
respectively \citep{Poglitsch2010, Griffin2010}. The
observations were performed in two orthogonal directions and the data
reduction has been performed using {\it HIPE} \citep{Ott2010} and {\it
  scanamorphos} \citep{Roussel2012}. A more detailed description of the data
reduction is given in \citet{Ragan2012}.

Out of the 45 {\it Herschel} EPoS high-mass sources we selected a subsample of 17
regions given in Table \ref{tab:obs_log} that cover each important evolutionary stage: promising high-mass starless core candidates, IRDCs with weak MIR and FIR sources, indicative of early ongoing star formation activity, and known
high-mass protostellar objects (HMPOs).

The protostellar core population was previously characterized in \citet{Ragan2012} using {\it Herschel} photometry, supplemented by Spitzer, IRAS, and MSX data. By modeling the spectral energy distributions (SEDs), the authors fit the temperature, luminosity, and mass of each protostellar core in the sample.

\subsection{Nobeyama 45m observations}
\input{obs_log_resolution}
Between April 7 and 12 2010 the BEam Array Receiver System (BEARS,
\citealt{Sunada2000}) on the Nobeyama Radio Observatory
(NRO\footnote{Nobeyama Radio Observatory is a branch of the National
  Astronomical Observatory of Japan, National Institutes of Natural
  Sciences.}) 45\,m telescope was used to map six of the regions in
\nhp; the details are given in Table \ref{tab:obs_log}. 
At a frequency of the \nhp (1-0) transition of 93.173\,GHz, the spatial
resolution of the NRO 45\,m telescope is 18\arcsec (HPBW) and the observing
mode with a bandwidth of 32\,MHz has a frequency resolution of
62.5\,kHz, or 0.2\,km/s. All observations were performed using
on-the-fly (OTF) mapping in varying weather conditions, with an
average system temperature of 280\,K and high precipitable water vapors between
3\,mm and 9\,mm. The pointing was made using the single-pixel receiver
  S40 tuned to SiO. As pointing sources we used the SiO masers of the
  late-type stars V468 Cyg, IRC+00363, and R Aql. Although the wind conditions
contributed to the pointing uncertainties, the pointing is better than a
third of the beam. 

For the data reduction we used NOSTAR \citep{Sawada2008}, a software
package provided by the NRO for OTF data. The data were sampled to a spatial
grid of 7.5\arcsec and smoothed to a spectral resolution of 0.5\,km/s. To
account for the different efficiencies of the 25 receivers in the BEARS array
we corrected each pixel to the efficiency of the S100 receiver, using
individual correction factors and a beam efficiency of $\eta$ = 0.46 to
calculate main-beam temperatures. The noisy edges due to less coverage were removed within NOSTAR by suppressing pixels in the final maps with an rms
noise above 0.15\,K. The resulting antenna temperature maps have an average rms noise between
0.12\,K and 0.13\,K per beam.

\subsection{MOPRA observations}
The 11 sources, listed in Table \ref{tab:obs_log} were mapped with the 22\,m
MOPRA radio telescope, operated by the Australia Telescope National Facility
(ATNF) in OTF mode. The observations were carried out in 2010, June 1 to 5 and 25
to 27, as well as July 7 through 9. High precipitable water vapors during
the observations result in system temperatures mostly of between 200\,K and
300\,K. Observations with system temperatures higher than
500\,K were ignored during the data reduction. 

We employed 13 of the MOPRA spectrometer (MOPS) zoom bands, each of 138\,MHz
width and 4096 channels, resulting in a velocity resolution of 0.11\,km/s at
90\,GHz. The spectral setup covered transitions of CH$_3$CCH, H$^{13}$CN,
\hdzco, SiO, C$_2$H, HNCO, HCN, \hco, HNC, HCCCN, CH$_3$CN, $^{13}$CS, and
\nhp in the 90\,GHz regime (for details on the transitions and their
excitation conditions see \citealt{Vasyunina2011}). At this wavelength, the
MOPRA beam FWHM is 35.5\arcsec and the beam efficiency is assumed to be
constant over the frequency range with $\eta$ = 0.49 \citep{Ladd2005}.
The data were reduced using LIVEDATA and GRIDZILLA, an
mapping analysis package provided by the ATNF. To
improve the signal-to-noise ratio we spatially smoothed the data to a beam
FWHM of 46\arcsec within (and as suggested by) {\it gridzilla}.  
The final maps were smoothed to a spectral resolution of 0.21\,km/s -
0.23\,km/s (depending on the transition frequency). Spectra with an rms noise above 0.12\,K have been
removed; this affected pixels at the edges. The resulting average rms noise
of the individual maps is then below 0.09\,K/beam. 

The observed regions of interest are dense but still
cold. Therefore, with the achieved sensitivity at the given spatial
resolution we did not detect the more complex or low-abundance
molecules. For example, although SiO\,(2-1) has been detected toward several
positions we mapped (\citealt{Sridharan2002, Sakai2010}),
the strongest SiO emitter found by Linz et al. (in prep., G28.34-2)
is at our noise level and therefore not detected. Commonly detected and
  reasonably well mapped are HCN\,(1-0), HNC\,(1-0), \hco\,(1-0), \hdzco\,(1-0), and \nhp\,(1-0).
As we discuss in Sect. \ref{sec:hyperfine_fitting}, we concentrate on \nhp, a
well-known cold dense gas tracer. 

\subsection{Dust continuum}
To trace the total cold gas we used the cold
dust emission as a tracer of the molecular gas. Because most of the selected sources lie
within the Galactic plane, the APEX 870\micron survey ATLASGAL
\citep{Schuller2009, Contreras2013} covers all but
two sources. Its beam size is 19.2\arcsec and its average rms noise is
50\,mJy/beam. IRDC\,18151 with a Galactic latitude of $\sim$\,1.7\degr is not covered by ATLASGAL. We used IRAM 30\,m MAMBO data instead 
\citep{Beuther2002}. At a
wavelength of 1.2\,mm, the beam width is 10.5\arcsec, and the rms noise in the
dust map is 17\,mJy/beam. In addition, ISOSS\,J20153 is outside the coverage
of ATLASGAL. Here we used 850\micron data from the SCUBA camera at
the JCMT\footnote{{\it The James Clerk Maxwell Telescope} is operated by the Joint
  Astronomy Centre on behalf of the Particle Physics and Astronomy Research
  Council of the United Kingdom, the Netherlands Association for Scientific
  Research, and the National Research Council of Canada.} \citep{Hennemann2008}. The beam width is 14\arcsec and the rms noise 14\,mJy/beam. 
A summary of the properties of the submm data is given in Table
\ref{tab:dust_prop}.

\input{dust_properties}

\subsection{\nhp hyperfine fitting}
\label{sec:hyperfine_fitting}
Our molecular line study focuses on the \nhp\,J=1-0 line as dense
molecular gas tracer. Using the Einstein A and collison coefficient for a
  temperature of 20\,K \citep{Schoier2005}, its critical density is
1.6\e{5}\,cm$^{-3}$, and its hyperfine structure allows one to
reliably measure its optical depth and thus the distribution over a wide range
of densities. In addition, the velocity and linewidth can be measured without
being affected by optical depth effects. Finally, it is detected toward both low- and high-mass star-forming regions of various evolutionary stages \citep{Schlingman2011}. Therefore, it is well-suited for studies of young high-mass star-forming regions. 

To extract the \nhp line parameters, we fit a \nhp hyperfine structure to each pixel using {\it class} from
the {\it GILDAS\footnote{http://www.iram.fr/IRAMFR/GILDAS}} package. For every
spectrum we calculated the rms and the peak intensity of the brightest component derived from
the fit parameters. When the peak was higher than three times the rms value, the
fit parameters peak velocity and the linewidth together with an integrated intensity were stored to a parameter
map. Otherwise, the pixel was left blank. The low detection threshold
of 3\,$\sigma$ is justified for two reasons. (1) For only a very limited number of
pixels that fulfill the 3\,$\sigma$ criterion, the fitted linewidth is
twice the channel width or smaller. Therefore, introducing an additional check on
the integrated intensity, such as a 5\,$\sigma$, does not improve the fit
reliability. (2) The resulting parameter maps only show 
a smooth transition in each
parameter relative to the neighboring pixels. For the same pixels, smoothing over
a larger area would increase the signal-to-noise ratio but would worsen the resolution. Therefore, the small-scale
structure would be lost. 
For our purposes, fitting the hyperfine structure even in low
signal-to-noise ratio maps provides reliable results. 

From the integrated intensity ($\int$\,T$_{mb}$), determined as the sum over
channels times the channel
width, and the fitted optical depth $\tau$, we calculated the column densities of \nhp. We
used the formula from \citet{Tielens2005},
\begin{eqnarray}
  N_J &=& 1.94\times10^3\, \frac{\nu^2 \int T_{mb}}{A_u}\times\frac{\tau}{1
    -e^{\tau}} \text{\ \ (for J+1 to J)}\\
  N_{tot} &=& \frac{Q}{g_u} \times N_J \times e^{E_u / kT_{ex}}\text{,}
\end{eqnarray}
where N$_J$ is the number of molecules in the the J-th level, and N$_{tot}$
the total number of molecules. A$_u$ is the Einstein A coefficient of the upper
level, E$_u$/k is the excitation energy of the upper level in K, both
adopted from \citet{Schoier2005} and \citet{Vasyunina2011}; Q is the partition
function of the given level, taken from the Cologne Database for Molecular
Spectroscopy \citep{Muller2005}, and g$_u$ is the degeneracy of the energy level.

The rms of our observations limits the detection of signal, but
the \nhp column density also depends on the measured opacity and assumed temperature. As we discuss in Sect. \ref{sec:comparing_intnhp_dustcont}, we used a
constant gas temperature of 20\,K. Varying the temperature by
up to 5\,K, the calculated column densities vary by less than
25\%. Taking the error on the integrated intensity and the partition
function into account as well, we assumed the error on the column density to be on the order of 50\%. With
the given rms toward the edge of the MOPRA data, our theoretical 5\,$\sigma$ \nhp detection
limit in the optically thin case is given by 1.5\e{12}\,cm$^{-2}$ for the velocity resolution of
0.2\,km/s. Since most sources have considerable optical depth, the lowest
measured \nhp column density is higher. The lowest calculated values are
given in Table \ref{tab:obs_log}. Similarly, for the Nobeyama 45\,m data the
5\,$\sigma$ detection limit is 2.1\e{12}\,cm$^{-2}$. Since some central
regions for instance of IRDC\,18223 have a much lower rms of only 0.05\,K instead
of 0.12\,K, here the lowest calculated values are even lower.

The uncertainties on the fit parameters velocity and linewidth are mainly constrained by the
signal-to-noise ratio and the line shape. Since even the broadest velocity
resolution within our sample of 0.5\,km/s resolves the lines, the
uncertainties on the linewidth are similar for all data. Spectra with a
signal-to-noise ratio better than 7\,$\sigma$ and Gaussian-shaped line profiles
have typical linewidth uncertainties of below 5\%, while non-Gaussian line
profiles and low signal-to-noise ratios may lead to uncertainties of up to
20\%. Instead, the recovery of the peak velocity shows an additional slight
velocity resolution dependency. Still, both line shape and signal-to-noise
ratio dominate, and down to a peak line strength of 3\,$\sigma$, the
uncertainties on the velocity are below 0.2\,km/s.

\subsection{Identification of dust peaks}
\label{sec:dust_clfind}
To set the molecular line data in context to its environment we used the dust
continuum to obtain gas column densities and masses. For the calculation of
column densities from fluxes we used
\begin{equation}
  N_{gas} = \frac{ R F_{\lambda} }{ B_{\lambda}(\lambda,T) \mu m_{H} \kappa \Omega}{\text ,}
\end{equation}
with the gas-to-dust mass ratio R\,=\,100, F$_{\lambda}$ the flux at the given
wavelength, B$_{\lambda}$($\lambda$,T) the blackbody radiation as a function
of wavelength and temperature, $\mu$ the mean molecular weight of the
ISM of 2.8,
m$_{H}$ the mass of a hydrogen atom, and the beam size $\Omega$. Assuming
typical beam-averaged volume densities in the dense gas of 10$^5$\,cm$^{-3}$
and dust grains with thin ice mantles, we interpolated the dust mass
absorption coefficient from \citet{Ossenkopf1994} to the desired wavelength. 
The corresponding dust opacities for the different wavelengths are listed in
Table \ref{tab:dust_prop}.

The gas and dust temperatures should be coupled at densities typical for
dense clump ($>$\,10$^5$\,cm$^{-3}$, \citealt{Goldsmith2001}), and have been measured to be between
15\,K and 20\,K \citep{Sridharan2005, Pillai2006, Peretto2010,
  Battersby2011, Wilcock2011, Vasyunina2011, Wienen2012, Wilcock2012}. Since
most regions in our survey already show signs of ongoing star formation, we
assumed a single temperature value of 20\,K for all clumps.

With the distance d as additional parameter, the mass can be calculated from the
integrated flux in a similar way as given above, 
\begin{equation}
  M_{gas} = \frac{  R d^2 F_{\lambda} }{ B_{\lambda}(\lambda,T) \kappa} \text{.}
  \label{eqn:mass}
\end{equation}

To identify emission peaks and their connected fluxes in the dust maps we used CLUMPFIND
\citep{Williams1994}. Since we aim to compare our results to the dense gas
measured by \nhp, we selected a lowest emission contour corresponding to
1\e{22}\,cm$^{-2}$ ($>$\,6\,$\sigma$ for ATLASGAL, $>$\,12\,$\sigma$ for SCUBA), or, in the case of ISOSS\,J20153,
2\e{22}\,cm$^{-2}$ ($>$\,3\,$\sigma$). Additional levels were added in steps
of 3\,$\sigma$, see Table \ref{tab:dust_prop}.
All clumps for which we mapped the peak position are listed
together with their column density and mass in Table \ref{tab:clump_list}. 
\input{EPOS_clumps_mvir_alt_noPACS}
For clumps that have common names in the
literature, we adopted their previous labeling. The clump names
and references are given in Table \ref{tab:clump_list} as well.

The uncertainties on both the gas column density and mass are dominated by the
dust properties. The flux calibration of the ATLASGAL data is reliable
within 15\%, and typical peak and clump-integrated fluxes are an order
of magnitude higher than the rms of the data. The uncertainties on the
dust properties are difficult to assess, but from comparison with other values
(e.g. \citealt{Hildebrand1983}) or using slightly different parameters within
the same model \citep{Ossenkopf1994}, we assumed them to be on
the order of a factor two. Together with the uncertainties from the dust
temperature, we estimate the total uncertainties on the column densities to be
a factor of $\sim$\,3. For the gas mass, the uncertainty of the distance of
$\sim$\,0.5\,kpc introduces an additional error of $\sim$\,50\%. Therefore, we estimate the total uncertainty of the gas mass to be on the order of a factor of five.


\subsection{Abundance ratio}
\label{sec:explain_abundance}
For positions where not only dust continuum but also \nhp has been detected we
calculated the abundance ratios. With a resolution of 18\arcsec and 19.2\arcsec,
the Nobeyama 45\,m data have almost the same resolution as
the ATLASGAL 870\micron data. Therefore we calculated the \nhp abundance ratio
by plain division, taking the different beam size into account, but did
not smooth the data. To calculate abundance ratios for sources observed with
MOPRA at a resolution of 46\arcsec, we applied a Gaussian smoothing to
the dust data to have both at a common resolution. 
We then calculated the \nhp abundance for the 46\arcsec beam.
For the analysis, we only considered dust measurements in the Gaussian-smoothed images above 3\,$\sigma$.

\section{Morphology of the dense gas}
In the following section we concentrate on the \nhp
observations. First we compare the \nhp with the cold dust distribution as
measured by ATLASGAL and set it in context with the PACS 70\micron
measurements. Then we describe the velocity and linewidth distribution
of the dense gas. 

\subsection{Comparing integrated \nhp and dust continuum emission}
\label{sec:comparing_intnhp_dustcont}
\begin{figure*}[tbp]
  \includegraphics[width=1.\textwidth]{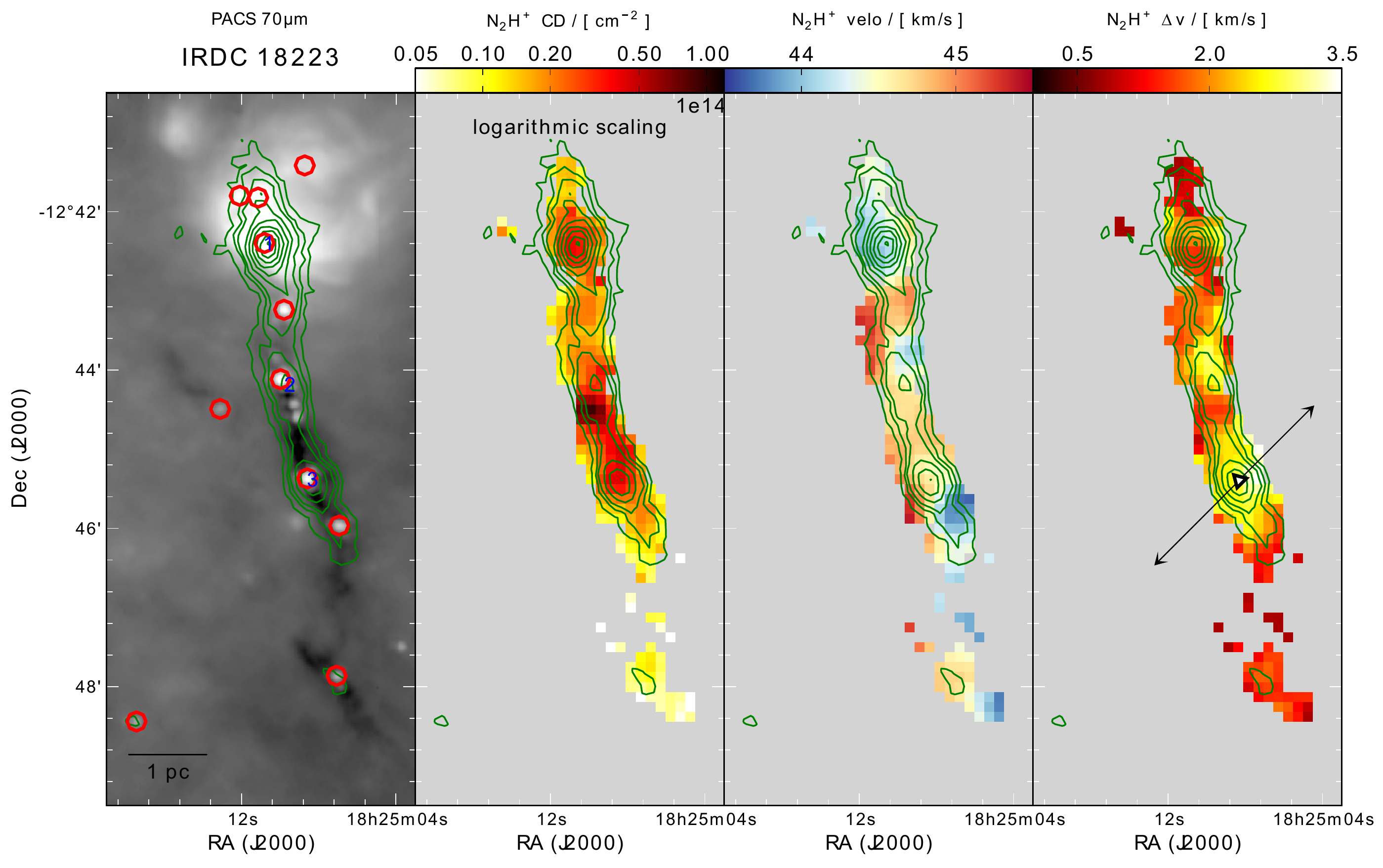}
  \includegraphics[width=1.\textwidth]{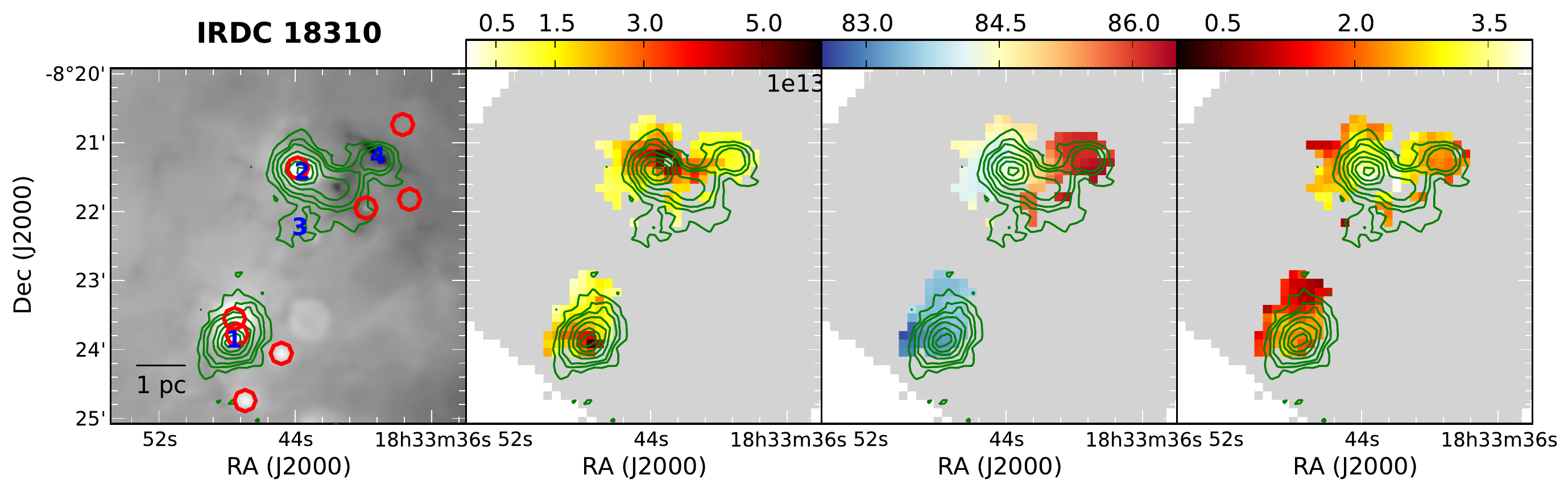}
  \includegraphics[width=1.\textwidth]{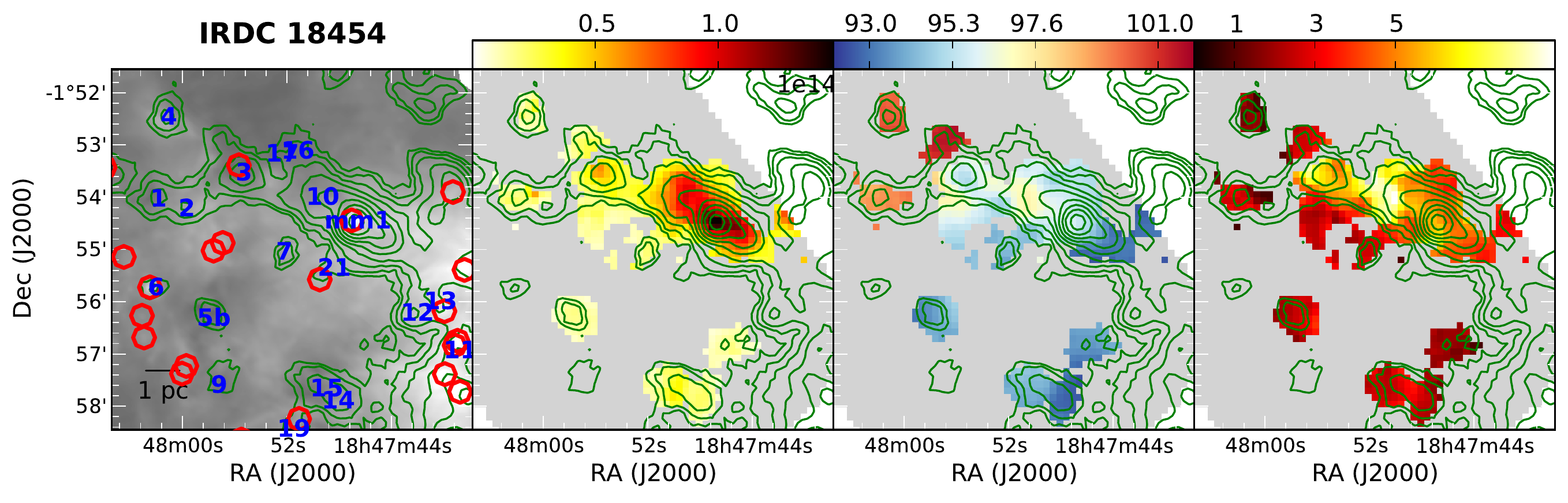}
  \caption{Parameter maps of the regions IRDC\,18223, IRDC\,18310, and
    IRDC\,18454 mapped with the Nobeyama 45\,m telescope, in top,
    middle, and bottom panel, respectively. The left panels of each row
    are the PACS 70\micron maps with the PACS point sources detected by
    \citet{Ragan2012} indicated by red circles, the blue numbers refer
    to the submm continuum peaks as given in Table \ref{tab:clump_list}. The second panels
    display the \nhp column density derived from fitting the
    full \nhp hyperfine structure. The third and fourth panels show
    the corresponding velocity and linewidth (FWHM) of each fit. For
    IRDC\,18223, and IRDC\,18310 the contours
    from ATLASGAL 870\micron are plotted with the lowest level
    representing 0.31\,Jy, and continue in steps of 0.3\,Jy. The contour
    levels for IRDC\,18454 are logarithmically spaced, with 10 levels
    between 0.31\,Jy and 31\,Jy. The column density scale of IRDC\,18223
    is logarithmic. The arrow in the fourth panel of IRDC\,18223 is
      taken from Fig. 4 of \citet{Fallscheer2009}, indicating the outflow
      direction.}
  \label{fig:n2hp_param_nobeyama}
\end{figure*}

\begin{figure*}[tbp]
  \includegraphics[width=1.\textwidth]{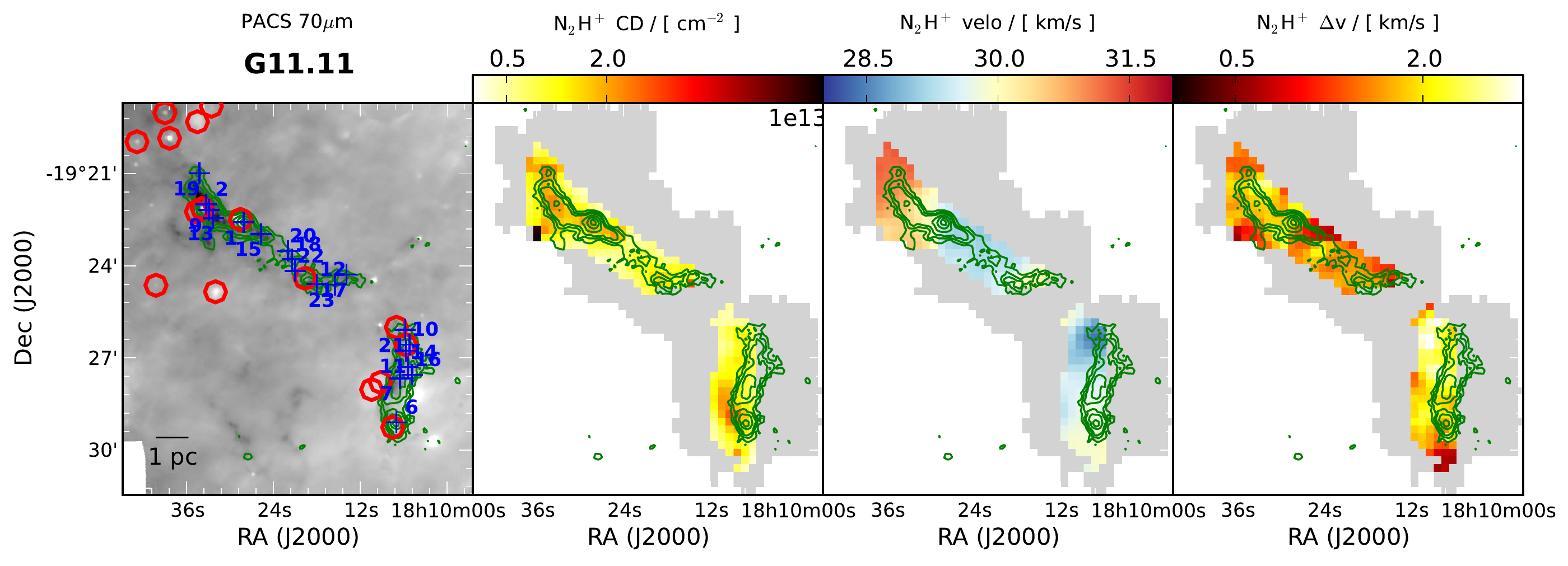}
  \includegraphics[width=1.0\textwidth]{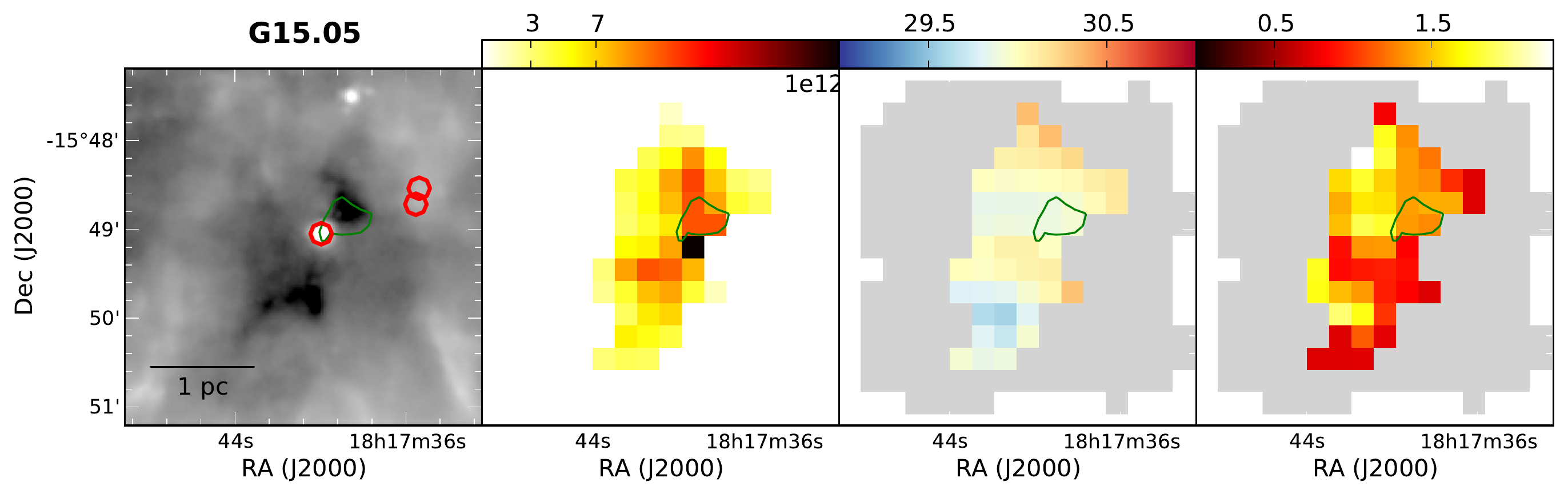}
  \includegraphics[width=1.0\textwidth]{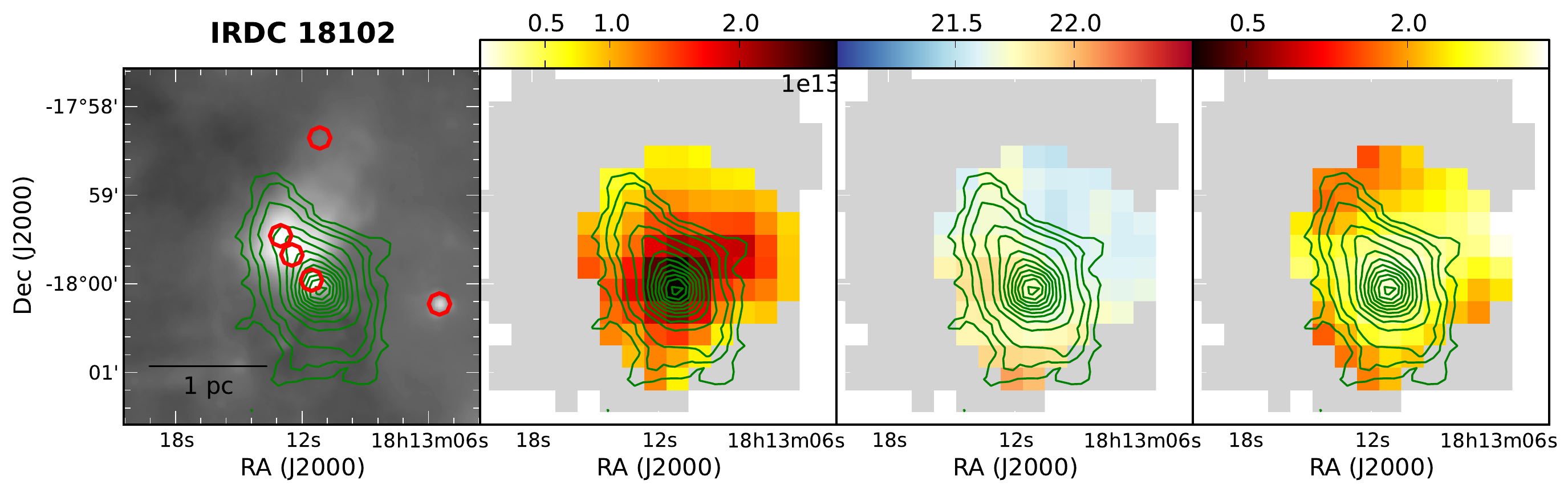}
  \caption{Parameter maps of the regions G11.11, G15.05, and
    IRDC\,18102, mapped with the MOPRA telescope. The left panels of each row
    are the PACS 70\micron maps with the PACS point sources detected by
    \citet{Ragan2012} indicated by red circles, the blue numbers refer
    to the submm continuum peaks as given in Table
    \ref{tab:clump_list}. The second panels
    display the \nhp column density derived from fitting the
    full \nhp hyperfine structure. The third and fourth panels show
    the corresponding velocity and linewidth (FWHM) of each fit. The
    green contours are from ATLASGAL 870\micron at 0.31\,Jy, 0.46\,Jy,
    and 0.61\,Jy, continuing in steps of 0.3\,Jy. The velocity resolution in the G15.05 map is
    smoothed to 0.4\,km/s to improve the signal-to-noise ratio and increase
    the number of detected \nhp positions.}
  \label{fig:n2hp_param_mopra1}
\end{figure*}

\begin{figure*}[tbp]
  \includegraphics[width=1.0\textwidth]{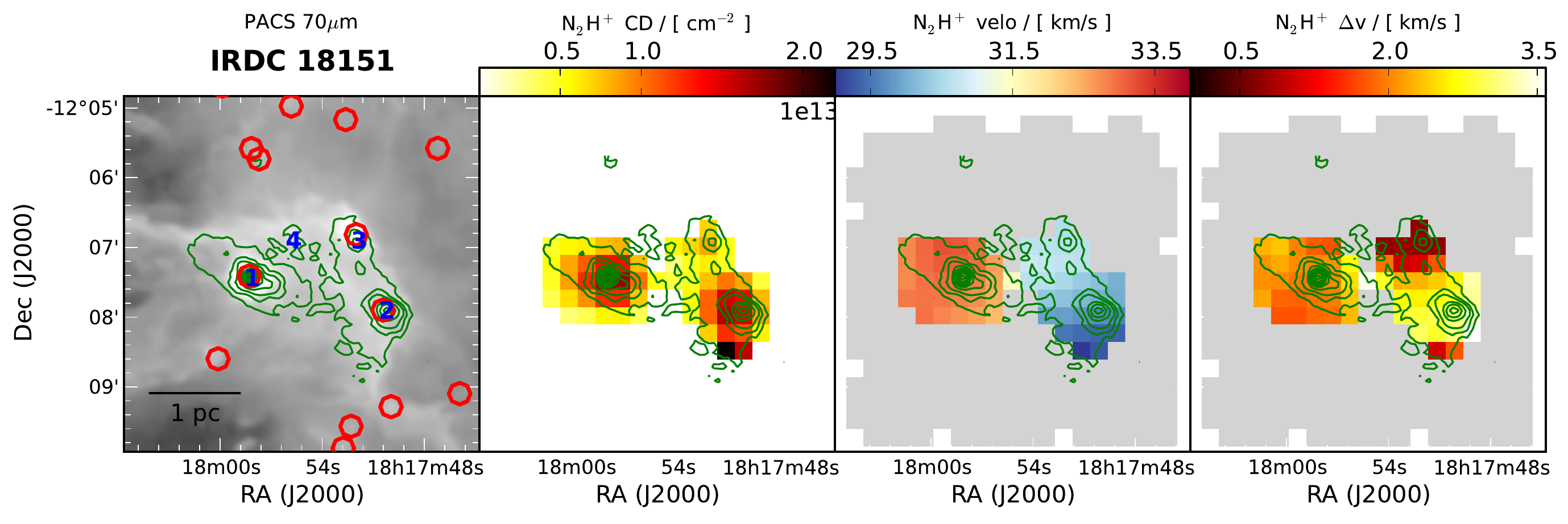}
  \includegraphics[width=1.0\textwidth]{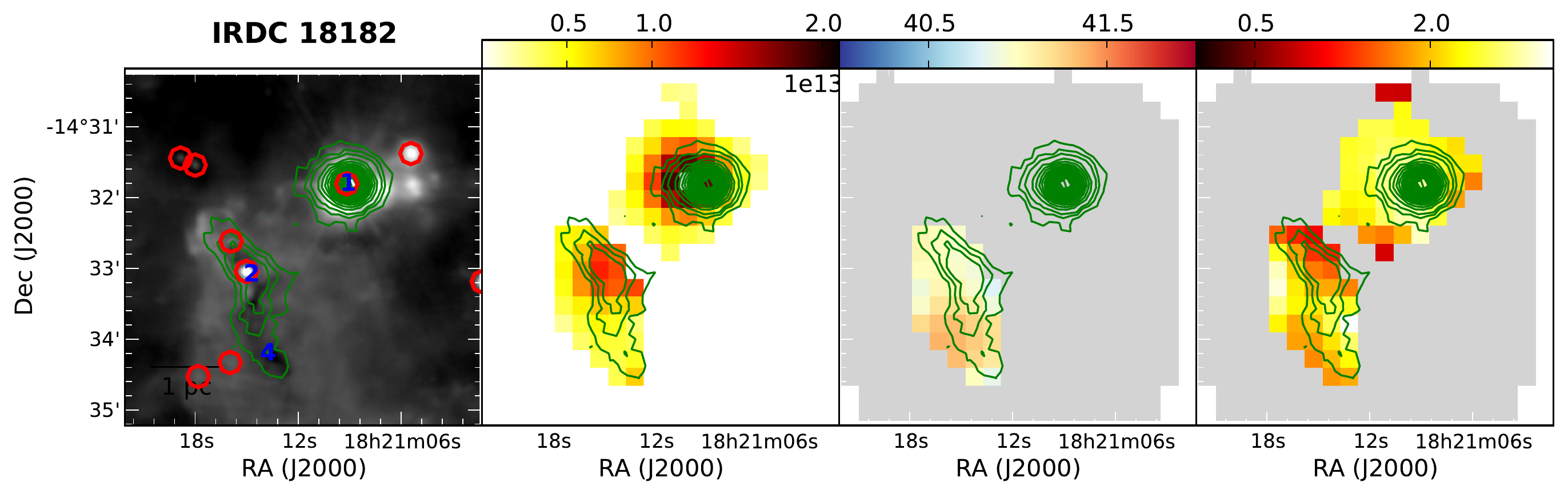}
  \includegraphics[width=1.0\textwidth]{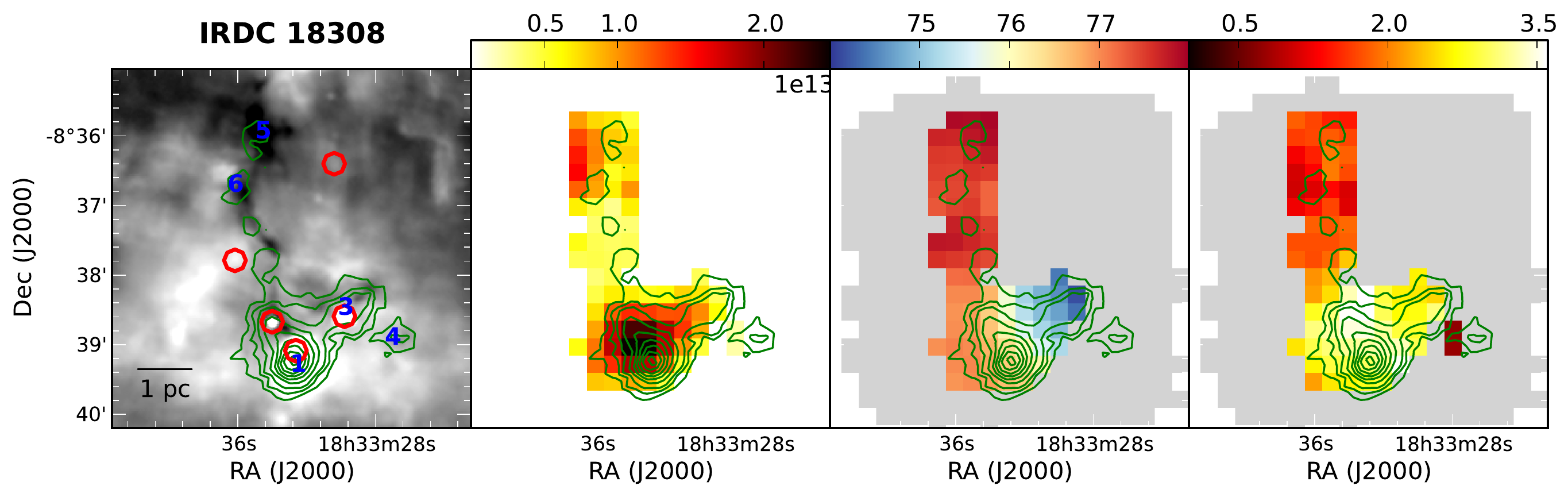}
  \caption{Parameter maps of the regions IRDC\,18151, IRDC\,18182, and
    IRDC\,18308, each mapped with the MOPRA telescope. The left panels
    of each row are the PACS 70\micron maps with the PACS point
    sources detected by \citet{Ragan2012} indicated by red circles,
    the blue numbers refer to the submm continuum peaks as given in
    Table \ref{tab:clump_list}. The second panels display the \nhp
    column density derived from fitting the full \nhp hyperfine
    structure. The third and fourth panels show the corresponding
    velocity and linewidth (FWHM) of each fit. The green contours are
    from ATLASGAL 870\micron at 0.31\,Jy, 0.46\,Jy, and 0.61\,Jy,
    continuing in steps of 0.3\,Jy. For IRDC\,18151 the contours are
    MAMBO 1.2\,mm observations, starting at 60\,mJy in steps of
    60\,mJy. In all three maps the velocity resolution is smoothed to
    0.4\,km/s.}
  \label{fig:n2hp_param_mopra2}
\end{figure*}

\begin{figure*}[tbp]
  \includegraphics[width=1.0\textwidth]{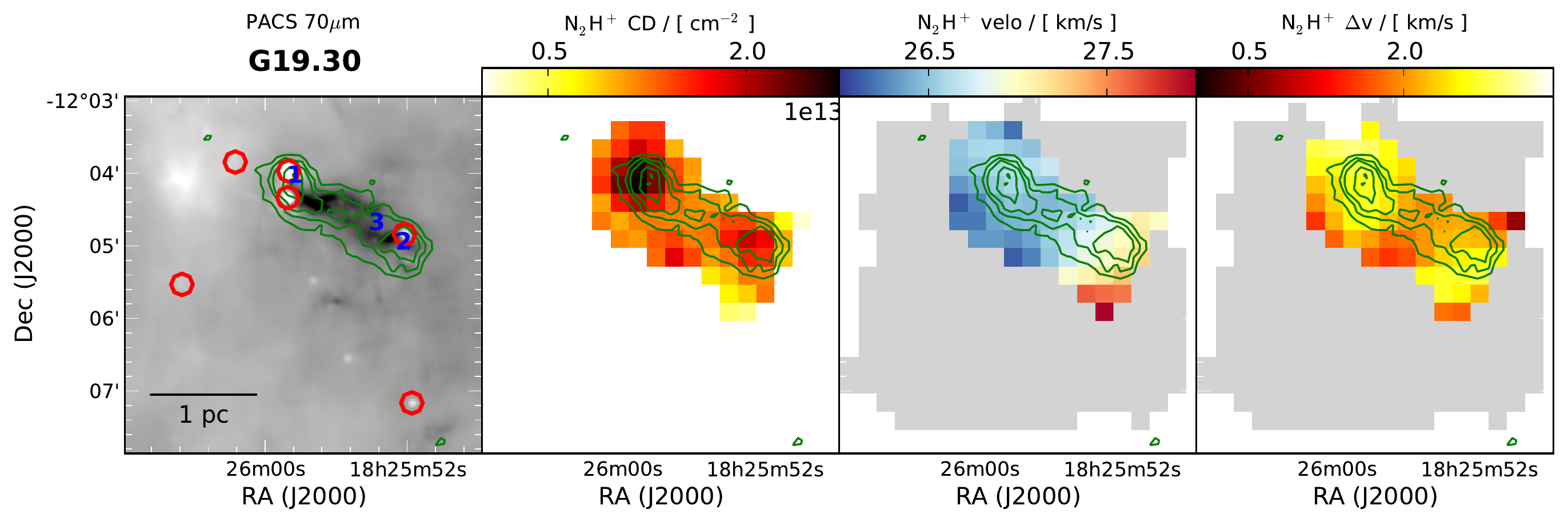}
  \includegraphics[width=1.0\textwidth]{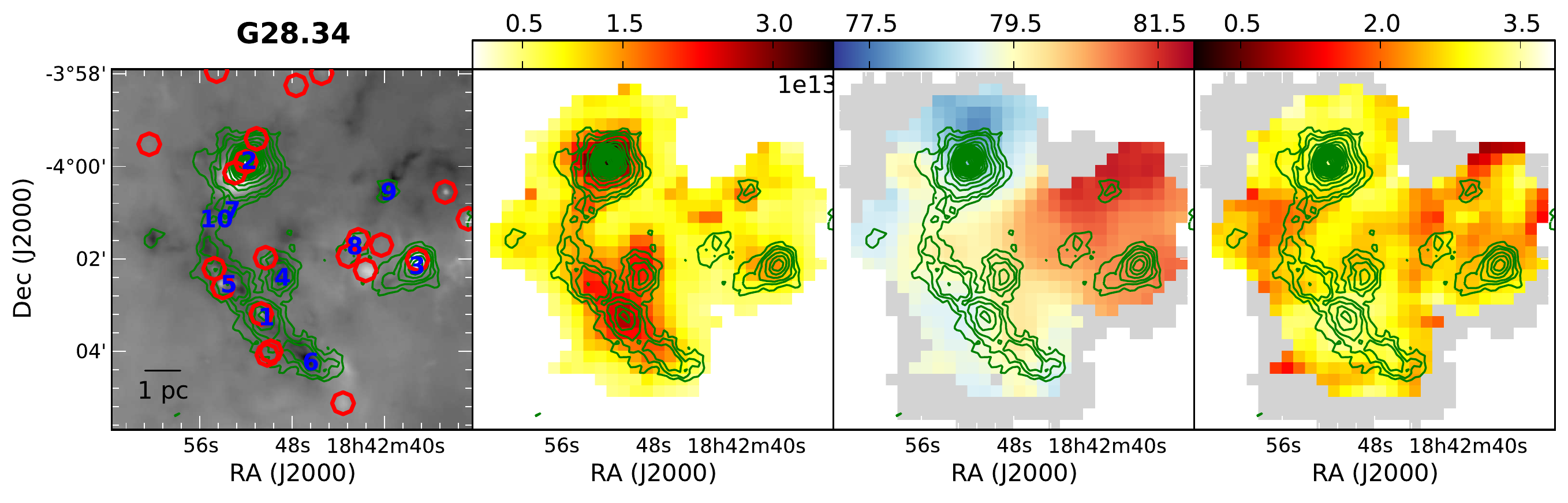}
  \includegraphics[width=1.0\textwidth]{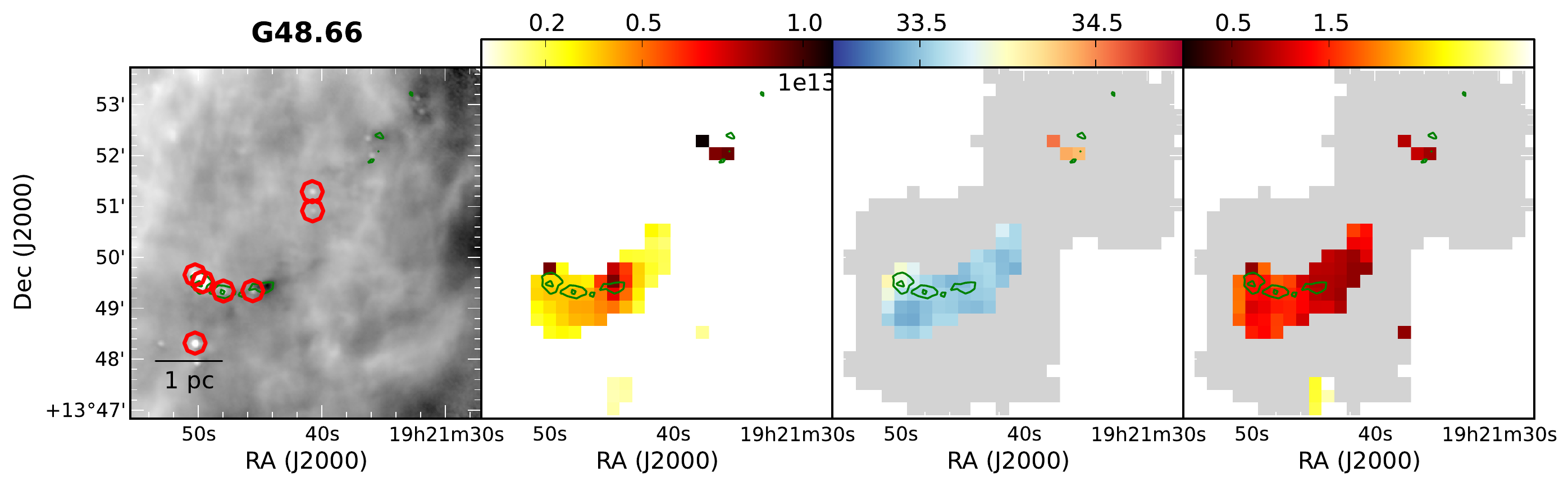}
  \caption{Parameter maps of the regions G19.30, G28.34, and G48.66,
    each mapped with the MOPRA telescope. The left panels of each row
    are the PACS 70\micron maps with the PACS point sources detected by
    \citet{Ragan2012} indicated by red circles, the blue numbers refer
    to the submm continuum peaks as given in Table
    \ref{tab:clump_list}. The second panels
    display the \nhp column density derived from fitting the
    full \nhp hyperfine structure. The third and fourth panels show
    the corresponding velocity and linewidth (FWHM) of each fit. The
    green contours are from ATLASGAL 870\micron at 0.31\,Jy, 0.46\,Jy,
    and 0.61\,Jy, continuing in steps of 0.3\,Jy. In all three maps the velocity resolution is smoothed to 0.4\,km/s.}
  \label{fig:n2hp_param_mopra3}
\end{figure*}

The left panels of Figs. \ref{fig:n2hp_param_nobeyama} through
\ref{fig:n2hp_param_mopra3} display the PACS 70\micron maps with
the long-wavelength dust
continuum contours on top, the second-left panel is the \nhp column density. 
They clearly show that the \nhp detection and column density agrees in general
with the measured dense gas emission, almost independent of the
evolutionary state of the clump. 

The southern component of IRDC\,G11.11 is peculiar, see
Fig. \ref{fig:n2hp_param_mopra1}. While for the northern component the molecular
gas traced by \nhp agrees quite well with the cold gas traced by thermal dust emission, the two dense gas tracers disagree for the southern part. Comparing the brightest peak in the
ATLASGAL data with the column density peak of the \nhp emission, we find a
positional difference of 37\arcsec, which is on the order of the beam
size. Since the northern and southern component have been observed
independently, a pointing error might explain the offset. However, before and
in between the OTF observations we checked the pointing and the offset
is considerably larger than the anticipated pointing uncertainty. Therefore,
we cannot explain the spatial offset of the southern map well.

For IRDC\,18182, the bright northwestern component is connected to
IRAS\,18182-1433 at a velocity of 59.1\,km/s \citep{Bronfman1996} and a
distance of 4.5\,kpc \citep{Faundez2004}. Instead, the region of interest is
the IRDC in the southeast at a distance of 3.44\,kpc with a velocity
of 41\,km/s \citep{Beuther2002a, Sridharan2005}. 

IRDC\,18308 has been selected within this sample for its infrared dark
cloud north of the HMPO IRAS\,18308-0841. At its distance of 4.4\,kpc we did not
detect the \nhp emission from the IRDC with the velocity resolution of
0.2\,km/s. To overcome the sensitivity issue we smoothed the velocity
resolution to 0.4\,km/s and were then able to trace the dense gas of the IRDC
within IRDC\,18308. For IRDC\,18306 the situation is similar,
we traced the HMPO, but not the IRDC. Unfortunately, even with a velocity
resolution of only 0.4\,km/s we were unable to detect \nhp from the
IRDC. Therefore, we excluded IRDC\,18306 from the discussion and show
its dense gas properties in Appendix \ref{fig:n2hp_param_18306at0.4kms}. 
To offer a better picture of the different regions we display in
Figs. \ref{fig:n2hp_param_nobeyama} through \ref{fig:n2hp_param_mopra3} the results from the smoothed maps, where
helpful. Because the coverage for many clumps is sufficient in
the higher resolution data, we used the 0.2\,km/s data for our
analysis.

The total gas peak column densities over
the 19.2\arcsec APEX 870\micron beam as given in Table \ref{tab:clump_list} range
from 1.4\e{22}\,cm$^{-2}$ to 8\e{23}\,cm$^{-2}$, and the median averaged peak
column density of clumps that have been mapped is
2.6\e{22}\,cm$^{-2}$. When we consider only clumps for which the peak
position has a detected \nhp signal, the median averaged peak column density
becomes 3.0\e{22}\,cm$^{-2}$. For the lower limit one needs to keep in mind
that we require a minimum column density threshold of
1.0\e{22}\,cm$^{-2}$ for a clump to be detected. The upper
limit is set by IRDC\,18454-mm1 (adopted from W43-mm1, \citealt{Motte2003, Beuther2012}), the brightest clump within
IRDC\,18454 and a well-known site of massive star formation. All other clumps with peak column densities
higher than 1\e{23}\,cm$^{-2}$ (IRDC\,18151-1, IRDC\,18182-1, and
G\,28.34-2) host evolved cores and could be warmer than
20\,K. Nevertheless, to calculate the column
densities we assumed a constant average dense gas temperature of
20\,K. While this is appropriate for most IRDCs in this sample with
ongoing early star formation (cf. point sources in
\citealt{Ragan2012}), using a higher temperature would decrease their peak column
densities. With the exception of W43-mm1, the upper limit of column
densities found within our survey's sources then becomes $\sim$\,1\e{23}\,cm$^{-2}$ on scales of the beam size.



\subsection{\nhp abundance}
\label{sec:results_nhp_abundance}
\begin{figure*}[tbp]
  \includegraphics[width=1.\textwidth]{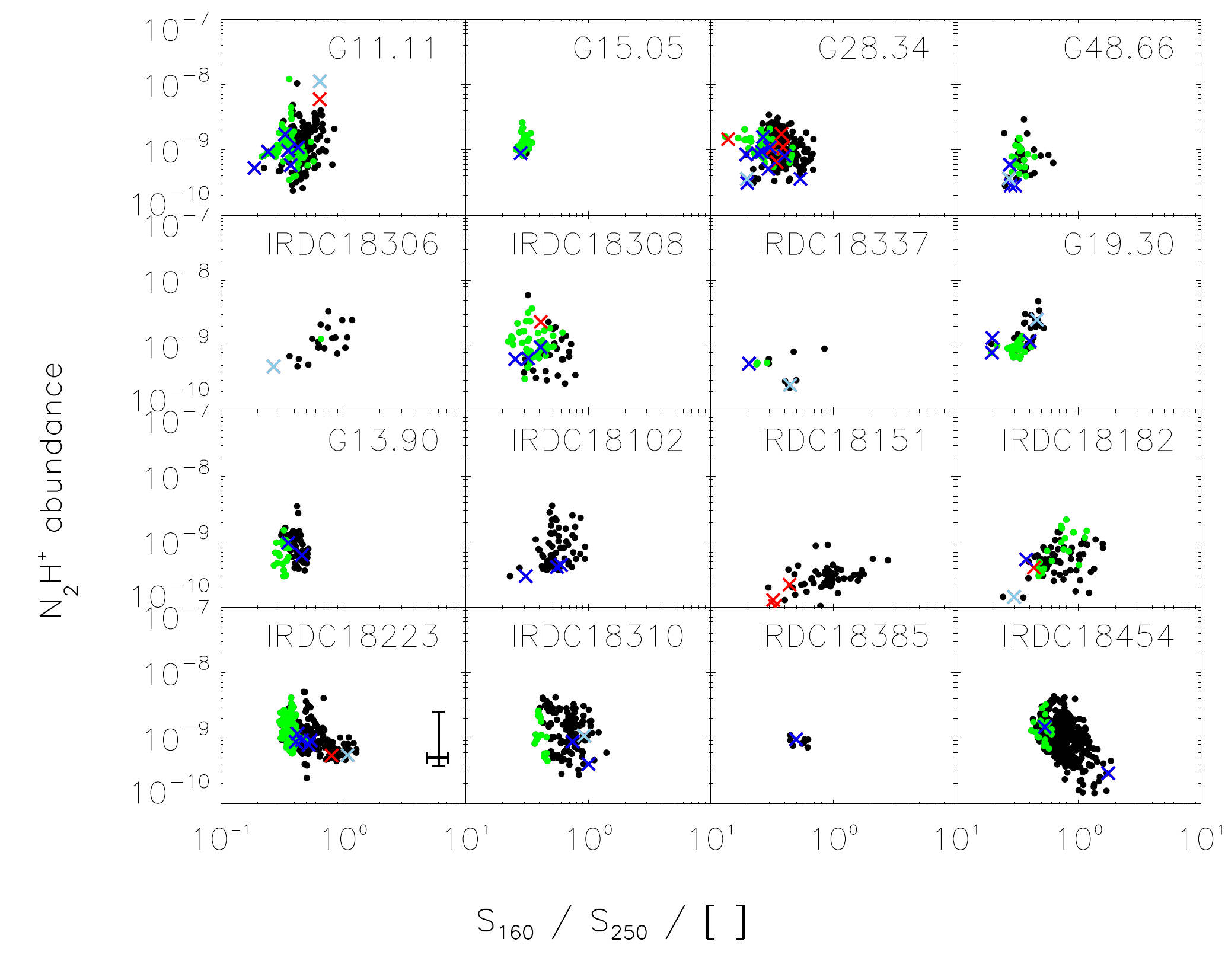}
  \caption{\nhp abundance ratio over the color index between
    160\micron and 250\micron. Marked by green dots are pixels that lie within
    IRDCs. Overplotted with red Xs are all mapped PACS sources. Blue crosses also have a 24\micron detection, while the
    light blue dots represent source that are saturated at
    24\micron. The uncertainties given for IRDC\,18223 are representative for
    all regions.}
  \label{fig:abundance_over_coltemp}
\end{figure*}

To study the details of the correlation between the dense gas and the
related \nhp column density, Fig. \ref{fig:abundance_over_coltemp}
shows the pixel-by-pixel
correlation between the \nhp abundance ratio versus the flux ratio between the
{\it Herschel} 160\micron and 250\micron bands. One should keep in mind that as explained in
Sect. \ref{sec:explain_abundance}, the abundance ratio refers to different beam
sizes, depending on the telescope that was used for the
observations. The flux ratio of the two PACS bands, or color temperature, can be considered as a
proxy of the dust temperature. For higher temperatures, the peak of the SED
moves to shorter wavelengths, the 160\micron becomes stronger relative to the
250\micron flux. Therefore, higher temperatures have higher FIR flux ratios. To derive proper temperatures, a pixel by pixel SED fitting is required, which
will be done in an independent paper (Ragan et al., in prep.). 
A known problem in the context of {\it Herschel} data are the unknown
background flux levels. Since we only discuss trends within individual
regions, we can safely neglect this problem. Thus, the flux ratios
between the different regions are not directly comparable. 

To allow a comparison between IRDCs and regions that do not show up in
extinction, we marked pixels that lie within regions of high extinction in
green. These regions were selected by visually identifying the 70\micron flux
levels below which extinction features are observed. Because the 70\micron dark
regions have no sharp boundaries, the chosen levels are not
unambiguous. 

Figure \ref{fig:abundance_over_coltemp} shows no strong overall correlation between the \nhp abundance
and the flux ratio. 
On the one hand, IRDC\,18223 is an example where there seems to be a correlation between the
temperature and the \nhp abundance. The highest abundance ratios are found at
FIR flux ratios of the bulge of the pixel distribution, while toward higher
temperatures the abundance seems to decrease systematically.
On the other hand, in G11.11 the \nhp abundance varies over two
orders of magnitude, but shows no correlation with the FIR flux
ratio. For example, in G11.11 and G28.34, there seems to be a trend
that the \nhp abundances become even larger toward the edges of our \nhp
mapping. However, in these regions we are limited by the sensitivity
of our dust measurements. 

Marked by an X in Fig. \ref{fig:abundance_over_coltemp} are the pixels
containing PACS point sources. In addition, we distinguished between sources that have only
been detected at 70\micron and longwards, MIPS-dark sources, and PACS sources with a
24\micron counterpart, so-called MIPS-bright sources. When the 24\micron image was saturated at the given
position, sources were considered as MIPS-bright as well. 
From the figure it can be seen that most embedded PACS sources have \nhp
abundance ratios below 1\e{-9}, but there seems to be no
correlation between the dust temperature on scales of the beam size
and the presence of embedded PACS point-sources.


\subsection{Large-scale velocity structure of clumps and
  filaments}
\label{sec:results_velo}
The velocity structure of the \nhp gas is shown in the third panel
(second from the right) of Figs. \ref{fig:n2hp_param_nobeyama} to
\ref{fig:n2hp_param_mopra3}. As explained in
Sect. \ref{sec:hyperfine_fitting}, we fit a single \nhp hyperfine
structure to every pixel and display the resulting peak velocity.

The southeastern region in the map of IRDC\,18182 is the IRDC in the EPoS sample. It is known that IRAS
18182-1433, originally targeted by \citet{Beuther2002a}, and the IRDC
have different velocities and therefore are spatially distinct. All
other mapped sources show velocity variations of only a few km/s and
are therefore coherent structures.

%

The source with the largest spread in velocity is IRDC\,18454\,/\,W43. The
mapped regions in the west, beyond W43-mm1 toward W43-main (which was not mapped), have the lowest velocities at below 93\,km/s, then there is a
velocity gradient across W43-MM1 ending east at 97.4\,km/s, and in the far
south there are two clumps at 100\,km/s. However, the velocity map was
derived by fitting a single \nhp hyperfine structure to each
spectrum. \citet{Beuther2007a} and \citet{Beuther2012} have shown that at least at high spectral and
spatial resolution, clump IRDC\,18454-1 has two
velocity components seperated by about 2\,km/s. Trying to fit each clump peak position with two \nhp
hyperfine structures, we find that for six continuum peaks, the \nhp
spectrum is better fitted by two independent components. For simplicity,
  we did not include the additional velocity component found toward
  IRDC\,18454 at $\sim$\,50\,km/s \citep{NguyenLuong2011}.

While a more detailed description of the double-line fits are
presented in Sect. \ref{sec:results_linewidth}, we here note
that the mapped line velocity represents either a single
component if one is much brighter than the other, or an average velocity of both. Therefore,
the uncertainties for IRDC\,18454 are significantly larger than for the
other regions. Nevertheless, the large-scale velocity gradient is no artifact, but
is evident in the individual spectra. 

Studying the velocity maps in Figs. \ref{fig:n2hp_param_nobeyama} through
\ref{fig:n2hp_param_mopra3} in more detail, we find two different
patterns of velocity structure. 
On the one hand, we find independent
clumps that lie within the same region and may interact now or in the future, but are currently separate entities in velocity. A good example is IRDC\,18151
plotted in Fig. \ref{fig:n2hp_param_mopra2}. Using 870\micron as dense
gas tracer we resolve two clump complexes separated by $\sim$ 1\,pc. Across each
clump there are only weak velocity variations, but the east and west
complex are separated in velocity space by 3.3\,km/s. Velocity
differences on the order of a few km/s between clumps within the same
structure are common and we consider such clumps as spatially connected.

On the other hand, we find smooth velocity gradients across larger
structures. Clumps that may be at different
velocities have connecting dense material with a continuous velocity
transition in between. To exclude overlapping, but independent clumps for
which either the spatial resolution or our mapping technique mimic a smooth
transition we searched for double-peaked velocities and a broadening
of the line width in such transition zones.

The velocity maps of G15.05, IRDC\,18102, IRDC\,18151, IRDC\,18182, and G48.66,
Figs. \ref{fig:n2hp_param_nobeyama} - \ref{fig:n2hp_param_mopra3}
immediately reveal that these complexes have no velocity
gradients in the gas above our detection limits given in Table \ref{tab:clump_properties} and therefore are of the first type. 
A summary of the clump classification is given in Table
\ref{tab:clump_properties}. \input{clump_properties_include_outflows}

For IRDC\,18310, shown in Fig. \ref{fig:n2hp_param_nobeyama}, the
velocity map shows that the IRAS source in the south has a velocity of
83.2\,km/s (see also Table \ref{tab:clump_list}), while the northern complex has
higher velocities. Nevertheless, the velocity spread suggests an association
between both clumps. In addition, the northern component itself has different
velocities toward the east and south, with 86.1\,km/s and 84.3\,km/s. In
between there is a narrow transition zone with a spatially associated increase
in the linewidth (see very right panel of
Fig. \ref{fig:n2hp_param_nobeyama}). The increase in linewidth suggests that
there is indeed an overlap of two independent velocity components and not
a large-scale velocity transition. Using the unsmoothed Nobeyama image at a
velocity resolution of 0.2\,km/s, the spectra suggest two independent
components. Therefore, IRDC\,18310 consists of three clumps, each
showing no resolved velocity structure.

IRDC\,18151, shown in Fig. \ref{fig:n2hp_param_mopra2}, consists of two clumps
at different velocities. While the velocities of the western clump agree
within 0.5\,km/s, the eastern clump has a velocity gradient from the
southeast to the northwest with a change in velocity of more than
1\,km/s. At the velocity resolution of 0.2\,km/s we did not detect the
lower column density transition region and cannot exclude a smooth transition
across both clumps. To overcome the sensitivity problem we smoothed the \nhp data
to a resolution of 0.4\,km/s. Still, only a single pixel with a good enough
signal-to-noise ratio connects the two dense gas clumps. As we will discuss in
Sect. \ref{sec:discuss_velo}, this pattern does not suggest a smooth transition.

For IRDC\,18454, IRDC\,18308, G11.11, G19.30, and G28.34 we found smooth
velocity gradients. One of the steepest smooth velocity gradients of the 
sample is found toward the southern part of IRDC\,18308, across the
HMPO. Although there is an increase in the linewidth map, even in the
unsmoothed higher resolution data we were unable to find two independent
components. Across 3.2\,pc the velocity changes by 2.4\,km/s, resulting
in a velocity gradient of 0.8\,km/s\,/pc. The change in velocity occurs parallel to the elongation
of the ATLASGAL 870\micron emission. 

The velocity gradient in the northern part of G11.11 is as
clear as for IRDC\,18308 and parallel to the extinction of G11.11. While the far
southern tip of the northern filament has a slightly different
velocity, up north it has an almost
constant velocity up to the point G11.11-1 and then shows a strong but
smooth gradient beyond.

As mentioned before, if considering the length and change in velocity alone, the samples
steepest velocity gradient is found for IRDC\,18454. Over a length of
8.4 pc the gradient is 0.9\,km/s\,/pc, but in between the two
endpoints the velocity is not increasing monotonically.

IRDC\,18223 shows significant changes in the velocity field, but
is not listed among the clumps with smooth velocity
gradients. The changes of the velocity are on 0.5\,pc - 1\,pc scales
and show no clear pattern. Nevertheless, at the given velocity
resolution of 0.5\,km/s we found no overlapping independent
\nhp components. Therefore, all the gas on the scales we trace seems
to be connected. It is worth noting that the two southern clumps, IRDC\,18223-2
and IRDC\,18223-3, have a gradient along the short axis of the filament, which
might be interpreted as evidence of rotation. Velocities in the east are higher than in the
west. In contrast, although less well mapped, the IRAS source in the north,
IRDC\,18223-1, has a velocity gradient along the short filament axis as well, but
in opposite direction; velocities in the west are higher than in east.

For G13.90, IRDC\,18385, IRDC\,18306, and IRDC\,18337 we lack the
sensitivity to draw a conclusion. While IRDC\,18385, and IRDC\,18306 are very poorly mapped, for G13.90 and IRDC\,18337 we mapped the main
  emission structures, but with the available sensitivity we did not trace the gas
  in between the dense clumps. In both sources, the detected clumps have
  different velocities, with a gradient across the clumps in
  IRDC\,18337. Since we did not trace the gas in between the dense
  clumps, we are unable to assess whether the velocity transitions are smooth, or if
  the clumps have no connection in velocity space.



\subsection{\nhp linewidth in the context of young PACS sources and
  column density peaks}
\label{sec:results_linewidth}
The right panels of Figs. \ref{fig:n2hp_param_nobeyama} to
\ref{fig:n2hp_param_mopra3} show the fitted linewidth (FWHM) for the mapped
regions. 
The distribution of the linewidth is very different for each region
and density peak. While it increases toward some of the submm
peaks (e.g. IRDC\,18102, IRDC\,18182-1), for others the peak of the
linewidth is on the edge of the submm clumps
(e.g. IRDC\,18223-1, IRDC\,18223-3, G19.30). The IRDCs
for which we detect \nhp and that have no embedded/detected PACS
source often have a narrower linewidth than other clumps of the same
region with embedded protostars. 

A brief description of the linewidth distribution of each region is given in
Appendix \ref{app:full_dv}. In the following we discuss a few
interesting or notable examples. 

While the linewidth in IRDC\,18223 significantly increases toward IRDC\,18223-2, the linewidth toward IRDC\,18223-1, a well-studied
HMPO \citep[][]{Sakai2010}, and
IRDC\,18223-3, an object known to drive a powerful outflow
\citep{Beuther2007a, Fallscheer2009}, increases toward the edges of the dust
continuum. Compared with other regions of IRDC\,18223, the linewidth at
IRDC\,18223-3 is broader, but it becomes even broader in the northwest. This aligns very well with the outflow found by
\citet{Fallscheer2009} and can be explained by
it (for the outflow direction see the first row of
  Fig. \ref{fig:n2hp_param_nobeyama}, right panel). 
IRDC\,18223-1 was originally identified as IRAS\,18223-1243 and
is bright at IR wavelengths (down to K band). However, typical tracers
of ongoing high-mass star formation such as cm emission, water and methanol
masers, or SiO-tracing shocks are not detected
\citep{Sridharan2002, Sakai2010}. Only the CO line wings found by
\citet{Sridharan2002} are indicative of outflows, which might
explain the bipolar broadening of the \nhp linewidth. Nevertheless, despite its prominence at IR wavelengths and with the
luminosity of the PACS point source at its peak of 2000\,L$_{\odot}$
\citep[point source 8 in][]{Ragan2012}, the
linewidths at the continuum peak are not exceptional within this
region. In contrast, although IRDC\,18223-2 is detected at near IR
wavelengths as well and the PACS point source at its center has a
luminosity of only 200\,L$_{\odot}$, the linewidth is 2.5\,kms/s
compared with 1.9\,km/s for IRDC\,18223-1. Because \citet{Beuther2007a}
found no SiO toward IRDC\,18223-1/2 we exclude a strong outflow, and the reason for the
line broadening is not clear at all. Thus one should keep in mind that IRDC\,18223-2 has not been
investigated in such great detail and we cannot entirely exclude an outflow.

\begin{figure}[tbp]
  \includegraphics[angle=-90,width=.5\textwidth]{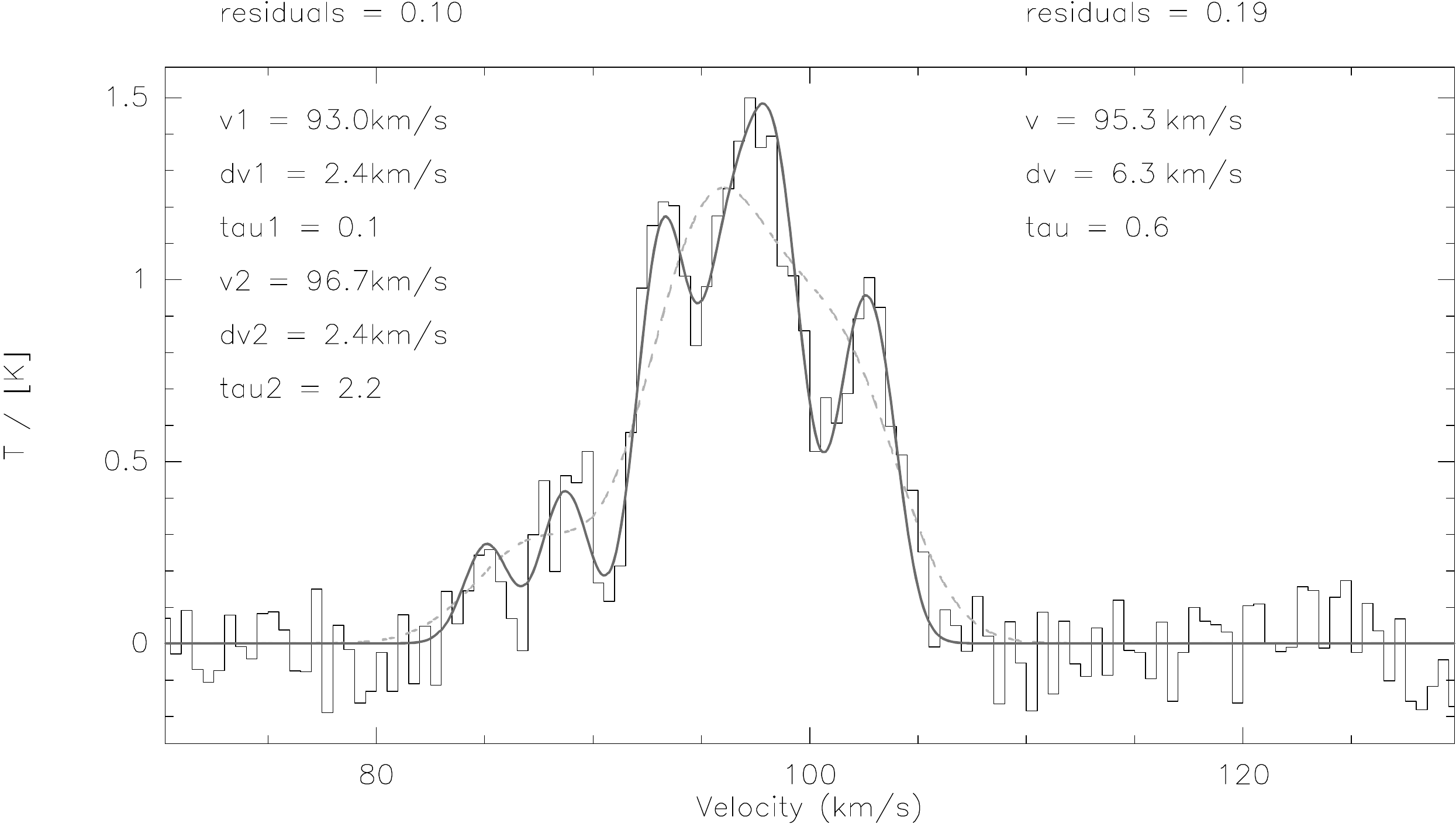}
  \caption{Spectra of IRDC\,18454-4 \citet{Beuther2007a, Beuther2012}. While the dashed line shows the single-component fit with the fitting parameters to the right, the solid
    line is the two-component fit with its fitting parameters to the
    left. The residuals are the results of the {\it minimize} task in {\it CLASS}.}
  \label{fig:double_spec_18454}
\end{figure}

We excluded IRDC\,18454 from the analysis of the linewidth because, as
mentioned in Sect. \ref{sec:results_velo}, we found multiple
velocity components toward several positions. 
Figure \ref{fig:double_spec_18454} displays an example of an \nhp spectrum that
compares a single-component fit with a double-component fit. Comparing the residuals of
the two different fits as calculated by {\it CLASS},  for the six
clumps in which we found two independent components the
residuals are on average reduced by 30\%. 
For all two-component fits, the linewidth decreases compared with a
single-component fit. 
Still, the linewidths are on average
broader than for the other clumps listed in Table
\ref{tab:clump_list}. 

Similar double velocity component fits toward the peaks are otherwise only
possible in G11.11. Here, eight of the clumps are fit better by two
independent \nhp components. Different from IRDC\,18454,
the linewidth of the two components becomes on average narrower than the
linewidth of other clumps in the sample. In addition, the improvement of the
residuals is only 20\%. Therefore it is unclear whether two
independent components are present or if the fit is simply improved
because of the larger number of free parameters.
A systematic study of the multiple components is beyond the
scope of this paper. 

For G13.90, IRDC\,18385, IRDC\,18306, and IRDC\,18337 the mapped areas
are not sufficient to draw conclusions.

\begin{figure*}[tbp]
  \includegraphics[width=1.\textwidth]{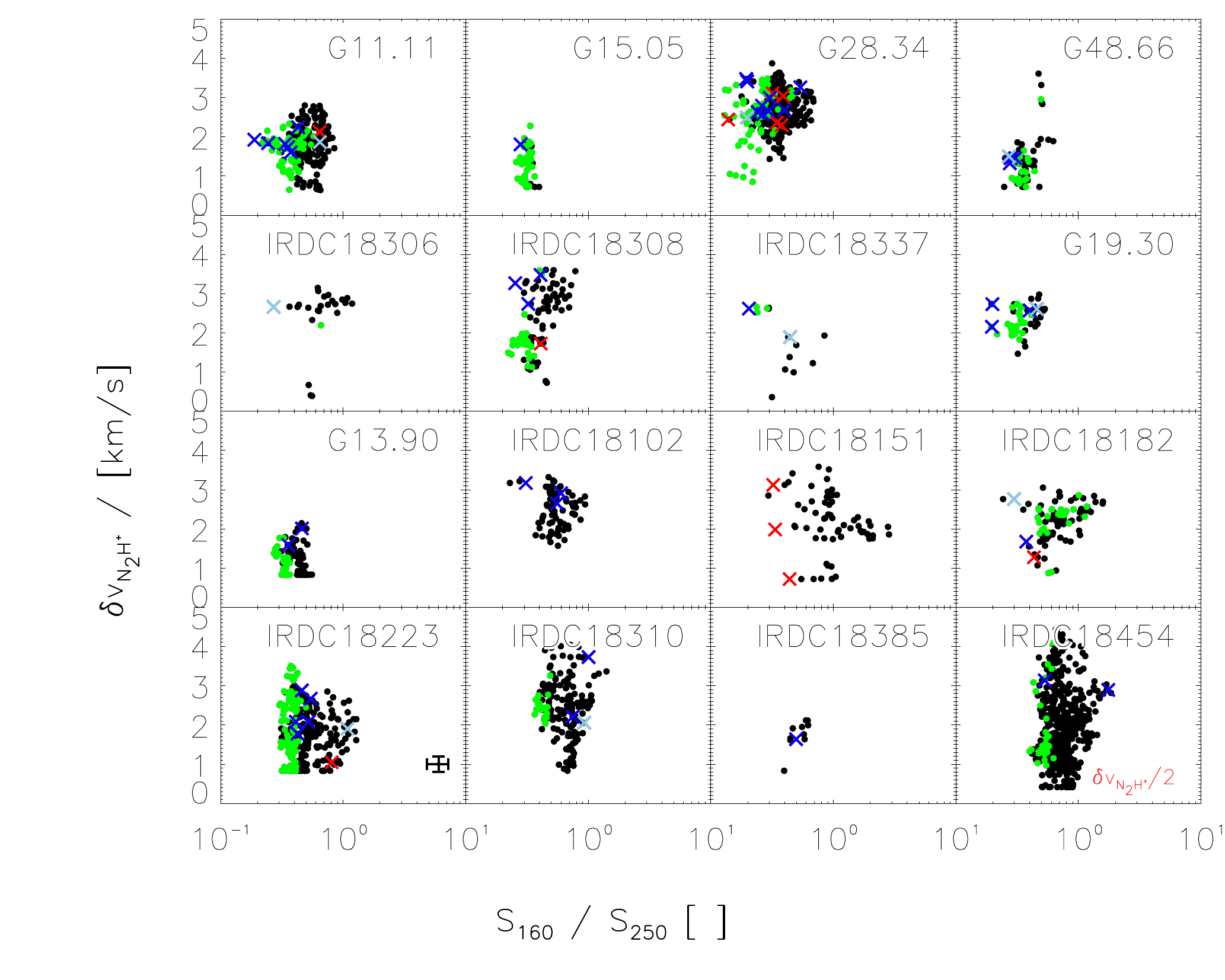}
  \caption{\nhp linewidth versus the color index for the
    160\micron over the 250\micron band. Marked by green dots are pixels that lie within
    IRDCs. Overplotted with red Xs are all PACS sources that
    were mapped. Blue dots also have a 24\micron detection, while the
    pale blue dots represent sources that are saturated at
    24\micron. The uncertainties given for IRDC\,18223 are
      representative for all regions. For IRDC\,18454 the linewidths were multiplied by a factor of 0.5 to fit the data points into the plotting range.}
  \label{fig:dv_over_coltemp}
\end{figure*}
\begin{figure*}[tbp]
  \includegraphics[width=1.\textwidth]{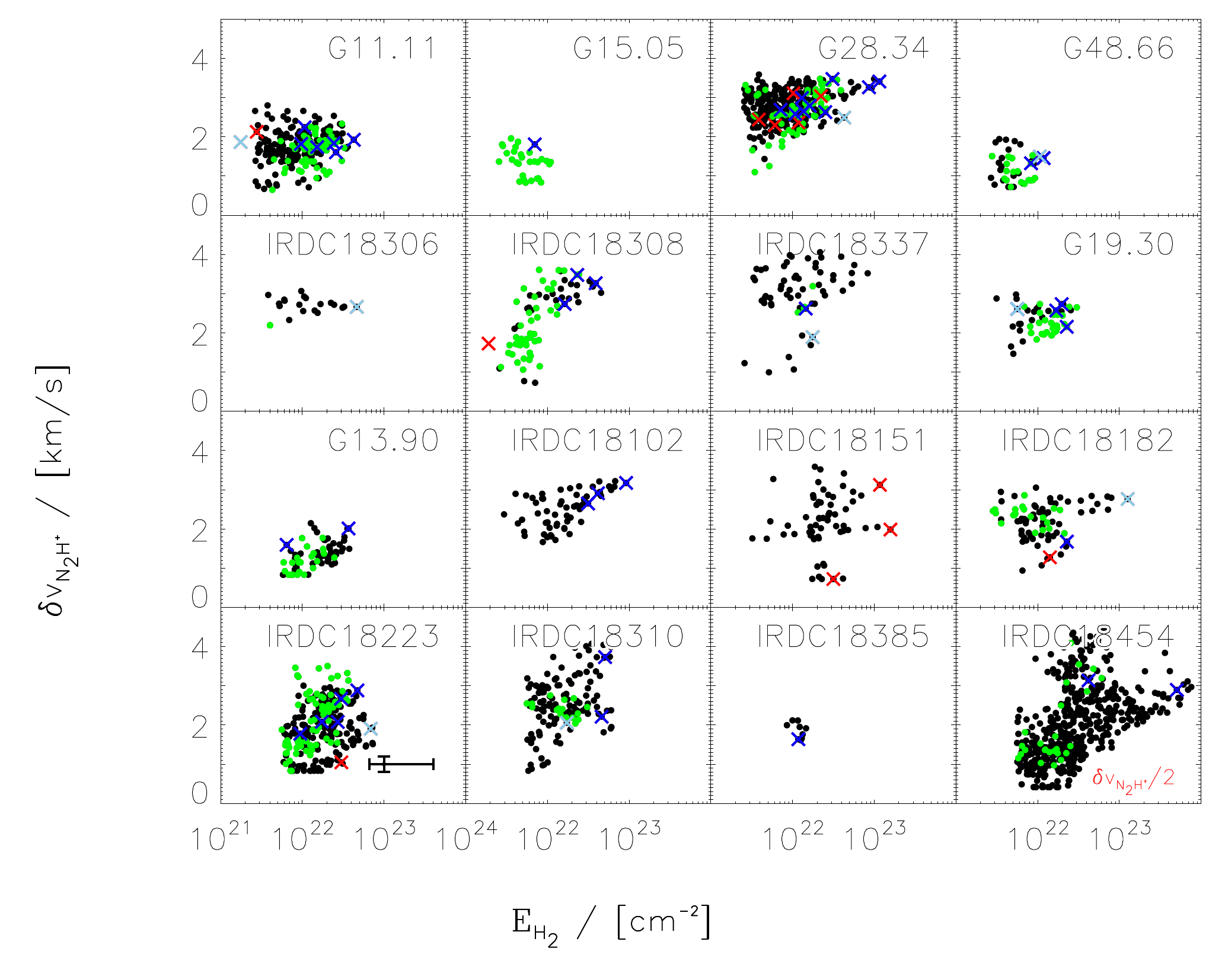}
  \caption{Plot of the \nhp linewidth versus the dust column
    density. Marked by green dots are pixels that lie within
    IRDCs. Overplotted with red Xs are all PACS sources that
    were mapped. Blue dots also have a 24\micron detection, while the
    pale blue dots represent sources that are saturated at
    24\micron. The uncertainties given for IRDC\,18223 are
      representative for all regions. For IRDC\,18454 the linewidths were multiplied by a factor of 0.5 to fit the data points into the plotting range.}
  \label{fig:dv_over_colden}
\end{figure*}
Similar to Fig. \ref{fig:abundance_over_coltemp},
Fig. \ref{fig:dv_over_coltemp} shows the relation between the linewidth (FWHM) of \nhp and the
color index. Because the color index is a proxy of the temperature, one might expect a
correlation between these two quantities, but there is no correlation at all. Figure \ref{fig:dv_over_colden} plots the \nhp
linewidth versus the \htwo column density, but again, we find no correlation. 

\begin{figure}[tbp]
  \includegraphics[width=.5\textwidth]{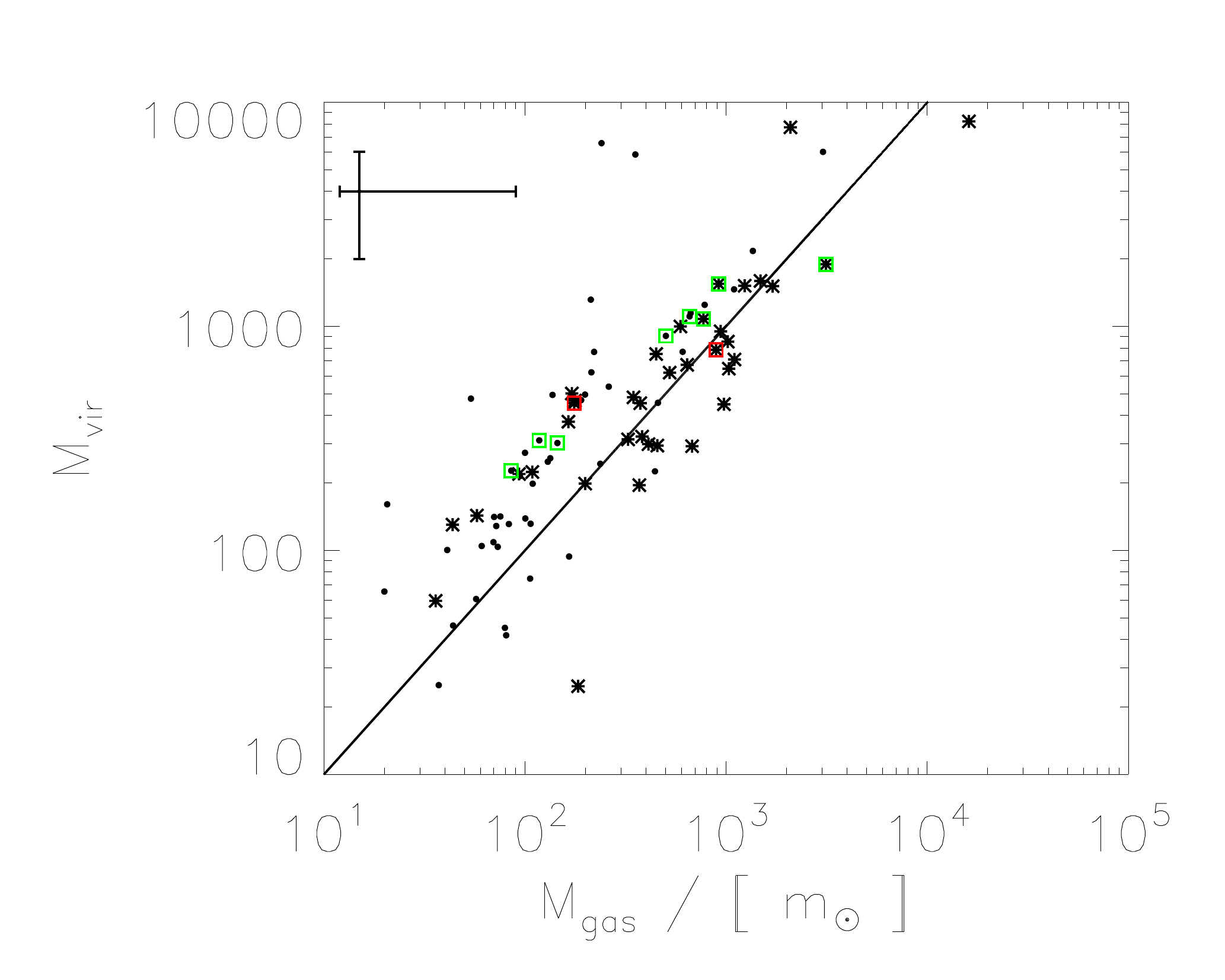}
  \includegraphics[width=.5\textwidth]{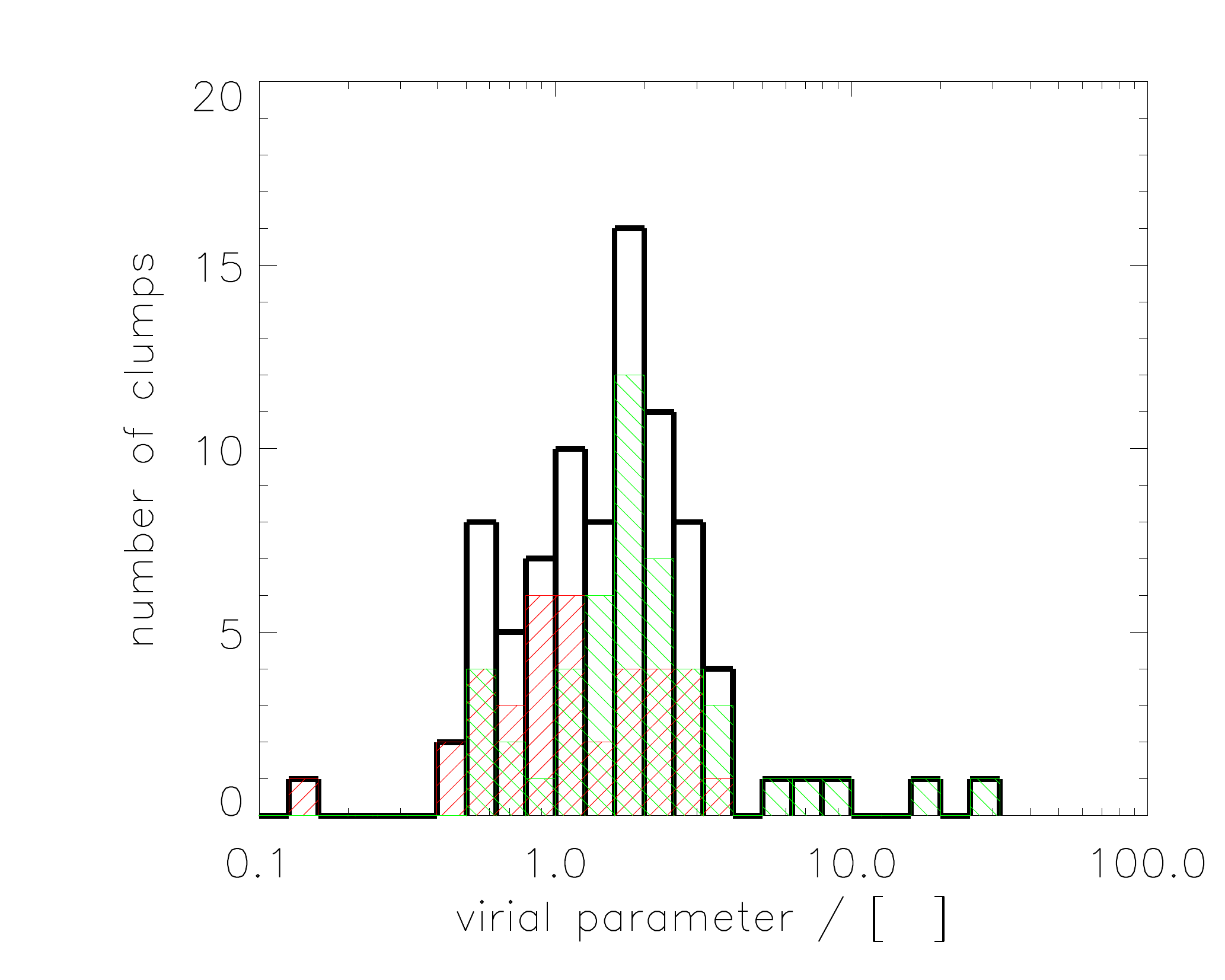}
  \caption{
Top panel: virial mass derived from the \nhp linewidth
    over the gas mass. The virial mass assumes a geometrical parameter of k=158, which is
    intermediate between k=126 for 1/$\rho^2$ and k=190 for 1/$\rho$. While the
    black dots indicate clumps without a PACS point source inside, the asterisks
    represent clumps with a PACS point source. Marked by green and red squares
    are the clumps of G28.34, with green boxes representing clumps in global
    infall, and red boxes representing clumps with signatures of outward moving
    gas (for details see \citealt{Tackenberg2013a}). The solid line indicates unity. Lower panel: histogram of the virial parameter
    $\alpha$. While the black histogram represents the full sample, the red and
    green histogram is the
    subset of clumps with and without a PACS point source, respectively.} 
  \label{fig:mvir_over_mgas}
\end{figure}
In the context of the linewidth and dust mass, the virial analysis can be used
to understand whether structures are gravitationally bound or are transient
structures. Following \citet{MacLaren1988}, we calculated the virial mass of
our clumps via M$_{vir}$\,=\,k\,R\,$\Delta$v$^2$. For the clump radius R we
used the effective radius calculated by CLUMPFIND. The geometrical
parameter k depends on the density distribution, with k\,=\,190 for
$\rho$\,$\sim$\,r$^{-1}$, and k\,=\,126 for
$\rho$\,$\sim$\,r$^{-2}$. \citet{Beuther2002a}, \citet{Hatchell2003}, and \citet{Peretto2006} found
typical density distributions in sites of massive star formation of $\rho$
$\propto$ r$^{\alpha}$ with $\alpha$\,$\sim$\,-1.6, in between both
parameters. While we list the virial mass for both parameters in Table
\ref{tab:clump_list}, we usde the intermediate value of k\,=\,158 in Fig. \ref{fig:mvir_over_mgas}.

The $\alpha$ parameter as defined in \citet{Bertoldi1992} is the ratio of the internal kinetic energy and the
gravitational energy. Their virial parameter as defined in
Eqn. 2.8a of \citet{Bertoldi1992} ($\alpha$\,=\,$\frac{5\sigma^2R}{GM}$)
without another geometrical parameter resembles a spherical distribution of
constant density. Because of the geometrical correction factor we applied to the mass
calculations, the presented virial parameters are smaller by a factor of
1.32. A histogram of the virial parameter is plotted in
Fig. \ref{fig:mvir_over_mgas}. 

If we assume the error on our linewidth to be lower than 15\%, the
uncertainties of the calculated virial mass are mainly determined by the
geometrical parameter k. The actual error on the given virial masses is
significantly larger since the calculation neglects all physical effects but
gravity and thermal motions (kinetic energy). For the conceptual quantity we
can neglect these effects and estimate the error to be $\sim$\,50\%. 

\section{Discussion of \nhp dense gas properties}
In the following, we discuss the kinematic properties of the sources we
mapped in \nhp, as described above.

\subsection{Dense clumps and cores}
The clump masses in the range of several tens of
\msun to a few thousands of \msun show that most regions have the
potential to form massive stars in the future, or show signs of
ongoing high-mass star formation. 
One should keep in mind that the
listed peak column densities are averaged over the beam. As has
been shown by \citet{Vasyunina2009} assuming an artifical r$^{-1}$
density profile, true peak column densities are higher by a factor of
20 to 40. This agrees with interferometric observations of
clumps within our sample \citep{Beuther2005a, Beuther2006,
  Fallscheer2011}. Therefore, all peak column densities become higher than
3\e{23}\,cm$^{-2}$, or 1\,g/cm$^2$. This reinforces the view that the mapped
clumps are capable of forming massive stars. 
The high column densities also agree with the detection of
\nhp as high-density gas tracer. 

\subsection{Abundance ratios}
To understand why the abundance of \nhp is expected to vary
with embedded sources or temperature, one needs to understand the formation mechanism. The
formation of \nhp works via H$_{3}^{+}$ which also builds the basis for the
formation of \hco from CO. Because of the high abundance of CO in cold
dense clouds, the production of \hco is initially dominant and consumes all
H$_{3}^{+}$. 
If during cloud contraction the temperatures become cold enough for
CO to freeze out, \nhp can be produced more efficiently and eventually becomes
more abundant than \hco.
The situation changes again when CO is released from the grains either due
to heating or due to shocks. The CO destroys the \nhp and forms \hco instead,
making \hco more abundant again. (For a more detailed discussion see
\citealt{Jorgensen2004}.) 

In summary, the early (more diffuse) cloud phase is
dominated by \hco, while the quiet dense clumps should be dominated by
\nhp. With the onset of star formation, \hco is becoming dominant again.

The EPoS sample mainly has been selected to cover regions of ongoing, but early star
formation. For this \nhp line survey, we selected regions covering all
evolutionary stages. Many of them have both infrared-quiet regions at the
wavelengths range covered previous to {\it Herschel} and well-known and luminous IRAS
sources. Together with the {\it Herschel} data, hardly any region of high column
density is genuinely infrared-dark. 

As a result of the \nhp evolution and the broad range of evolutionary
stages covered, we expect a wide range of \nhp abundance ratios. As has
been discussed in Sect. \ref{sec:results_nhp_abundance},
Fig. \ref{fig:abundance_over_coltemp} shows the correlation between the \nhp
abundance and the 160\micron to 250\micron flux ratio as a proxy of the
temperature. For all regions, the bulk of all pixels has \nhp abundances
ratios of 1\e{-9}. This agrees well with earlier studies of high-mass
star-forming regions \citep[][and references therein]{Vasyunina2011}. At the
same time, several regions (e.g. G11.11, G28.34, IRDC\,18454) show abundance
variations of two orders of magnitude. While this is a result of the various
evolutionary stages within each region, it is worth noting that it seems to
be uncorrelated to the flux ratio of 160\micron over 250\micron.

To correlate some areas with an evolutionary stage, in
Fig. \ref{fig:abundance_over_coltemp} we mark regions
that show up in extinction at 70\micron by green dots. 
As
Fig. \ref{fig:abundance_over_coltemp} shows, these regions are among
the coldest within each region. Nevertheless, high \nhp
abundances are found not only in IRDCs or cold regions.
In contrast to IRDC pixels which mark the earliest
and coldest evolutionary stages, the PACS-only sources mark regions in
which star formation is about to start (red), and the MIPS bright PACS
sources indicate ongoing star formation (blue). All pixels connected
to a PACS source have low \nhp abundances. Whether this is due to an
increase in temperature or probably to shocks is unclear. 

It has been shown in \citet{Ragan2012} that sources with a detected 24\micron counterpart are on
average warmer, more luminous, and more massive and that therefore a 24\micron counterpart
is indicative of a more evolved source. Nevertheless, the PACS core properties
in \citet{Ragan2012} show a large overlap between MIPS bright and dark
sources. Therefore, one cannot draw a clear conclusion on the evolutionary stage
(temperature, luminosity, or mass) based on a 24\micron detection alone. This
easily explains the exceptions, for instance, in G11.11. 


\subsection{Signatures of overlapping dense cores within clumps}
\label{sec:discuss_double_line}
We have described in Sect. \ref{sec:results_velo} two independent
velocity components toward six of the IRDC\,18454 continuum peaks, as
well as seven clumps in G11.11 with double-peaked \nhp lines. The two components
have velocity offsets of only a few km/s. Since the hyperfine
structure of \nhp includes an optically thin component, we can exclude opacity
and self-absorption effects, a common feature in dense star-forming regions. The two independent
velocity components within IRDC\,18454 have previously been reported by
\citet{Beuther2007a, Beuther2012} and Ragan et al. (in prep.) found
multiple velocity components toward G11.11. Combining our \nhp Nobeyama
data with PdBI observations at $\sim$\,4\arcsec, \citet{Beuther2013} revealed multiple independent velocity components
toward IRDC\,18310-4. These are not resolved within the Nobeyama data alone
at the spatial resolution of 18\arcsec. Similar multicomponent velocity signatures have
been found in high spatial resolution images of dense cores in
Cygnus-X \citep{Csengeri2011, Csengeri2011a} and toward IRDCs by \citet{Bihr2013}. Therefore, it seems to be a common feature in high-mass
star-forming regions. 

Using radiative transfer calculations of collapsing high-mass star-forming
regions, \citet{Smith2013} showed that such double-peaked line
profiles may be produced by the superposition of infalling dense cores. Therefore,
in high-resolution studies, which filter out the large-scale emission, multiple
cores along the line of sight can be detected. However, comparing our beam
sizes of $\sim$\,0.5\,pc for IRDC\,18454 and $\sim$\,0.8\,pc for G11.11 to
typical sizes of cores below 0.1\,pc, the larger-scale clump gas is expected to probably
dominate our signal. Therefore it is clear that multiple velocity components
due to cores are more likely to be identified in high spatial resolution
imaging. 
While IRDC\,18454 is at the intersection of the spiral arm and
the Galactic bar, and therefore exceptional in many aspects, G11.11 is most likely
a more typical high-mass star-forming region, similar to what has been
simulated by \citet{Smith2009} and \citet{Smith2013}. If the double-peaked
line profiles originate from two dense cores within our beam, as suggested by
\citet{Smith2013}, the cores within G11.11 would need to be extremely dense
or large.
Instead, it seems more realistic that we detect the gas of the clump as one
velocity component, and the second component is produced by an embedded single
core of high-density contrast moving relative to its parent clump.
For IRDC\,18454 we found double velocity spectra even inbetween the
peak positions. This suggests that the components are coming from two
overlapping sheets that are close in velocity. It is unclear whether these
sheets are interacting or not.



\subsection{Accretion flows along filaments?}
\label{sec:discuss_velo}
In Sect. \ref{sec:results_velo} we presented the velocity structure of
the 16 observed high-mass star-forming regions. As we described, five
complexes have no velocity structure, while six regions have smooth
large-scale velocity gradients. The velocity structure of IRDC\,18223 is more
complex and does not fit into either of these categories. For four regions we lack the sensitivity
to draw a conclusion.


Despite the two general appearances, the large-scale velocity structure of the
clumps is very diverse. In general, structures larger than 1\,pc usually
show some velocity fluctuations. These can be either steady and smooth, or
pointing to separate entities. It is worth noting that the
  physical resolution of the \nhp observations ranges from 0.1\,pc to
  1.0\,pc, with an average of 0.3\,pc for the 18\arcsec Nobeyama beam,
  and 0.7\,pc for the 46\arcsec with MOPRA. Therefore, we are unable
  to resolve smaller structures, and the 1\,pc limit is
  observationally set. In fact, velocity fluctuations on smaller
  scales are still likely. However, the observations show that on the
  clump scale, some clumps do show gas motions, while others are
  kinematically more quiescent. 
  High-resolution studies, for example that of \citet{Ragan2012a}, have proven for some regions that gas motions continue on smaller scales.


\begin{figure}[tbp]
  \includegraphics[width=.5\textwidth]{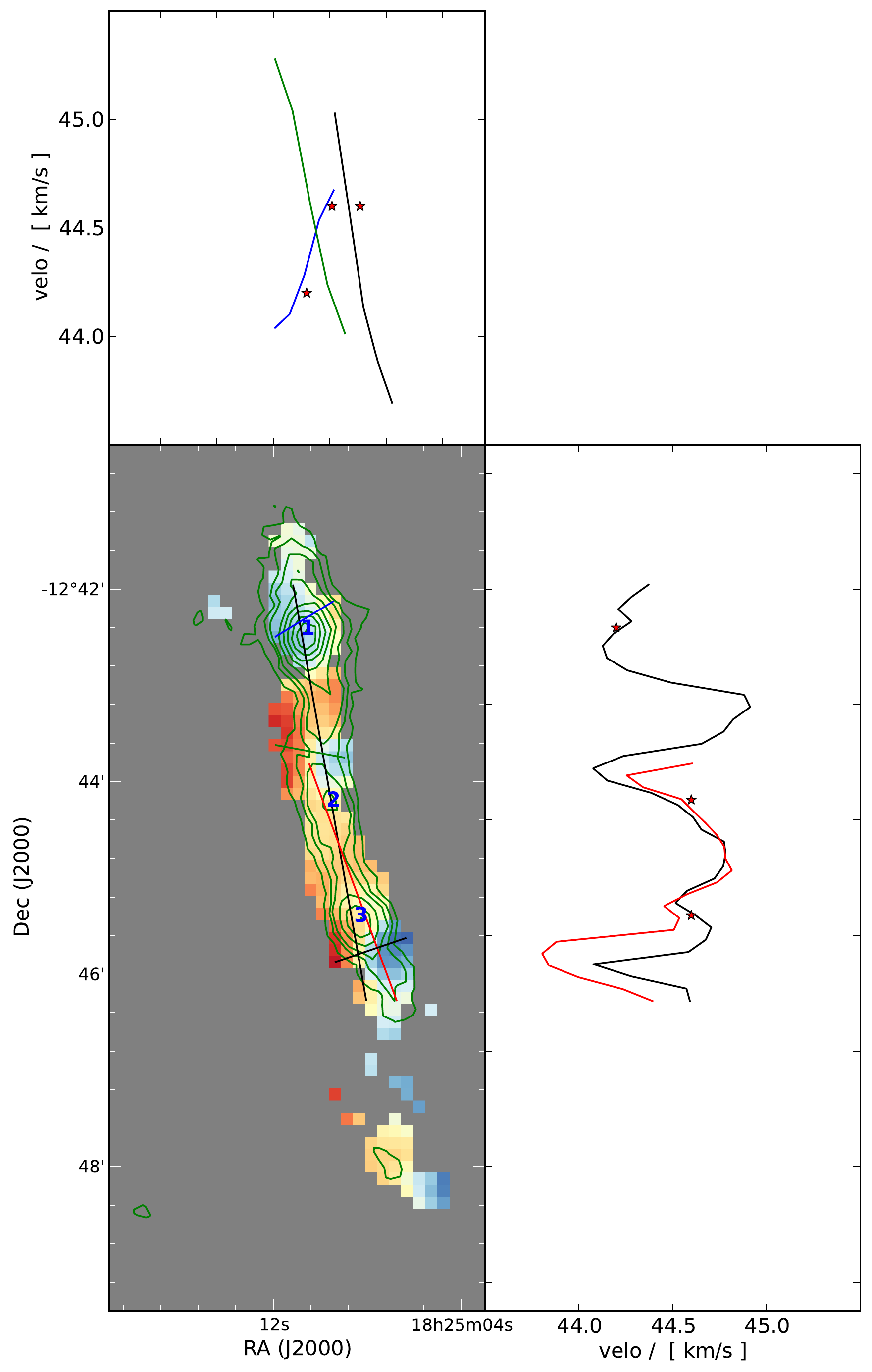}
  \caption{Profile of the \nhp velocity of IRDC\,18223. The left panel
    shows the velocity map with contours from ATLASGAL superimposed (see also
    Fig. \ref{fig:n2hp_param_nobeyama}). The
    right and top panels show the velocity cuts along the lines marked on
    the velocity map. The stars mark the velocities of the clump peaks.}
  \label{fig:velo_profile_18223}
\end{figure}
\begin{figure}[tbp]
  \includegraphics[width=.5\textwidth]{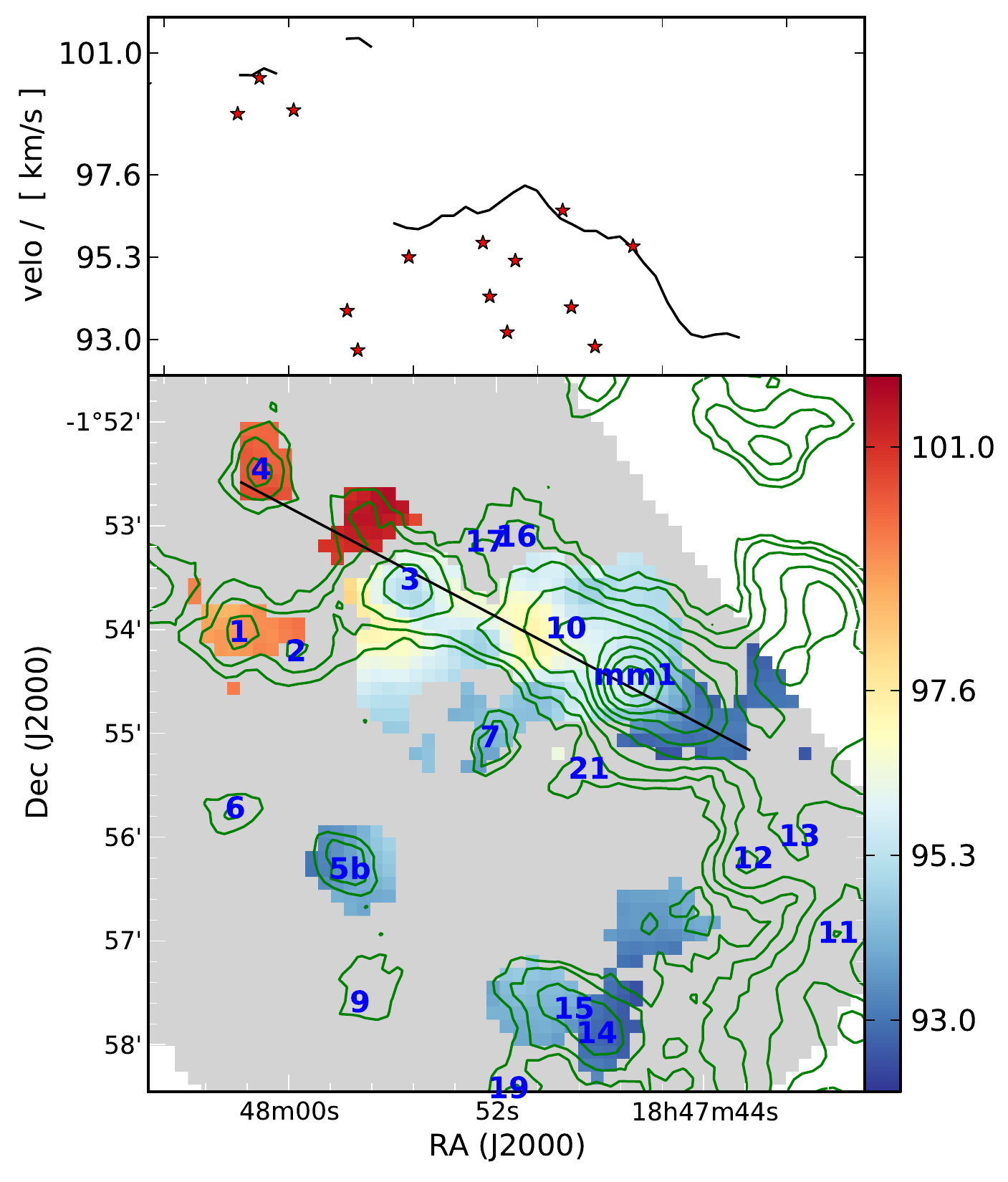}
  \caption{Profile of the \nhp velocity of IRDC\,18454. The left panel
    shows the velocity map with contours from ATLASGAL superimposed (see also
    Fig. \ref{fig:n2hp_param_nobeyama}). The
    top panel shows the velocity cut along the line marked on the
    velocity map. The stars mark the velocities of the clump peaks.}
  \label{fig:velo_profile_18454}
\end{figure}
\begin{figure}[tbp]
  \includegraphics[width=.5\textwidth]{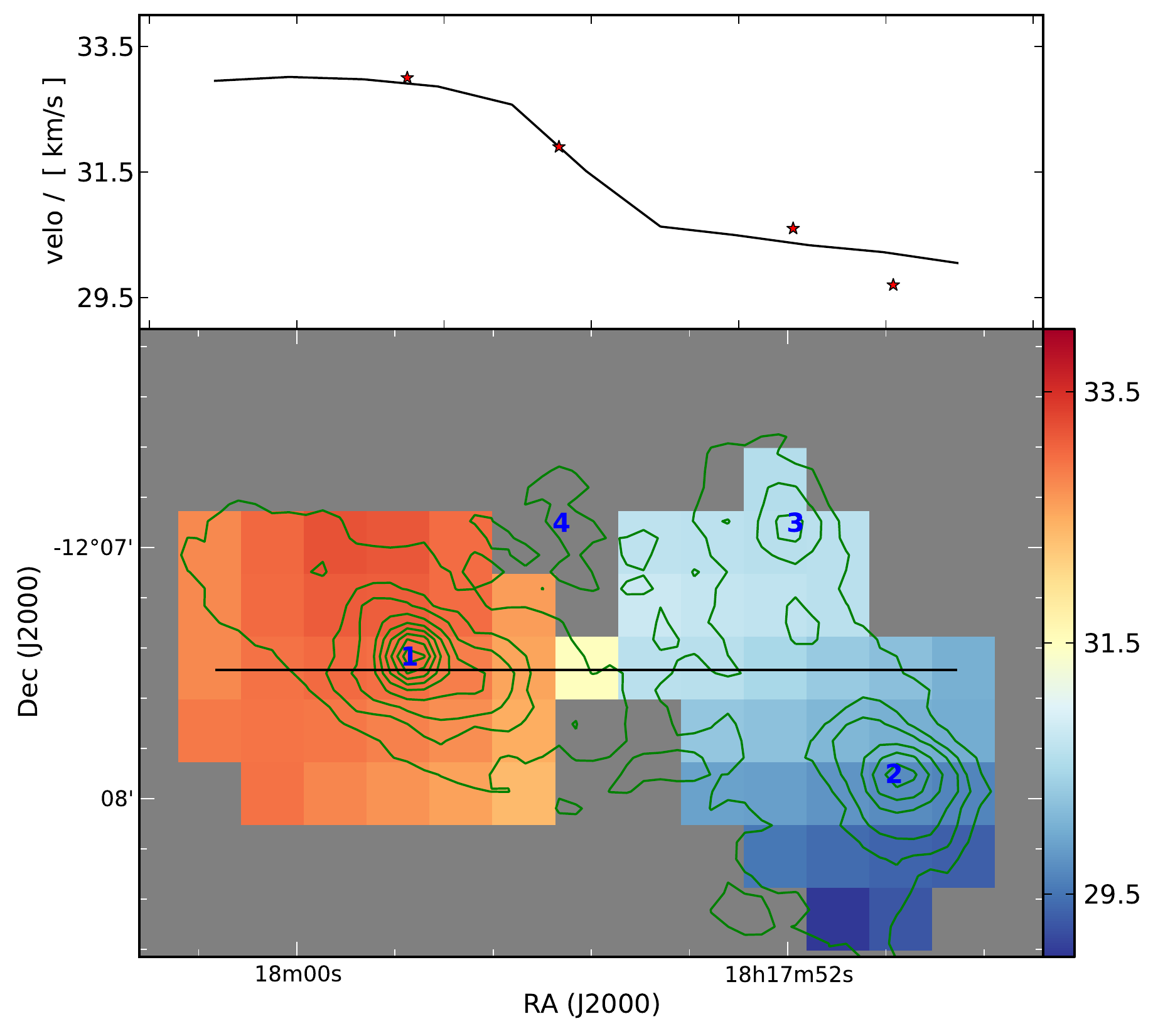}
  \caption{Profile of the \nhp velocity of IRDC\,18151. The left panel
    shows the velocity map with contours from ATLASGAL superimposed (see also
    Fig. \ref{fig:n2hp_param_mopra2}). The
    top panel shows the velocity cut along the line marked on the
    velocity map. The stars mark the velocities of the clump peaks.}
  \label{fig:velo_profile_18151}
\end{figure}
\begin{figure}[tbp]
  \includegraphics[width=.5\textwidth]{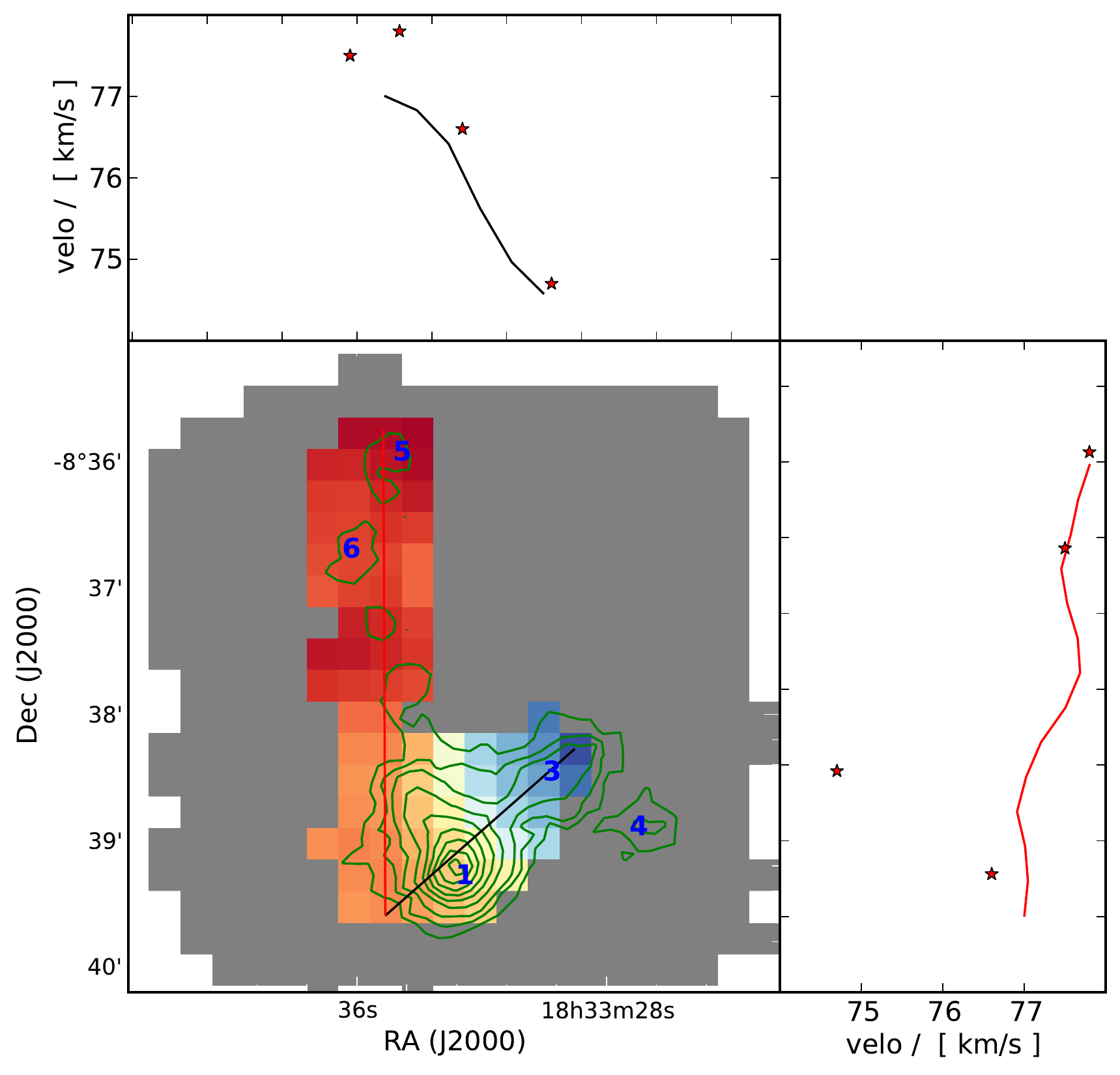}
  \caption{Profile of the \nhp velocity of IRDC\,18308. The left panel
    shows the velocity map with contours from ATLASGAL superimposed (see also
    Fig. \ref{fig:n2hp_param_mopra2}). The
    top and right panels show the velocity cuts along the lines marked
    on the velocity map. The stars mark the velocities of the clump peak.}
  \label{fig:velo_profile_18308}
\end{figure}
\begin{figure}[tbp]
  \includegraphics[width=.5\textwidth]{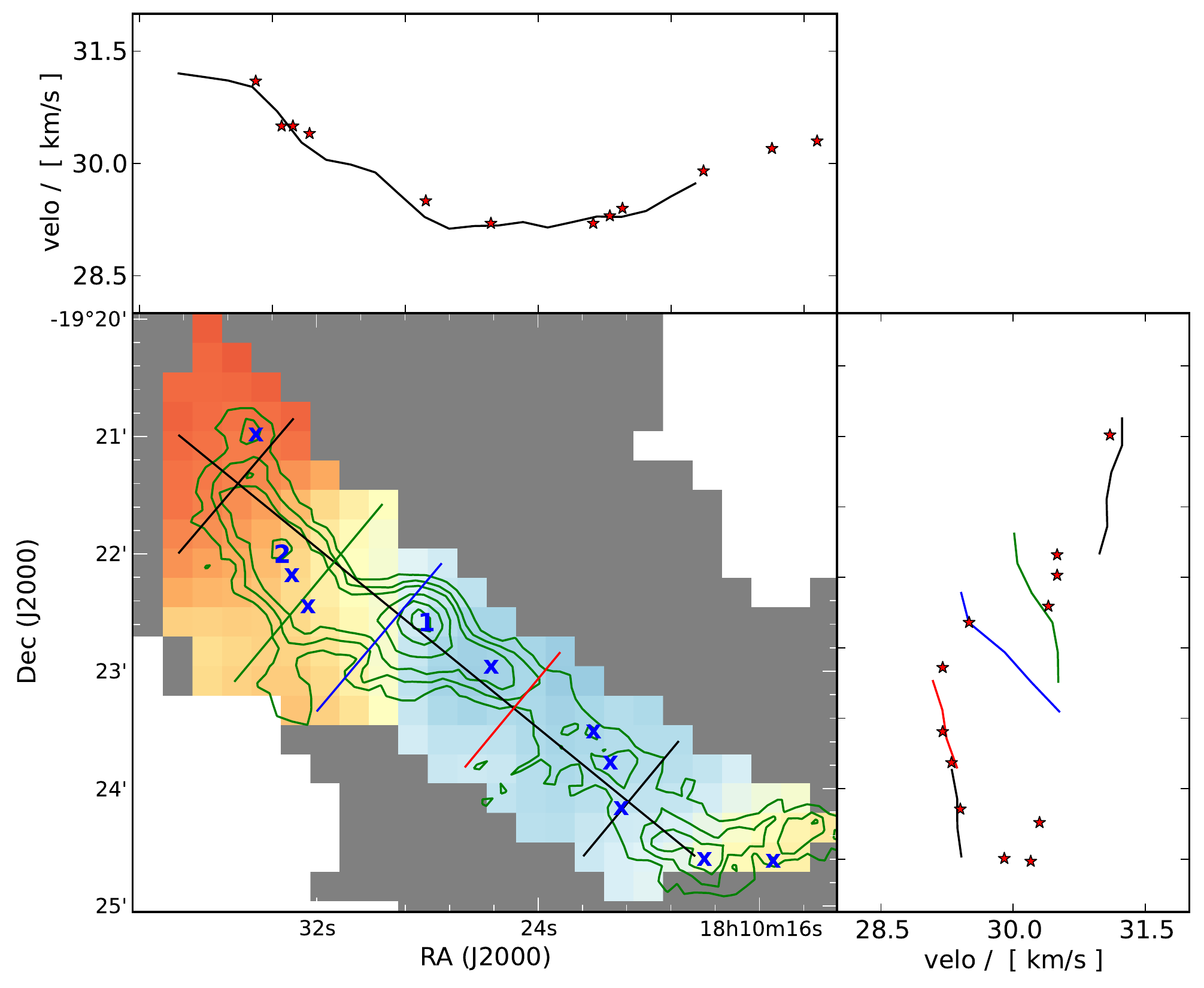}
  \caption{Profile of the \nhp velocity of the northern part of G11.11. The left panel
    shows the velocity map with contours from ATLASGAL superimposed (see also
    Fig. \ref{fig:n2hp_param_mopra1}). The
    right and top panels show the velocity cuts along the lines marked
    on the velocity map. The stars mark the velocities of the clump peak.}
  \label{fig:velo_profile_G11}
\end{figure}
\begin{figure}[tbp]
  \includegraphics[width=.5\textwidth]{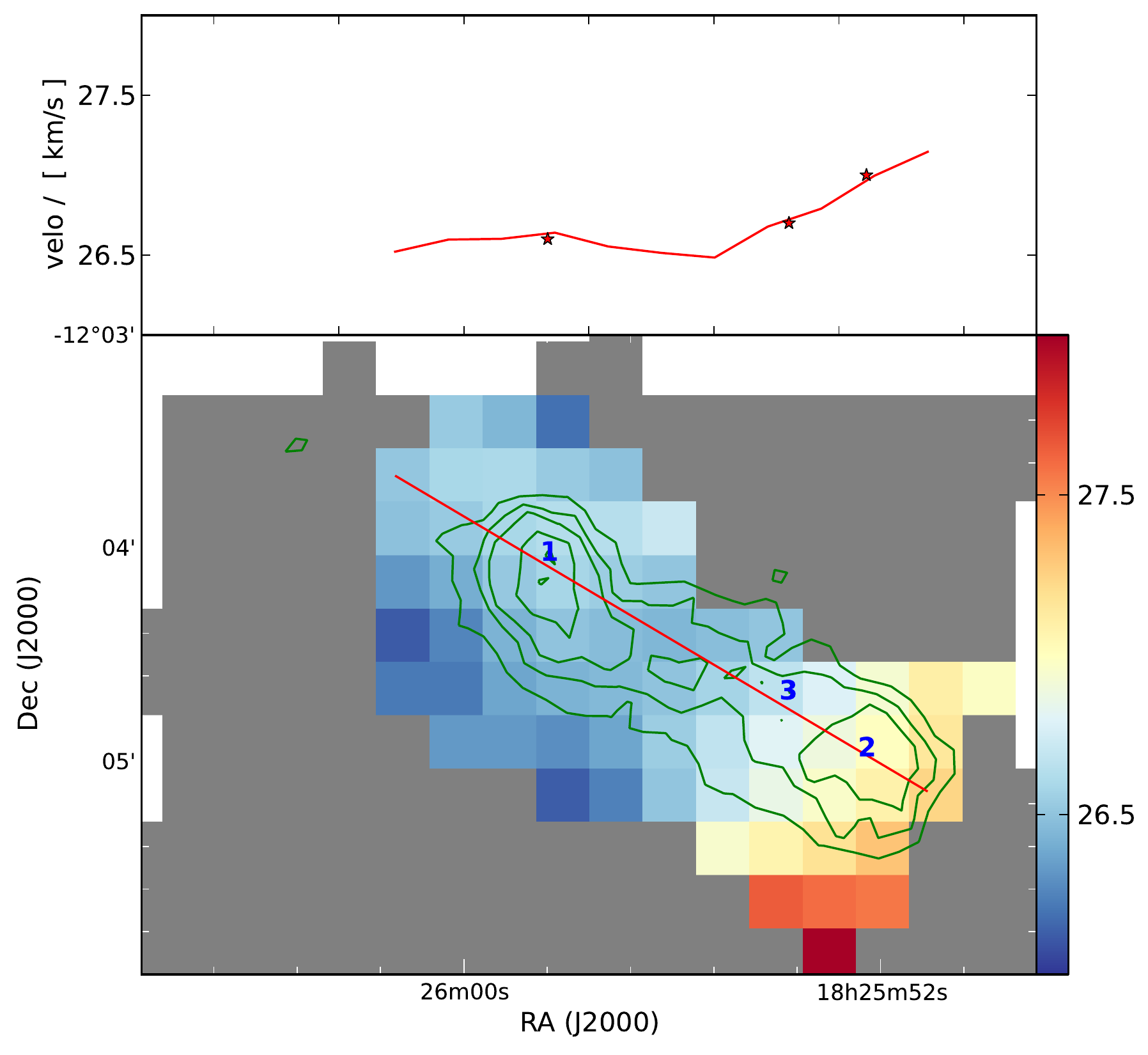}
  \caption{Profile of the \nhp velocity of G19.30. The left panel
    shows the velocity map with contours from ATLASGAL superimposed (see also
    Fig. \ref{fig:n2hp_param_mopra3}). The
    top panel shows the velocity cut along the line marked on the
    velocity map. The stars mark the velocities of the clump peak.}
  \label{fig:velo_profile_G19}
\end{figure}
\begin{figure}[tbp]
  \includegraphics[width=.5\textwidth]{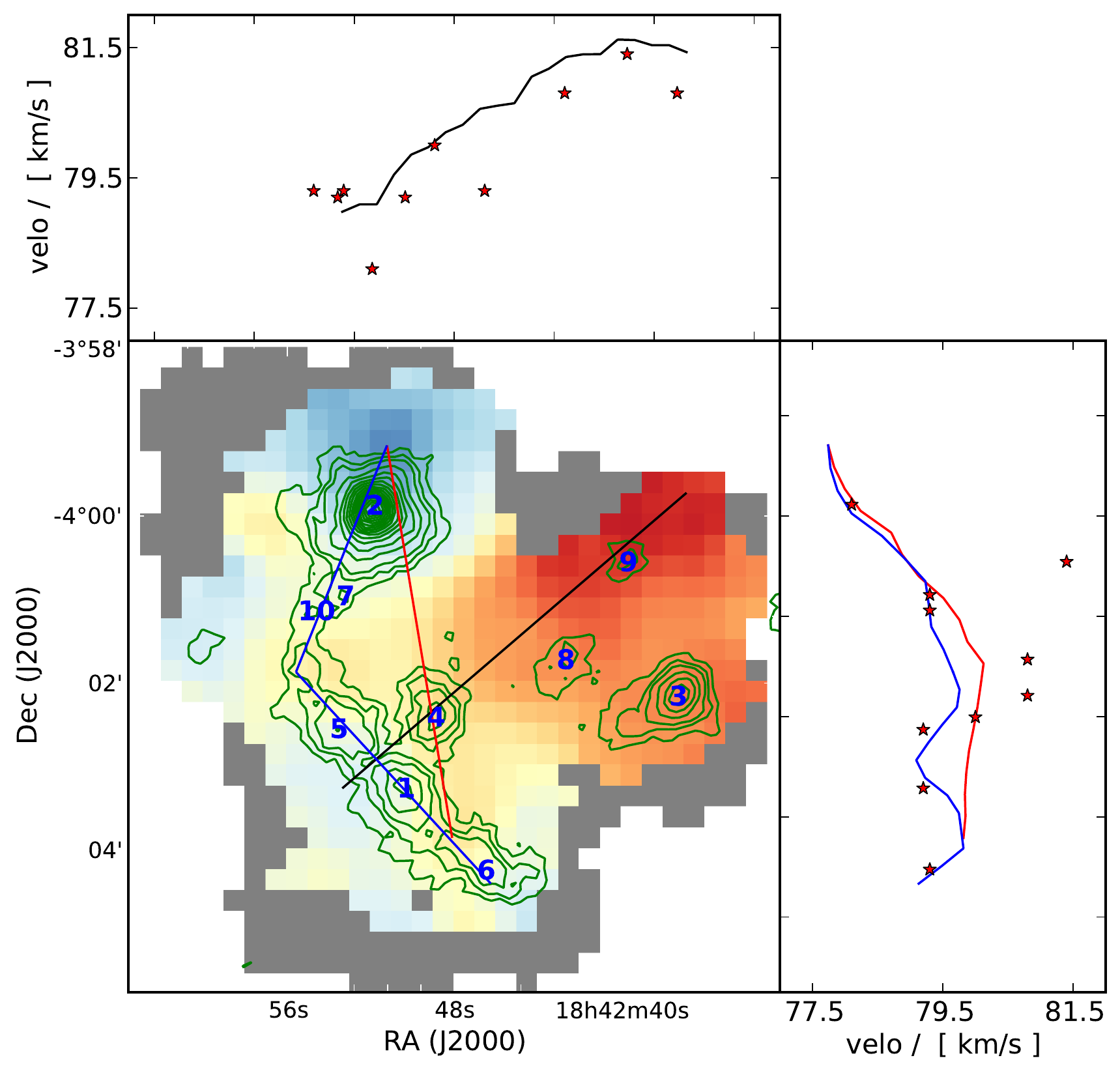}
  \caption{Profile of the \nhp velocity of G28.34. The left panel
    shows the velocity map with contours from ATLASGAL superimposed (see also
    Fig. \ref{fig:n2hp_param_mopra3}). The
    top and right panels show the velocity cuts along the lines marked
    on the velocity map. The stars mark the velocities of the clump peaks.}
  \label{fig:velo_profile_G28}
\end{figure}
To understand the velocity structure of complexes with smooth
velocity transitions, Figs \ref{fig:velo_profile_18223} through
\ref{fig:velo_profile_G28} visualize the velocity gradients along given
lines. As discussed in Sect. \ref{sec:results_velo}, our velocity map of
IRDC\,18151 consists of two larger structures, IRDC\,18151-1 in the east, and
IRDC\,18151-2 and IRDC\,18151-3 in the west. The overall changes within the
eastern and western clump are $\sim$\,0.5\,km/s and $\sim$\,1\,km/s,
respectively. While the velocity cut through the eastern clump
shows hardly any variation, the western clump has a noticeable velocity
gradient. To detect at least part of the gas at intermediate velocities, we smoothed the \nhp
to a velocity resolution of 0.4\,km/s. Figure \ref{fig:velo_profile_18151}
shows the velocity profile across both clumps. While the western clump shows a
slight velocity increase toward the east, the eastern clump shows no
velocity gradient. In particular, the two gradients seem not to match, and if they interact
dynamically, the transition zone would need to be short. Therefore we
conclude that the two structures are individual components, but in the
context of other dense gas tracers they seem to be embedded within the same
cloud. If seen from a 
slightly different angle, the double velocity components
discussed in Sect. \ref{sec:discuss_double_line} could well originate from such
a structure. 

\subsubsection{Flows along G11.11}
We found a clear smooth transition of the velocity toward the northern part of
G11.11. Shown by the top profile in Fig. \ref{fig:velo_profile_G11}, between G11.11-1 and
G11.11-12, the differences in velocity are below
0.5\,km/s. Along the profile just south of G11.11-1, the velocity starts to
increase, with higher velocities toward G11.11-2 and beyond. On the other
hand, the profiles perpendicular to the filament (right panel of
Fig. \ref{fig:velo_profile_G11}) have almost constant velocities. Only the
profile closest to G11.11-1 has a velocity gradient, but the filament has
a bend exactly at the position of the profile. A profile perpendicular to the
actual shape of the IRDC would have no velocity gradient. Therefore, we
conclude that the velocity gradient occurs solely along the filament. 

Both \citet{Tobin2012} (observationally), and \citet[][numerically]{Smith2013} suggested large-scale accretion flows along filaments on, and
probably producing, 
central cores. They described the expected observational signatures for
filaments that are inclined from the plane of the sky. Imagine
a cylinder with a central core and material flowing onto the core from
both sides at a constant velocity. For simplicity and without loss of
generality, we set the central core at rest. Then the gas has a constant velocity along the filament on each side of the central
object. Because the gas is
flowing in from both sides onto the core, a constant velocity is observed for
both directions. The angle of the filament to the line of sight
determines the observed velocity component. While \citet{Tobin2012} accounted
for a gravitationally accelerated gas flow that is expected to have a velocity jump
at its center, the synthetical observations of high-mass star-forming regions
performed by \citet{Smith2013} have a smooth transition. These authorts also found
local velocity variations connected to smaller substructures. 


Recalling the velocity structure we found for G11.11, our observations may be
explained by such accretion flows along the filament. The filament would need
to have an angle such that the southeast is farther away from us than the
northwest. The almost constant velocity over $\sim$\,3\,pc would be material
moving toward G11.11-1, the most massive clump in the region. Just before
G11.11-1, the velocity starts to increase and we observe the transition across
the clump. Beyond G11.11-1, the gravitational potential of G11.11-2, the second-most massive clump in this region, accretes material on its own, and
accelerates the gas even farther beyond the position of G11.11-1.

The scales we trace are an order of magnitude larger than what has been
discussed by \citet{Smith2013} and our resolution is an order of magnitude
lower. Because of the second dense clump, we do not observe the theoretically
predicted pattern. The increase in velocity may also be explained
by solid-body rotation of part of the filament. Nevertheless, we propose an
accretion flow along the filament as a possible explanation for the observed
velocity pattern in G11.11. This view is supported by the fact that
star-formation is most active at the center of the potential infall.

Consequently, if high-mass star formation is actively ongoing
  within G11.11-1, the material flow along the filament suggests
  continuous feeding of the mass reservoir from which forming stars can
  accrete.

\subsubsection{Flows along IRDC\,18308}
A similar scenario might explain the velocity pattern along the IR-dark part
of IRDC\,18308. The cut along the IRDC ending at the HMPO (Fig. \ref{fig:velo_profile_18308}) shows only minor changes in velocity across
the IRDC of $\sim$\,2.5\,pc length. In the vicinity of IRDC\,18308-1, the velocity changes by
  almost 1\,km/s on a short physical scale of only $\sim$\,0.6\,pc. The cut does leave a gap to the
HMPO and does not fully close up in velocity. As described in
Sect. \ref{sec:results_velo}, across the HMPO 
we found one of the steepest velocity gradients in our sample, but the origin
is unclear. One possible explanation that would produce a similar velocity
profile is solid-body rotation. In this picture, the knees at both ends of the
profile would be caused by a transition from solid-body rotation to viscous
rotation because of the lower densities in the outer regions. A full
explanation would require a combination of hydrodynamic simulations with
radiative transfer calculations. This is beyond the scope of this paper. 

\subsubsection{Flows along G19.30?}
In G19.30, along the northeastern part of the IRDC, the velocity is constant
over $\sim$\,1\,pc and then rises toward its other end. This suggests
that the gas is flowing across G19.30-1, at the northeastern end, through
G19.30-3 toward G19.30-2. Interestingly, G19.30-1 is the most massive clump and therefore
is potentially building the center of gravity. Therefore,
the flow is opposite to the gravitational potential. The dust temperatures
derived for the cores within G19.30 by \citet{Ragan2012} are higher for the
more massive clump G19.30-1, which increases our uncertainties on the
mass. Nevertheless, even if we assume the higher dust temperature of 25\,K for
the whole G19.30-1 clump, and a temperature of 17\,K for G19.30-2, the two masses
become of the same order. The additional clump G19.30-3 close to G19.30-2, is
of similar mass as G19.30-2 and therefore increases the gravitational
potential of the southwestern end. But even if our interpretation of a
gas flow along the filament were correct, we found no evidence that is it
driven by gravity. Instead, this example might be interpreted as an indication
for a primordial origin of the flows, which leads to the formation of the
clumps. 
Similar as for G11.11, the flows along the filaments
  within G19.30 and IRDC\,18308 support the idea that the mass
  reservoir in high-mass star formation is continuously replenished.


\subsubsection{The peculiar case of IRDC\,18223}
\label{sec:discuss_velo_18223}
IRDC\,18223 is filamentary, but with a more complex velocity structure than
the previously discussed regions. The velocity seems to be oscillating along the filament. The emission peaks of the southern clumps, IRDC\,18223-2 and
IRDC\,18223-3, are at the same velocity of 44.6\,km/s and in between the
variations are minor. The peak of the clump IRDC\,18223-1, harboring the HMPO
IRAS\,18223-1243, has a velocity of 44.2\,km/s. 

Perpendicular to the filament, we found three extreme velocity gradients, shown
in Fig. \ref{fig:velo_profile_18223}. Although they do not pass exactly
across the dust continuum emission peaks, each seems to be associated with a
clump. A straightforward interpretation would be solid-body rotation along the filament
axis. But as mentioned in Sect. \ref{sec:results_velo}, the two lower
profiles indicate rotation in the direction opposite to the uppermost profile. In
Sect. \ref{sec:results_linewidth} we discussed the possible influence of the
powerful outflow within IRDC\,18223-3 on the linewidth distribution. The
same outflow could also alter the velocity distribution in its direct
vicinity. Nevertheless, a change in rotation orientation along a single
filament seems counterintuitive. For the massive filament DR21 within Cygnus
X, \citet{Schneider2010} found three velocity gradients perpendicular to the
filament axis, with alternating directions. Suggesting turbulent colliding
flows as origin of the filament, these authors interpreted the velocity
pattern as a remnant of the external flow motions.

Within IRDC\,18223-3, on scales of 7\arcsec, \citet{Fallscheer2009}
found a velocity gradient in \nhp and CH$_3$OH, a high-density and
shocked-gas tracer. They successfully modeled the CH$_3$OH velocity
gradient with a rotating and infalling toroid that was significantly larger
than the massive-disk candidates (e.g. \citealt{Cesaroni2005}; for an example of the opposite see \citealt{Boley2012}). However, the-small
scale velocity gradient presented in \citet{Fallscheer2009} not
only has a slightly different orientation (by $\sim$\,45\degr), but is
also rotating in the oppsite direction  to the gradient
presented in Fig. \ref{fig:velo_profile_18223}. This implies that the two
gradients are independent and have different origins; the small-scale
infalling toroid does not connect the large-scale envelope to a
possibly even smaller-scaled disk. To first order this
  contradicts our previous statement of continuously fed clumps because in this particular case, the
  large-scale mass reservoir, or envelope, seems to be disconnected
  from the embedded HMPO. On the one hand, the velocity gradient seen in the
  large-scale gas might be confused with the outflow. As we discussed in
  Sec. \ref{sec:outflow_turbulence}, there is a clear correlation between the outflow and the
  linewidth distribution. Although \nhp is not known to be commonly entrained
  in outflows, a similar correlation between the outflow and the velocity gradient
  seems probable and might explain the change in velocity. Here we need to
  emphasize that the outflow direction and velocity gradient have an angle of
  almost 90\degr. On the other
  hand, it may well be that the accretion phase of IRDC\,18223-3 has already
  stopped, and the large-scale envelope and the inner toroidal structure are
  decoupled. But apart from its youth, it is
  counterintuitive to assume that the core first accreted
  from a rotating structure and then developed an internal source rotating in
  the opposite direction. Therefore, we cannot offer a good explanation.

\subsubsection{G28.34}
A quite large and complex region is G28.34. Most prominent and clear, it
has a strong but smooth east-to-west velocity gradient. 
Along the velocity cut given in the top panel of
Fig. \ref{fig:velo_profile_G28}, the velocity changes by 2.6\,km/s over
7.1\,pc. This results in a velocity gradient of 0.4\,km/s/pc. 
The clump emission peaks within the southeastern component of G28.34, namely
G28.34-6, G28.34-1, and G28.34-5, all have the same velocity. As the
velocity cut along them shows, the velocity of the dense gas between them
seems to be oscillating, similar as found for IRDC\,18223. Even farther north,
the emission peaks of G28.34-7, and G28.34-10 are still at the same velocity as
the more southern clumps. Only north of them the velocity changes
monotonically to lower velocities on the order of a parsec.

From the smooth velocity transition from east to west and from the absence of
double-peaked \nhp profiles, we conclude that all dense gas within the region is
physically connected. To fully explain the observed velocity
signatures of the complex dense gas morphology in G28.34, numerical
simulations are most likely required as well.

\subsubsection{IRDC\,18454}
The velocity analysis of IRDC\,18454 is hampered because, as
discussed in Sect. \ref{sec:discuss_double_line}, several clumps show
two independent velocity components at their peak position. Therefore, the
mapped single velocities can either be the stronger of the two components, or
an average of both. However, the large-scale velocity gradient in east-to-west
direction is clear and no artifact. 

\subsection{Discussion of the linewidth distribution}

\subsubsection{Outflow-induced turbulence}
\label{sec:outflow_turbulence}

As we have described in Sect. \ref{sec:results_linewidth}, we found no
correlation between the linewidth and the temperature or \htwo
column density. 

The lack of a correlation between the temperature and the
linewidth can be explained by the narrow thermal linewidth. At 20\,K, the
thermal line-broadening of \nhp is $\sim$\,0.18\,km/s (in FWHM), and
$\sim$\,0.16\,km/s at 15\,K. Compared with the observed linewidths of a few
km/s, the nonthermal contribution dominates by far and temperature variations
of a few K are weaker than our uncertainties of the measured FWHM.

Another possible contribution to the linewidth stems from turbulence. Because
the clouds are close to virial equilibrium (see
Fig. \ref{fig:mvir_over_mgas}), the linewidth is expected to scale with the column
density. As mentioned in Sect. \ref{sec:results_linewidth}, we found no
correlation between these two quantities either. 

Instead, the spatial resolution of the data prefers molecular outflows as the
dominating contribution to the measured linewidths. Examining the \nhp
linewidth maps of each region, we found clear imprints of known outflows onto
the linewidth distribution. A good example in this context is IRDC\,18223-3,
shown in Fig. \ref{fig:n2hp_param_nobeyama}. As we described
in Sect. \ref{sec:results_linewidth}, \citet{Fallscheer2009} reported the
outflow direction in IRDC\,18223-3 to be $\sim$\,135\degr east of north. As
Fig. \ref{fig:n2hp_param_nobeyama} shows, we observe a line broadening along the
same axis, with the highest velocities toward the edge. 
Due to the complex overall velocity structure it is unclear whether
the aligned velocity gradient is connected to the outflow as
well. This would suggest the origin of the increased linewidth to be
an unresolved velocity gradient. In contrast, shocks connected
to bipolar outflows could enhance the turbulence, leading to broader
lines. Another possible origin of a linewidth-broadening toward the
edge of clumps is provided by gravo-turbulent fragmentation, as
described in \citet{Klessen2005}. Therefore, we cannot assess the physical origin of the
linewidth distribution. In this particular case the alignment
of outflow and linewidth broadening suggests a direct connection.

A similarly
clear correlation between the linewidth and outflow we found for
IRDC\,18102, for which the outflow direction has been observed by
Beuther (priv. communication). For G11.11-1, \citet{Gomez2011}
observed an outflow in east-west direction, or 90\degr east of north, while we found a
significant line-broadening similar to a bipolar outflow with an angle
of $\sim$\,150\degr east of north. This means that the angle between
  outflow and increase in linewidth is on the order of 60\degr. Therefore, a
  direct correlation between the two observations is not mandatory, but possible in the
  context of the other regions.

Although the outflow has not been spatially resolved, the presence of
SiO suggests the existence of an outflow toward the northern part of
G15.05, while the southern part is SiO-quiet (Linz et al., in
prep.). Even though both the northern and southern part are IR-dark even at
70\micron and the mass distribution peaks in between, the linewidth distribution in the north is significantly
larger than in the south. A possible outflow might explain this
linewidth distribution. Using SiO as a general tracer of outflows, we
found a similar correspondence between increased linewidth and the
presence of outflows for IRDC\,18151-2, G19.30-1,
and G19.30-2, and the eastern tip of G48.66.

No SiO was found toward the IR dark part of IRDC\,18182,
IRDC\,18182-2, and -4, for which the linewidths are among narrowest within
the region. In between the two clumps the linewidth
broadens without the presence of known outflows. Therefore, additional
factors have to be dominating here, for example, superposition of the two
independent velocity components.

The four pointings toward G28.34 that cover SiO emission (Linz et al., in
prep.) agree with the picture
of an outflow-dominated linewidth distribution. Nevertheless, due to
the generally broad lines in that region, the variations are less
pronounced than in all other regions.

A particular case seems to be IRDC\,18151-1. \citet{Fallscheer2011}
reported an outflow connected to it with an angle of 315\degr east of
north, but the linewidth is narrower than in IRDC\,18151-2, 1.9\,km/s compared with 3.1\,km/s,
respectively. In addition, it shows hardly any linewidth
structure, and therefore differs from the other clumps discussed so
far. 

\citet{Sridharan2002} and \citet{Sakai2010} found no
SiO toward the two northern clumps of IRDC\,18223, IRDC\,18223-1, and
IRDC\,18223-2, but \citet{Sridharan2002} interpreted the detection of CO line wings
as an indication for an outflow. Therefore, the bipolar line-broadening of the
edges of IRDC\,18223-1 (see Sect. \ref{sec:results_linewidth}) might be caused by an outflow. In contrast, the linewidth broadens toward the center of
IRDC\,18223-2. Because the large-scale velocity gradient across
IRDC\,18223-2 is twice as steep as for IRDC\,18223-1, the
line-broadening toward the center could be explained by unresolved small-scale
velocity gradients. An alternative explanation for the broadening of the
linewidth could be colliding flows, as discussed in
Sec. \ref{sec:discuss_velo_18223}. However, the velocity gradients are not
perfectly aligned with the linewidth broadening, which would be expected for
colliding flows. 
%
%
We have no information on the SiO or other outflow tracers
for IRDC\,18310, and IRDC\,18308.

\subsubsection{The particular case IRDC\,18454}
The line-width map of IRDC\,18454 for a single-component fit (Fig. \ref{fig:n2hp_param_nobeyama}) provides a combination of the two
detected velocity components (see Sec. \ref{sec:discuss_double_line}). Therefore, it quantifies the internal motions within each
beam. If the two velocity components originate from two sheets
  that are interacting, a combined linewidth is appropriate for the virial analysis. However, resolving the individual velocity components at the
peak positions significantly reduces the line width of the clumps. They become more similar to the other regions within
this survey, but still, the resulting line widths are broader
than the average line width that we found for all other
clumps. 

Combining our Nobeyama 45\,m \nhp single-dish data with observations
from the Plateau de Bure interferometer, \citet{Beuther2012} found
two independent velocity components toward IRDC\,18454-1 as
well. While one component is consistent in both width
and velocity with one of our velocity components, the other component is
offset by $\sim$\,1\,km/s and twice as wide in the single-dish data. This difference can be
explained by the better spatial resolution of $\sim$\,3.5\arcsec of the PdBI
observations. Although the two fragments, which \citet{Beuther2012} resolved
within IRDC\,18454-1, have very similar velocities, Fig. 7 in their paper
shows a velocity spread of a few km/s over the Nobeyama beam size of 18\arcsec. As already
apparent from Fig. \ref{fig:n2hp_param_nobeyama}, the linewidth
toward IRDC\,18454-1 is particularly narrow compared with almost all other
clumps within IRDC\,18454. This can partly be explained by the absence of any
70\micron PACS source and its physical proximity to W43-mm1. 

\subsection{Discussion of the virial analysis}
\label{sec:discuss_mvir}
In the context of the virial analysis, bound clumps are expected because
most clumps show signs of ongoing early star formation or already have
formed dense cores. Nevertheless, in globally unbound regions collapsing
fragments might exist as well and form stars and clusters. As shown in
Fig. \ref{fig:mvir_over_mgas}, most clumps have larger virial masses than
gas masses, or a slightly positive $\alpha$ index. Still, within the errors
all clumps might be gravitationally bound. Marked in the upper panel of
Fig. \ref{fig:mvir_over_mgas} are all clumps associated with G28.34. As we
will discuss in a second paper (Tackenberg et al, 2013), G28.34 is in a state of global
infall. Nevertheless, despite their global contraction, their virial
parameters are not different from the bulk of the distribution. This
again emphasizes that a virial mass larger than the gas mass does not
necessarily lead to dissolving clumps.

From Fig. \ref{fig:mvir_over_mgas} it appears that clumps without a PACS point source have a higher $\alpha$ parameter than clumps with a PACS point
source. Furthermore, more massive clumps tend to have lower $\alpha$
parameters. Here it is worth noting that clumps with PACS point sources are on
average more massive than clumps without embedded sources. 
In the context of both, flows along the filaments as discussed here and global infall as discussed in Tackenberg et al. (2013), the
mass of clumps increases over time. The difference between
the populations with and without a PACS point source can accordingly be explained to
first order by two simple
scenarios. {\bf (1)} The dense material that we observe today accumulated and fragmented
on a very short time-scale. Suggested by the competitive-accretion scenario,
some clump fragments accrete mass faster than others. Therefore, they became
dense enough to start forming embedded sources earlier.
{\bf (2)} In an alternative scenario, not all clumps formed at the same time.
Assuming a similar mass-accretion rate for all clumps, the oldest clumps are the densest ones today, and have already started to form stars.

\section{Conclusions}
Complementary to the existing {\it Herschel}/EPoS data, we mapped 17 massive
star-forming regions in \nhp. 
Using CLUMPFIND on submm data, mainly from the ATLASGAL survey at 870\micron,
we extracted the clump population that our observations
cover. Assuming a constant dust temperature of 20\,K and the distances 
given in \citet{Ragan2012}, we calculated both peak column densities
and clump masses. With only a few exceptions, the \nhp and total gas column
density distribution agrees very well spatially, but the \nhp abundance varies in IRDCs
by two orders of magnitude.

While five complexes show no velocity structure, six regions have smooth
velocity gradients. 
For three regions the velocity structure is consistent with
accretion flows along a filament. For DR\,21 \citet{Schneider2010} found a
velocity pattern very similar to what we found for IRDC\,18223 and suggested
that the velocity gradients perpendicular to the filament are remnants of
colliding flow motions from which the filaments formed. For G28.34 and
IRDC\,18454, the velocity structure is very complex and a detailed modeling is
beyond the scope of this work. 

The linewidth distribution among the sources is very diverse; in particular,
IRDC\,18454 stands out. Even after resolving double-peaked line profiles and
treating them carefully, its average linewidth is twice the value we found for
all other regions. This might be explained by the location of IRDC\,18454,
which is associated with W43, a mini-starburst at the junction of a spiral arm
and the Galactic bar. Sources for which the outflow direction is known show a 
line-broadening along the outflow. In addition, all clumps that have measured
SiO show an increased linewidth, while the other clumps show hardly any
variations. Therefore, the linewidth on the scale of clumps probably is
dominated by outflows and unresolved velocity gradients. 

In conclusion, this study supports a very dynamic star-formation
scenario. Clumps and cores accrete mass from the dense
filaments. It is unclear whether the cores and filaments form from primordial flows
or if the clump gravitational potentials are the origin of the flows.

\begin{acknowledgements}
We are grateful to the referee for her/his constructive
comments. S. E. R. is supported by grant RA 2158/1-1, which is part of the Deutsche Forschungsgemeinschaft priority program 1573 (“Physics of the Interstellar
Medium”).
PACS has been developed by a consortium of institutes led by MPE (Germany) and
including UVIE (Austria); KU Leuven, CSL, IMEC (Belgium); CEA, LAM (France);
MPIA (Germany); INAF-IFSI/OAA/OAP/OAT, LENS, SISSA (Italy); IAC (Spain). This
development has been supported by the funding agencies BMVIT (Austria),
ESA-PRODEX (Belgium), CEA/CNES (France), DLR (Germany), ASI/INAF (Italy), and
CICYT/MCYT (Spain). SPIRE has been developed by a consortium of institutes led by Cardiff University (UK) and including Univ. Lethbridge (Canada); NAOC (China); CEA, LAM (France); IFSI, Univ. Padua (Italy); IAC (Spain); Stockholm Observatory (Sweden); Imperial College London, RAL, UCL-MSSL, UKATC, Univ. Sussex (UK); and Caltech, JPL, NHSC, Univ. Colorado (USA). This development has been supported by national funding agencies: CSA (Canada); NAOC (China); CEA, CNES, CNRS (France); ASI (Italy); MCINN (Spain); SNSB (Sweden); STFC (UK); and NASA (USA).
\end{acknowledgements}

\bibliographystyle{aa}
\bibliography{lib}


\begin{appendix}
  \section{Description of the linewidth distribution}
  \label{app:full_dv}
  \noindent{\it IRDC\,18223:} We found a clear increase of the linewidth toward
  the center of IRDC\,18223-2, but IRDC\,18223-2 and -3 show an
  increase toward the edge.
  \\
  \noindent{\it IRDC\,18310:} In IRDC\,18310-1 the linewidth increases not
  toward the submm peak, but toward the detected PACS point sources
  within. IRDC\,18310-4 is a 70\micron dark region and  the
  linewidth increases toward its edge. In between IRDC\,18310-2,
  -3, and -4 the measured increase in linewidth is due to two
  independent, overlapping components.
  \\
  \noindent{\it G11.11:} While for the two main peaks of G11.11, G11.11-1, and G11.11-2,
  the linewidth increases prototypically toward its edges, the other
  clumps of the northern part shown in Fig. \ref{fig:n2hp_param_mopra1}
  show no such clear variation of the linewidth. The systematic offset
  in the mapping of the southern part of G11.11 does not allow a
  systematic study of the line parameters. Nevertheless, it is worth
  noting that the linewidth of the southern part is significantly broader 
  than it is for the northern mapped part.
  \\
  \noindent{\it G15.05:} No clumps detected at given column density threshold.
  \\
  \noindent{\it IRDC\,18102:} We found an increase in linewidth toward
  the submm peak with its connected PACS point source. An additional broadening of the linewidth from east to south is prominent as well. 
  \\
  \noindent{\it IRDC\,18151:} In IRDC\,18151, IRDC\,18151-2 has the widest
  linewidth with 3.1\,km/s. However, it is known to have an
  outflow \citep{Beuther2007a}, which migh explain the broad
  linewidth. Both brighter at PACS 70\micron (the associated PACS point
  source is with 4026\,L$_{\odot}$ $\sim$\,10 times more luminous than the one connected to IRDC\,18151-2) and with a collimated
  outflow \citep{Fallscheer2011} as well, IRDC\,18151-1 has a
  linewidth of only 1.9\,km/s. Along the dust continuum emission away
  from the center the linewidth increases again. However, the reported
  outflow is perpendicular to the dust continuum emission and therefore
  cannot explain the broadening of the linewidth. 
  Despite its still luminous PACS point source of 184\,L$_{\odot}$,
  IRDC\,18151-3 has the narrowest linewidth in this region with
  0.7\,km/s. That is consistent with the fact that
  \citet{Lopez-Sepulcre2011} found no outflows toward that source. 
  \\
  \noindent{\it IRDC\,18182:} The well-studied HMPO in the north with four
  outflows connected to it \citep{Beuther2006} has a broadening
  linewidth toward its center. The region of interest, the IRDC in the
  southeast has its narrowest linewidth at the peak of
  IRDC\,18182-2. Although there is a PACS continuum source at its peak,
  \citet{Beuther2007a} detected no SiO toward that peak. The southern part
  of the IRDC, IRDC\,18182-4 has a slightly broader linewidth.
  \\
  \noindent{\it IRDC\,18308:} While the IR dark filament, connected to IRDC\,18308-5 and
  IRDC\,18308-6, has
  a similar linewidth all along between 1\,km/s and 2\,km/s, the southern complex of
  IRDC\,18308 has linewidths broader than 2.5\,km/s. However, the
  peak of the linewidth is offset from the submm peaks, but is 
  north of IRDC\,18308-1. It is interesting to note that the \nhp column
  density is not aligned with the submm continuum peak IRDC\,18308-1,
  and that the \nhp linewidth peak is offset from the \nhp column density
  peak. While the offset between the continuum and the \nhp emission
  peak of $\sim$\,15\arcsec might be explained by the pointing
  uncertainties, the offset between the column density peak and the
  linewidth peak of $\sim$ 42\arcsec are not. Since the linewidth and
  \nhp column density are measured from the same data, positional
  uncertainties are not an explanation for their difference of 30\arcsec. 
  As described in \ref{sec:results_velo}, we checked for the peak of the
  linewidth map whether its spectrum is fit better by two independent
  \nhp velocity components, but only founnd a single component. Because of the
  high-velocity resolution of 0.1\,km/s for that check we can exclude an
  additional component as the cause of this broadening. 
  \\
  \noindent{\it G19.30:} The linewidth distribution in G19.30 peaks
  toward the two more massive submm peaks G19.30-1 and -2 with an additional increase in linewidth north of G19.30-1. In between P1
  and P2, the linewidth is narrowest within G19.30-3, but still broad
  with $\sim$\,2\,km/s. 
  \\
  \noindent{\it G28.34:} The linewidth in G28.34 is in general very
  broad with $\Delta$v\,$>$\,2\,km/s with only few exceptions. It peaks
  toward the two main peaks G28.34-1 and -2 with linewidths of up to
  3.5\,km/s. The narrowest linewidth is found to be in between the two
  main peaks at G28.34-10 at a linewidth of 2.1\,km/s. Using the maps
  smoothed to a velocity resolution of 0.4\,km/s, we found even narrower
  linewidths along the IR-dark filament beyond G28.34-9. 
  \\
  \noindent{\it G48.66:} The linewidths in G48 are narrower than those of
  other regions mapped within this sample. For the part where \nhp is
  detected, the region with the highest absorption and the embedded but detected
  PACS sources, the velocity dispersion is broadest and decreases along the
  filament to lower column densities.

\end{appendix}
\begin{appendix}
  \section{Parameter maps of omitted regions}
  \label{app:omitted_regions}
  \subsection{Nobeyama 45\,m data}
  \begin{figure*}[tbp]
    \includegraphics[width=1.\textwidth]{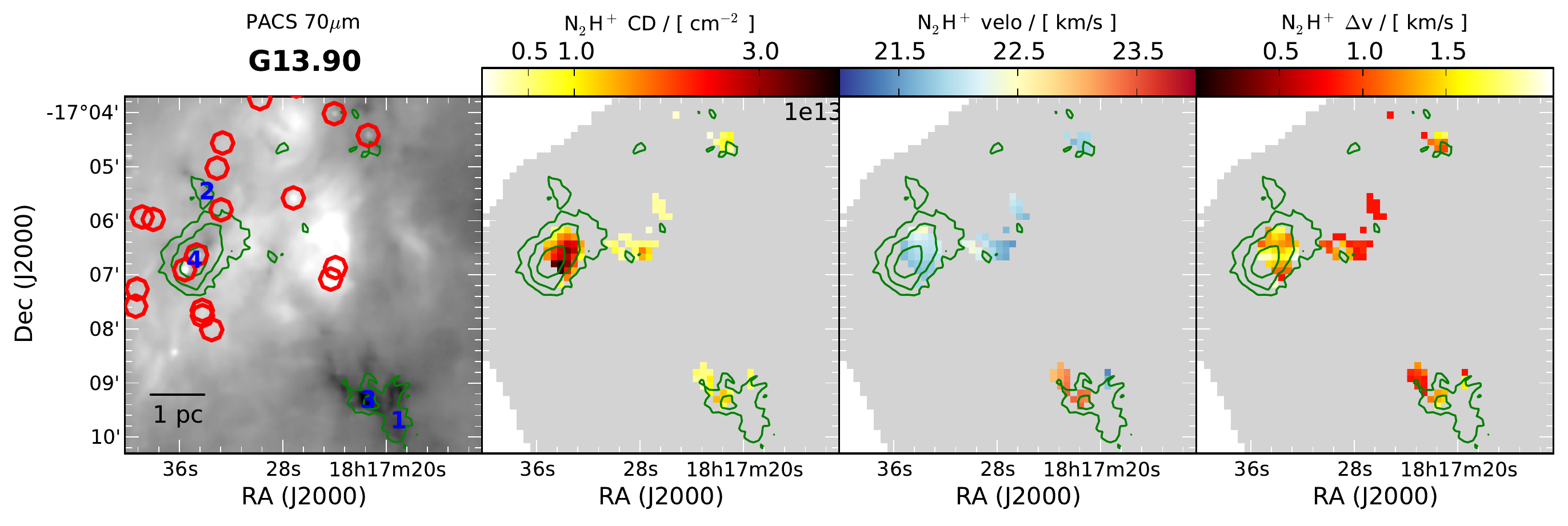}
    \caption{G13.90}
    \label{fig:n2hp_param_13}
  \end{figure*}

  \subsection{MOPRA data}
  \begin{figure*}[tbp]
    \includegraphics[width=1.0\textwidth]{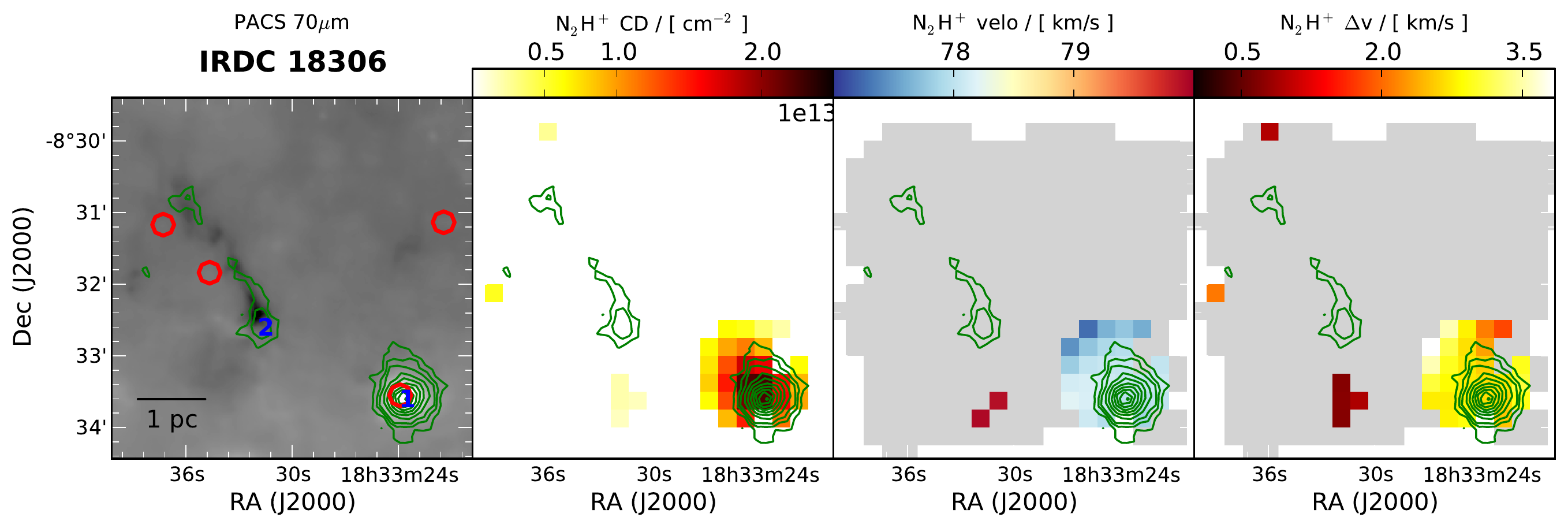}
    \caption{18306}
    \label{fig:n2hp_param_18306at0.4kms}
  \end{figure*}
  \begin{figure*}[tbp]
    \includegraphics[width=1.0\textwidth]{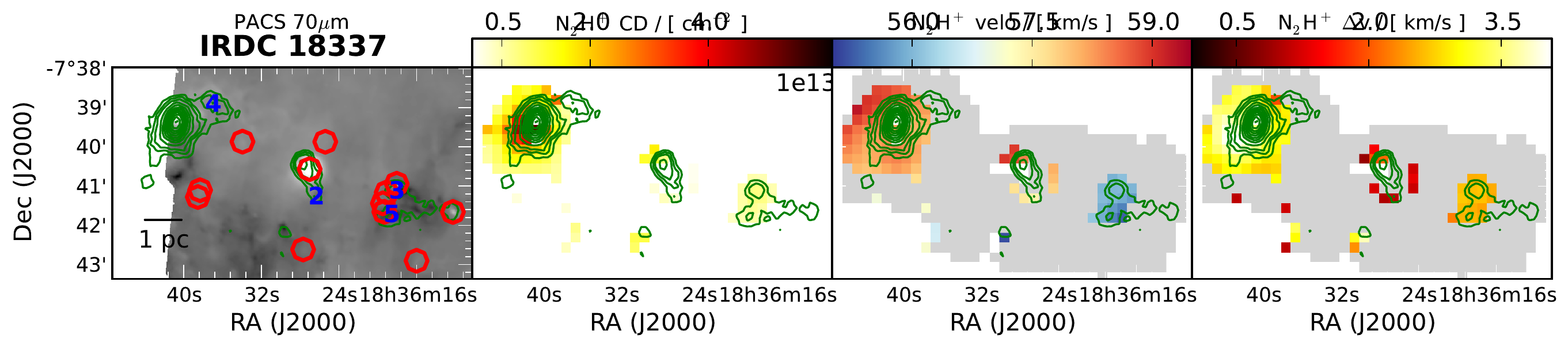}
    \caption{18337}
    \label{fig:n2hp_param_18337at0.4kms}
  \end{figure*}


\end{appendix}


\end{document}

%% file: obs_log_resolution.tex
\begin{table*}[htbp]
  \centering
  \caption{Observed IRDCs. \label{tab:obs_log}} 
  \begin{tabular}{|c|rr
      >{\centering\arraybackslash}*{2}{p{1.5cm}}
      >{\centering\arraybackslash}*{1}{p{1.cm}}
      >{\centering\arraybackslash}*{1}{p{1.cm}}
      >{\centering\arraybackslash}*{2}{p{2.cm}}|}
    \hline
    Source  & RA(J2000) & DEC(J2000) & Gal.\- longitude & Gal.\- latitude &
    Distance & {\bf beam size} & minimum \nhp detection & minimum \nhp detection \@ 0.4\,km/s\\
    name & [ hh mm ss.s ] & [ dd:mm:ss ] & [ \degr ] & [ \degr ] & [ kpc ] & [ pc ] &
    \e{12}\,cm$^{-2}$ & \e{12}\,cm$^{-2}$ \\
    \hline
    \multicolumn{8}{|l|}{\bf Observed with Nobeyama 45m}\\
    IRDC\,18223 &               18 25 10.7 & -12 45 12 &      18.613 &      -0.081 &     3.5  & 0.3 &     0.8 &     \\
    IRDC\,18310 &               18 33 44.7 & -08 22 36 &      23.467 &       0.085 &     4.9  & 0.4 &     3.9 &     \\
    IRDC\,18385 &               18 41 17.1 & -05 09 15 &      27.189 &      -0.098 &     3.1  & 0.3 &     1.2 &     \\
    IRDC\,18454 &               18 47 58.1 & -01 54 41 &      30.835 &      -0.100 &     5.3  & 0.5 &     4.1 &     \\
    ISOSS\,J20153 &               20 15 21.4 & +34 53 52 &      72.953 &      -0.027 &      1.2  & 0.1 &     8.1 &     \\
    IRDC\,13.90 &               18 17 26.1 & -17 05 26 &      13.906 &      -0.473 &     2.6  & 0.2 &     1.4 &     \\
    \hline                                                                                                   
    \multicolumn{8}{|l|}{\bf Observed with Mopra}\\                                                          
    IRDC\,18102 &               18 13 12.2 & -17 59 34 &      12.632 &      -0.016 &       2.6  & 0.6 &    5.7 &    3.3 \\
    IRDC\,18151 &               18 17 55.3 & -12 07 29 &      18.335 &       1.778 &       2.7  & 0.6 &    3.9 &    2.8 \\
    IRDC\,18182 &               18 21 12.2 & -14 32 46 &      16.577 &      -0.069 &       3.4  & 0.7 &    3.2 &    2.2 \\
    IRDC\,18306 &               18 33 29.9 & -08 32 07 &      23.298 &       0.066 &       3.6  & 0.8 &    1.5 &    1.6 \\
    IRDC\,18308 &               18 33 33.1 & -08 37 46 &      23.220 &       0.011 &       4.4  & 1.0 &    2.9 &    2.1 \\
    IRDC\,18337 &               18 36 29.6 & -07 40 33 &      24.402 &      -0.197 &       3.7  & 0.8 &    1.8 &    1.1 \\
    IRDC\,19.30 &               18 25 58.1 & -12 05 09 &      19.293 &       0.060 &       2.4  & 0.5 &    7.6 &    1.3 \\
    IRDC\,11.11 &               18 10 20.0 & -19 25 05 &      11.056 &      -0.107 &       3.4  & 0.7 &    2.4 &    1.6 \\
    IRDC\,15.05 &               18 17 40.4 & -15 49 12 &      15.052 &       0.080 &       3.0  & 0.7 &    3.4 &    1.4 \\
    IRDC\,28.34 &               18 42 48.1 & -04 01 51 &      28.361 &       0.080 &       4.5  & 1.0 &    4.5 &    3.2 \\
    IRDC\,48.66 &               19 21 40.7 & +13 50 32 &      48.666 &      -0.263 &       2.6  & 0.6 &    2.3 &    0.8 \\
    \hline
    \end{tabular}\\
    \footnotesize{Position columns (2-5) give the center coordinate of the
      maps. The actual areas that were mapped are displayed in Fig.
      \ref{fig:n2hp_param_nobeyama} through Fig. \ref{fig:n2hp_param_mopra3}. The
      distances are adopted from \citep{Ragan2012} with references
      therein. The detected \nhp minimum is the minimum plotted in Figs. \ref{fig:n2hp_param_nobeyama} through \ref{fig:n2hp_param_mopra3}. For sources observed with MOPRA we also give
      the improved minimum detection when smoothing the spectra to a velocity
      resolution of 0.4\,km/s. Also observed with the NRO 45m telescope, but not detected, were IRDC20081, ISOSSJ19357, ISOSSJ19557, and ISOSSJ20093.}
\end{table*}

%% file: dust_properties.tex
\begin{table*}[htbp]
  \begin{tabular}{|c|c|c|c|c|c|c|}
    \hline
    Dust data & wavelength & beam FWHM & rms noise & $\kappa_{dust}$ &
    lowest contour & column
    density threshold\\
    & [ $\mu$m ] & [ \arcsec ] & [ mJy\,beam$^{-1}$ ] & [ cm$^2$\,g$^{-1}$ ] & [
      mJy\,beam$^{-1}$ ] & [ cm$^{-2}$ ]\\
    \hline
    ATLASGAL/APEX & 870 & 19.2 & 50 & 1.42 & 310 & 1\e{22}\\
    MAMBO/IRAM 30\,m & 1200 & 10.5 & 17 & 0.79 & 60 & 2\e{22} \\
    SCUBA/JCMT & 850 & 14.0 & 14 & 1.48 & 176 & 1\e{22} \\
    \hline
  \end{tabular}
  \caption{Summary of the different dust data and the corresponding
    properties. The column densities were calculated under the assumptions
    given in Sec. \ref{sec:dust_clfind} for 20\,K. The last column is
    the column density corresponding to the lowest emission contour used within {\it CLUMPFIND}.} 
  \label{tab:dust_prop}
\end{table*}

%% file: EPOS_clumps_mvir_alt_noPACS.tex
\begin{longtab}
  \begin{longtable}{p{1.5cm}*{2}{l}*{2}{p{.7cm}} p{.7cm} *{2}{p{1.cm} p{1.cm} p{1.cm}}} 
    \caption{870\micron ATLASGAL clump properties \label{tab:clump_list}} \\
    \hline
    IRDC ident & RA & DEC & N$_{H_2}$ & M$_{dust}$ & R$_{eff}$ & N$_2$H$^+$
  abundance & v$_{N_2H^+}$ & $\Delta$v$_{N_2H^+}$ & aver. $\Delta$v$_{N_2H^+}$ & M$_{vir,k=128}$ & M$_{vir,k=190}$ \\
    & [\,$^{\circ}$\,] & [\,$^{\circ}$\,] & [\,10$^{22}$ cm$^{-2}$\,] & [\,\msun\,] & [\,\arcsec\,] & [\,10$^{12}$\,] & [\,km/s\,] & [\,km/s\,] & [\,km/s\,] & [\,\msun\,] & [\,\msun\,] \\
    \hline
    \endfirsthead
    \caption{continued.} \\
    \hline
    IRDC ident & RA & DEC & N$_{H_2}$ & M$_{dust}$ & R$_{eff}$ & N$_{N_2H^+}$
    & v$_{N_2H^+}$ & $\Delta$v$_{N_2H^+}$ & aver. $\Delta$v$_{N_2H^+}$ & M$_{vir,k=128}$ & M$_{vir,k=190}$ \\
    & [\,$^{\circ}$\,] & [\,$^{\circ}$\,] & [\,10$^{22}$ cm$^{-2}$\,] & [\,\msun\,] & [\,\arcsec\,] & [\,10$^{12}$\,] & [\,km/s\,] & [\,km/s\,] & [\,km/s\,] & [\,\msun\,] & [\,\msun\,] \\
    \hline
    \endhead
    \hline
    \endfoot
               18102&                   18 13 11.2&  -18 00 06&           11.2&          1019.&           57.1&            3.0&           21.7&            3.2&            2.7&       680&      1026\\
         18151-1$^1$&                   18 17 58.2&  -12 07 26&           32.2&           677.&           34.5&            1.1&           33.0&            1.9&            2.0&       232&       350\\
         18151-2$^1$&                   18 17 50.3&  -12 07 54&           20.4&           524.&           32.5&            1.3&           29.7&            3.1&            3.1&       495&       747\\
         18151-3$^1$&                   18 17 51.9&  -12 06 54&            6.6&           184.&           23.7&            ---&            ---&            ---&            0.7&        19&        29\\
               18151-4&                   18 17 55.7&  -12 06 54&            2.6&            21.&            9.0&            ---&            ---&            ---&            3.0&       128&       193\\
               18182-1$^1$&                   18 21 09.2&  -14 31 47&           17.2&          1032.&           36.3&            1.4&           59.7&            2.7&            2.6&       515&       777\\
               18182-2$^7$&                   18 21 14.8&  -14 33 04&            2.9&           382.&           33.7&            5.5&           41.0&            1.6&            1.9&       256&       387\\
               18182-4$^7$&                   18 21 13.8&  -14 34 11&            1.6&           100.&           20.3&            3.3&           41.2&            1.9&            2.3&       217&       327\\
         18223-1$^1$&                   18 25 10.6&  -12 42 24&            8.1&           975.&           43.1&            4.7&           44.2&            1.9&            2.0&       358&       53\\
         18223-3$^1$&                   18 25 08.3&  -12 45 23&            4.9&           449.&           32.6&            9.2&           44.6&            2.9&            2.9&       599&       90\\
         18223-2$^1$&                   18 25 09.5&  -12 44 11&            3.3&           374.&           31.6&           10.1&           44.6&            2.5&            2.3&       361&       54\\
               18306-1&                   18 33 23.6&  -08 33 36&            9.0&           640.&           32.3&            4.8&           78.0&            2.7&            2.7&       536&       809\\
               18306-2&                   18 33 31.6&  -08 32 36&            2.0&           107.&           18.8&            ---&            ---&            ---&            1.6&       104&       158\\
         18308-1$^1$&                   18 33 32.6&  -08 39 16&            8.2&          1483.&           40.3&            3.9&           76.6&            3.0&            3.4&      1271&      1916\\
         18308-3$^1$&                   18 33 29.8&  -08 38 27&            2.7&           592.&           33.7&            6.5&           74.7&            2.7&            3.0&       796&      1201\\
         18308-4$^1$&                   18 33 27.0&  -08 38 53&            1.6&            81.&           14.0&            ---&            ---&            ---&            0.9&        33&        50\\
               18308-5&                   18 33 34.6&  -08 35 55&            1.4&            61.&           12.7&           10.6&           77.8&            1.8&            1.6&        83&       125\\
               18308-6&                   18 33 36.2&  -08 36 41&            1.4&            57.&           12.2&            ---&            ---&            ---&            1.2&        48&        72\\
         18310-1$^1$&                   18 33 47.7&  -08 23 51&            6.5&          1100.&           33.2&            9.0&           83.2&            1.9&            2.4&       566&       854\\
         18310-2$^1$&                   18 33 43.7&  -08 21 25&            6.3&          1236.&           33.5&            6.0&           84.3&            3.7&            3.5&      1209&      1824\\
         18310-4$^1$&                   18 33 39.2&  -08 21 11&            2.6&           346.&           23.2&            9.4&           86.1&            2.4&            2.3&       384&       579\\
         18310-3$^1$&                   18 33 43.8&  -08 22 13&            1.5&           221.&           21.1&            ---&            ---&            ---&            3.1&       612&       924\\
               18337-4&                   18 36 37.0&  -07 38 53&            2.1&           214.&           26.0&            ---&            ---&            ---&            2.9&       497&       749\\
         18337-3$^1$&                   18 36 18.2&  -07 41 06&            1.9&           177.&           24.9&            5.4&           56.0&            2.6&            2.6&       371&       560\\
               18337-5&                   18 36 18.7&  -07 41 43&            1.7&           137.&           22.4&            ---&            ---&            ---&            2.8&       394&       594\\
         18337-2$^1$&                   18 36 26.5&  -07 41 15&            1.5&            44.&           12.7&            2.9&           57.4&            1.1&            1.1&        36&        55\\
           18454-mm1$^4$&                   18 47 46.7&  -01 54 26&           80.8&         16117.&           50.6&            2.6&           95.6&            5.4&            6.3&      6552&      9880\\
               18454-10&                   18 47 49.4&  -01 53 59&            8.4&          3032.&           40.1&           12.5&           96.6&            7.3&            6.0&      4787&      7219\\
               18454-11&                   18 47 38.9&  -01 56 55&            8.0&          2641.&           38.2&            ---&            ---&            ---&            ---&       ---&       ---\\
         18454-3$^2$&                   18 47 55.4&  -01 53 31&            6.4&          2089.&           41.2&            8.9&           95.3&            6.0&            6.8&      6161&      9290\\
               18454-12&                   18 47 42.2&  -01 56 12&            5.3&          1358.&           35.3&            ---&            ---&            ---&            3.9&      1729&      2608\\
               18454-13&                   18 47 40.4&  -01 55 59&            5.3&          1704.&           29.9&            ---&            ---&            ---&            3.5&      1204&      1816\\
               18454-14&                   18 47 48.2&  -01 57 53&            4.0&          1095.&           36.8&            6.3&           92.8&            2.7&            3.1&      1165&      1757\\
               18454-15&                   18 47 49.1&  -01 57 39&            3.6&           782.&           27.7&            7.5&           93.9&            3.2&            3.3&       994&      1499\\
         18454-1$^2$&                   18 48 02.0&  -01 54 01&            3.4&           608.&           25.6&            6.9&           99.3&            2.6&            2.7&       613&       925\\
         18454-4$^2$&                   18 48 01.1&  -01 52 27&            3.3&           443.&           22.4&            6.3&          100.3&            1.8&            1.6&       179&       270\\
         18454-2$^2$&                   18 47 59.8&  -01 54 12&            3.3&           666.&           27.5&            ---&            ---&            ---&            3.2&       913&      1377\\
               18454-16&                   18 47 51.3&  -01 53 06&            2.5&           354.&           20.6&            ---&            ---&            ---&            8.3&      4660&      7028\\
        18454-5b$^2$&                   18 47 57.7&  -01 56 18&            2.4&           261.&           18.2&            7.5&           93.8&            2.6&            2.7&       428&       646\\
         18454-7$^2$&                   18 47 52.3&  -01 55 02&            2.4&           190.&           15.9&           10.4&           94.2&            2.6&            2.7&       373&       563\\
               18454-17&                   18 47 52.5&  -01 53 09&            2.1&           240.&           18.5&            ---&            ---&            ---&            9.3&      5237&      7897\\
         18454-6$^2$&                   18 48 02.1&  -01 55 43&            1.9&           115.&           13.1&            ---&            ---&            ---&            ---&       ---&       ---\\
               18454-19&                   18 47 51.6&  -01 58 25&            1.8&           171.&           16.6&            ---&            ---&            ---&            2.7&       399&       603\\
         18454-9$^2$&                   18 47 57.3&  -01 57 35&            1.6&           199.&           18.2&            ---&            ---&            ---&            2.6&       395&       596\\
               18454-21&                   18 47 48.5&  -01 55 20&            1.6&           212.&           18.8&            ---&            ---&            ---&            4.1&      1048&      1581\\
                1930-1&                   18 25 58.4&  -12 04 01&            4.0&           164.&           30.1&           10.8&           26.6&            2.7&            2.6&       299&       451\\
                1930-2&                   18 25 52.3&  -12 04 56&            2.9&            93.&           23.9&            7.9&           27.0&            2.2&            2.2&       174&       263\\
                1930-3&                   18 25 53.8&  -12 04 40&            2.0&            94.&           26.2&            8.9&           26.7&            2.1&            2.1&       171&       257\\
        G11.11-1$^3$&                   18 10 28.1&  -19 22 35&            5.5&           455.&           30.1&            5.3&           29.5&            1.9&            1.9&       234&       352\\
        G11.11-6$^3$&                   18 10 07.0&  -19 29 06&            4.3&           411.&           31.4&            4.7&           29.8&            1.7&            1.9&       237&       358\\
        G11.11-7$^3$&                   18 10 06.5&  -19 27 40&            3.9&           458.&           31.0&            2.9&           29.4&            2.3&            2.4&       363&       547\\
        G11.11-2$^3$&                   18 10 33.3&  -19 22 00&            3.2&           325.&           29.3&            8.0&           30.5&            1.9&            2.0&       249&       376\\
              G11.11-9&                   18 10 32.9&  -19 22 11&            3.0&            83.&           14.0&            8.0&           30.5&            1.9&            1.9&       104&       157\\
              G11.11-10&                   18 10 05.2&  -19 26 34&            2.8&           109.&           18.2&            ---&            ---&            ---&            2.2&       178&       268\\
              G11.11-11&                   18 10 05.7&  -19 26 46&            2.7&           130.&           19.1&            ---&            ---&            ---&            2.2&       198&       299\\
              G11.11-12&                   18 10 18.0&  -19 24 36&            2.6&           199.&           25.1&            7.8&           29.9&            1.7&            1.7&       158&       238\\
              G11.11-13&                   18 10 32.3&  -19 22 27&            2.5&           237.&           26.9&            7.9&           30.4&            1.8&            1.9&       194&       292\\
              G11.11-14&                   18 10 05.4&  -19 27 18&            2.5&           134.&           19.1&            ---&            ---&            ---&            2.3&       205&       309\\
              G11.11-15&                   18 10 25.7&  -19 22 58&            2.5&           166.&           22.2&           10.7&           29.2&            1.0&            1.3&        74&       112\\
              G11.11-16&                   18 10 04.8&  -19 27 33&            2.5&           109.&           18.5&            ---&            ---&            ---&            2.0&       158&       238\\
              G11.11-17&                   18 10 13.9&  -19 24 17&            2.1&           106.&           19.7&            3.9&           30.3&            1.4&            1.2&        59&        89\\
              G11.11-18&                   18 10 21.4&  -19 23 47&            1.8&            70.&           16.3&            8.5&           29.3&            1.6&            1.6&        86&       130\\
              G11.11-19&                   18 10 34.2&  -19 20 59&            1.7&            73.&           16.9&           17.4&           31.1&            1.7&            1.5&        82&       124\\
              G11.11-20&                   18 10 22.0&  -19 23 31&            1.7&           100.&           20.3&            8.2&           29.2&            1.6&            1.6&       110&       166\\
              G11.11-21&                   18 10 05.8&  -19 26 05&            1.6&            75.&           17.9&            7.7&           28.6&            1.9&            1.7&       112&       170\\
              G11.11-22&                   18 10 21.0&  -19 24 10&            1.6&            72.&           17.3&           10.7&           29.4&            1.6&            1.7&       102&       154\\
              G11.11-23&                   18 10 15.5&  -19 24 37&            1.5&            37.&           12.2&            7.4&           30.2&            1.5&            0.9&        19&        30\\
           G13-4$^5$&                   18 17 34.9&  -17 06 43&            4.0&           370.&           42.4&            4.7&           21.7&            2.0&            1.5&       155&       235\\
           G13-3$^5$&                   18 17 21.5&  -17 09 18&            2.0&            70.&           21.4&            6.6&           23.3&            1.4&            1.8&       112&       169\\
                 G13-1&                   18 17 19.2&  -17 09 41&            1.6&            54.&           19.7&            ---&            ---&            ---&            3.5&       379&       571\\
                 G13-2&                   18 17 34.0&  -17 05 27&            1.5&            20.&           12.2&            ---&            ---&            ---&            1.6&        52&        78\\
               G1505&                   18 17 39.0&  -15 48 53&            1.4&            36.&           14.8&           11.1&           29.9&            1.3&            1.3&        47&        71\\
         G2823-2$^6$&                   18 42 51.9&  -03 59 52&           15.9&          3136.&           51.1&            3.1&           78.1&            3.4&            3.3&      1509&      2276\\
               G2823-3&                   18 42 37.3&  -04 02 09&            5.7&           890.&           35.2&            3.6&           80.8&            2.5&            2.5&       626&       944\\
         G2823-1$^6$&                   18 42 50.4&  -04 03 15&            4.6&           919.&           37.4&            7.7&           79.2&            3.4&            3.5&      1234&      1861\\
               G2823-4&                   18 42 48.9&  -04 02 24&            3.5&           502.&           28.9&            8.1&           80.0&            2.7&            3.0&       722&      1089\\
               G2823-5&                   18 42 53.6&  -04 02 33&            3.0&           771.&           36.3&            8.8&           79.2&            2.6&            2.9&       860&      1298\\
               G2823-6&                   18 42 46.5&  -04 04 14&            2.8&           658.&           34.2&            5.4&           79.3&            3.0&            3.1&       883&      1331\\
               G2823-7&                   18 42 53.3&  -04 00 57&            1.9&           118.&           16.3&           13.5&           79.3&            2.2&            2.3&       247&       372\\
               G2823-8&                   18 42 42.7&  -04 01 43&            1.8&           176.&           20.0&            5.1&           80.8&            2.6&            2.6&       362&       546\\
               G2823-9&                   18 42 39.7&  -04 00 33&            1.8&            85.&           14.0&           13.5&           81.4&            2.2&            2.2&       180&       272\\
               G2823-10&                   18 42 54.7&  -04 01 08&            1.7&           145.&           18.2&           11.6&           79.3&            2.1&            2.2&       240&       362\\
\end{longtable}
\footnotesize{Notes: The columns are as follows: clump identifier, as
  displayed in the figues; names are adopted from $^1$ \citet{Beuther2002}, $^7$ \citet{Beuther2007}, $^2$ \citet{Beuther2012}, $^3$
  \citet{Johnstone2003}, $^4$ \citet[][]{Motte2003}, $^5$
  \citet{Vasyunina2009}, and $^6$ \citet{Wang2008}.}
\end{longtab}

%% file: clump_properties_include_outflows.tex
\begin{table*}
  \centering
  \caption{Selected clump properties. \label{tab:clump_properties}} 
  \begin{tabular}{|c| >{\centering\arraybackslash}*{1}{p{1.3cm}}  >{\centering\arraybackslash}*{1}{p{1.6cm}}
      >{\centering\arraybackslash}*{2}{p{2.2cm}}
      >{\centering\arraybackslash}*{1}{p{2.2cm}}
      >{\centering\arraybackslash}*{1}{p{1.8cm}} |}
    \hline
    Source  & Velocity gradient & Double component & Increase in $\Delta$v
    towards peak & Decrease in $\Delta$v towards peak & Outflow dominated $\Delta$v
     & Flow along filament\\ 
    \hline
    IRDC\,18223 & \cmark$^{1}$ & & 18223-2 & 18223-1 18223-3 & 18223-1$^3$
    18223-3& ?  \\
    IRDC\,18310 & \xmark & & & 18310-4 & &  \\
    IRDC\,18454 & \cmark & \cmark & & & & ?  \\
    IRDC\,18102 & \xmark & & 18102 & & 18102 &  \\
    IRDC\,18151 & \xmark & & 18151-2 & 18151-1 & 18151-2$^3$& \\
    IRDC\,18182 & \xmark & & 18182-1 & 18182-2 & & \\
    IRDC\,18308 & \cmark & & & 18308-1 & & \cmark \\
    IRDC\,19.30 & \cmark & & G19.30-1 & G19.30-2 & G19.30-1$^3$ G19.30-2$^3$ & \cmark \\
    IRDC\,11.11 & \cmark & \cmark & & G11.11-1 G11.11-2 & G11.11-1$^2$ & \cmark  \\
    IRDC\,15.05 & \xmark & & & & G15.05$^3$ & \\
    IRDC\,28.34 & \cmark & & G28.34-1 G28.34-2 & G28.34-10 & ? \\
    IRDC\,48.66 & \xmark & & & & G48.66$^3$ & \\
    \hline
  \end{tabular}\\
  \footnotesize{The columns are as follows: full region name; flag
    indicating whether we found a smooth velocity gradient along the region;
    flag indicating the presence of resolved independent velocity components
    along a line of sight; two columns for clumps for which we found a clear increase in
    linewidth toward the center and toward the edge clumps, respectively;
    flag indicating that the velocity profile is consistent with flows along
    the filament. Notes: (1) For 18223 we found several velocity gradients, both
    along and perpendicular to the filament. (2) The angle between the outflow
    and the linewidth broadening does not match exactly. (3) Indirect evidence
    for outflow (mainly from SiO), but the direction of the outflow is
    unknown.}
\end{table*}

%% file: n2hp_mapping.bbl
\begin{thebibliography}{90}
\expandafter\ifx\csname natexlab\endcsname\relax\def\natexlab#1{#1}\fi

\bibitem[{{Andr{\'e}} {et~al.}(2010){Andr{\'e}}, {Men'shchikov}, {Bontemps},
  {K{\"o}nyves}, {Motte}, {Schneider}, {Didelon}, {Minier}, {Saraceno},
  {Ward-Thompson}, {di Francesco}, {White}, {Molinari}, {Testi}, {Abergel},
  {Griffin}, {Henning}, {Royer}, {Mer{\'{\i}}n}, {Vavrek}, {Attard},
  {Arzoumanian}, {Wilson}, {Ade}, {Aussel}, {Baluteau}, {Benedettini},
  {Bernard}, {Blommaert}, {Cambr{\'e}sy}, {Cox}, {di Giorgio}, {Hargrave},
  {Hennemann}, {Huang}, {Kirk}, {Krause}, {Launhardt}, {Leeks}, {Le Pennec},
  {Li}, {Martin}, {Maury}, {Olofsson}, {Omont}, {Peretto}, {Pezzuto}, {Prusti},
  {Roussel}, {Russeil}, {Sauvage}, {Sibthorpe}, {Sicilia-Aguilar}, {Spinoglio},
  {Waelkens}, {Woodcraft}, \& {Zavagno}}]{Andre2010}
{Andr{\'e}}, P., {Men'shchikov}, A., {Bontemps}, S., {et~al.} 2010, \aap, 518,
  L102

\bibitem[{{Barnard}(1919)}]{Barnard1919}
{Barnard}, E.~E. 1919, \apj, 49, 1

\bibitem[{{Battersby} {et~al.}(2011){Battersby}, {Bally}, {Ginsburg},
  {Bernard}, {Brunt}, {Fuller}, {Martin}, {Molinari}, {Mottram}, {Peretto},
  {Testi}, \& {Thompson}}]{Battersby2011}
{Battersby}, C., {Bally}, J., {Ginsburg}, A., {et~al.} 2011, \aap, 535, A128

\bibitem[{{Benjamin} {et~al.}(2003){Benjamin}, {Churchwell}, {Babler}, {Bania},
  {Clemens}, {Cohen}, {Dickey}, {Indebetouw}, {Jackson}, {Kobulnicky},
  {Lazarian}, {Marston}, {Mathis}, {Meade}, {Seager}, {Stolovy}, {Watson},
  {Whitney}, {Wolff}, \& {Wolfire}}]{Benjamin2003}
{Benjamin}, R.~A., {Churchwell}, E., {Babler}, B.~L., {et~al.} 2003, \pasp,
  115, 953

\bibitem[{{Bertoldi} \& {McKee}(1992)}]{Bertoldi1992}
{Bertoldi}, F. \& {McKee}, C.~F. 1992, \apj, 395, 140

\bibitem[{{Beuther} {et~al.}(2007){Beuther}, {Churchwell}, {McKee}, \&
  {Tan}}]{Beuther2007}
{Beuther}, H., {Churchwell}, E.~B., {McKee}, C.~F., \& {Tan}, J.~C. 2007,
  Protostars and Planets V, 165

\bibitem[{{Beuther} {et~al.}(2013){Beuther}, {Linz}, {Tackenberg}, {Henning},
  {Krause}, {Ragan}, {Nielbock}, {Launhardt}, {Bihr}, {Schmiedeke}, {Smith}, \&
  {Sakai}}]{Beuther2013}
{Beuther}, H., {Linz}, H., {Tackenberg}, J., {et~al.} 2013, \aap, 553, A115

\bibitem[{{Beuther} {et~al.}(2002{\natexlab{a}}){Beuther}, {Schilke}, {Menten},
  {Motte}, {Sridharan}, \& {Wyrowski}}]{Beuther2002a}
{Beuther}, H., {Schilke}, P., {Menten}, K.~M., {et~al.} 2002{\natexlab{a}},
  \apj, 566, 945

\bibitem[{{Beuther} {et~al.}(2002{\natexlab{b}}){Beuther}, {Schilke}, {Menten},
  {Walmsley}, {Sridharan}, \& {Wyrowski}}]{Beuther2002}
{Beuther}, H., {Schilke}, P., {Menten}, K.~M., {et~al.} 2002{\natexlab{b}}, in
  Astronomical Society of the Pacific Conference Series, Vol. 267, Hot Star
  Workshop III: The Earliest Phases of Massive Star Birth, ed. {P.~Crowther},
  341--+

\bibitem[{{Beuther} \& {Sridharan}(2007)}]{Beuther2007a}
{Beuther}, H. \& {Sridharan}, T.~K. 2007, \apj, 668, 348

\bibitem[{{Beuther} {et~al.}(2005){Beuther}, {Sridharan}, \&
  {Saito}}]{Beuther2005a}
{Beuther}, H., {Sridharan}, T.~K., \& {Saito}, M. 2005, \apjl, 634, L185

\bibitem[{{Beuther} {et~al.}(2012){Beuther}, {Tackenberg}, {Linz}, {Henning},
  {Krause}, {Ragan}, {Nielbock}, {Launhardt}, {Schmiedeke}, {Schuller},
  {Carlhoff}, {Nguyen-Luong}, \& {Sakai}}]{Beuther2012}
{Beuther}, H., {Tackenberg}, J., {Linz}, H., {et~al.} 2012, \aap, 538, A11

\bibitem[{{Beuther} {et~al.}(2006){Beuther}, {Zhang}, {Sridharan}, {Lee}, \&
  {Zapata}}]{Beuther2006}
{Beuther}, H., {Zhang}, Q., {Sridharan}, T.~K., {Lee}, C.-F., \& {Zapata},
  L.~A. 2006, \aap, 454, 221

\bibitem[{{Bihr, S., Beuther, H., Linz, H., et al.}(2013)}]{Bihr2013}
{Bihr, S., Beuther, H., Linz, H., et al.} 2013, \aap, submitted

\bibitem[{{Boley} {et~al.}(2012){Boley}, {Linz}, {van Boekel}, {Bouwman},
  {Henning}, \& {Sobolev}}]{Boley2012}
{Boley}, P.~A., {Linz}, H., {van Boekel}, R., {et~al.} 2012, \aap, 547, A88

\bibitem[{{Bonnell} {et~al.}(2004){Bonnell}, {Vine}, \& {Bate}}]{Bonnell2004}
{Bonnell}, I.~A., {Vine}, S.~G., \& {Bate}, M.~R. 2004, \mnras, 349, 735

\bibitem[{{Bronfman} {et~al.}(1996){Bronfman}, {Nyman}, \&
  {May}}]{Bronfman1996}
{Bronfman}, L., {Nyman}, L.-A., \& {May}, J. 1996, \aaps, 115, 81

\bibitem[{{Carey} {et~al.}(1998){Carey}, {Clark}, {Egan}, {Price}, {Shipman},
  \& {Kuchar}}]{Carey1998}
{Carey}, S.~J., {Clark}, F.~O., {Egan}, M.~P., {et~al.} 1998, \apj, 508, 721

\bibitem[{{Carey} {et~al.}(2009){Carey}, {Noriega-Crespo}, {Mizuno}, {Shenoy},
  {Paladini}, {Kraemer}, {Price}, {Flagey}, {Ryan}, {Ingalls}, {Kuchar},
  {Pinheiro Gon{\c c}alves}, {Indebetouw}, {Billot}, {Marleau}, {Padgett},
  {Rebull}, {Bressert}, {Ali}, {Molinari}, {Martin}, {Berriman}, {Boulanger},
  {Latter}, {Miville-Deschenes}, {Shipman}, \& {Testi}}]{Carey2009}
{Carey}, S.~J., {Noriega-Crespo}, A., {Mizuno}, D.~R., {et~al.} 2009, \pasp,
  121, 76

\bibitem[{{Cesaroni} {et~al.}(2005){Cesaroni}, {Neri}, {Olmi}, {Testi},
  {Walmsley}, \& {Hofner}}]{Cesaroni2005}
{Cesaroni}, R., {Neri}, R., {Olmi}, L., {et~al.} 2005, \aap, 434, 1039

\bibitem[{{Clark} {et~al.}(2012){Clark}, {Glover}, {Klessen}, \&
  {Bonnell}}]{Clark2012}
{Clark}, P.~C., {Glover}, S.~C.~O., {Klessen}, R.~S., \& {Bonnell}, I.~A. 2012,
  \mnras, 424, 2599

\bibitem[{{Contreras} {et~al.}(2013){Contreras}, {Schuller}, {Urquhart},
  {Csengeri}, {Wyrowski}, {Beuther}, {Bontemps}, {Bronfman}, {Henning},
  {Menten}, {Schilke}, {Walmsley}, {Wienen}, {Tackenberg}, \&
  {Linz}}]{Contreras2013}
{Contreras}, Y., {Schuller}, F., {Urquhart}, J.~S., {et~al.} 2013, \aap, 549,
  A45

\bibitem[{{Csengeri} {et~al.}(2011{\natexlab{a}}){Csengeri}, {Bontemps},
  {Schneider}, {Motte}, \& {Dib}}]{Csengeri2011}
{Csengeri}, T., {Bontemps}, S., {Schneider}, N., {Motte}, F., \& {Dib}, S.
  2011{\natexlab{a}}, \aap, 527, A135

\bibitem[{{Csengeri} {et~al.}(2011{\natexlab{b}}){Csengeri}, {Bontemps},
  {Schneider}, {Motte}, {Gueth}, \& {Hora}}]{Csengeri2011a}
{Csengeri}, T., {Bontemps}, S., {Schneider}, N., {et~al.} 2011{\natexlab{b}},
  \apjl, 740, L5

\bibitem[{{Egan} {et~al.}(2003){Egan}, {Price}, {Kraemer}, {Mizuno}, {Carey},
  {Wright}, {Engelke}, {Cohen}, \& {Gugliotti}}]{Egan2003}
{Egan}, M.~P., {Price}, S.~D., {Kraemer}, K.~E., {et~al.} 2003, VizieR Online
  Data Catalog, 5114, 0

\bibitem[{{Egan} {et~al.}(1998){Egan}, {Shipman}, {Price}, {Carey}, {Clark}, \&
  {Cohen}}]{Egan1998}
{Egan}, M.~P., {Shipman}, R.~F., {Price}, S.~D., {et~al.} 1998, \apjl, 494,
  L199

\bibitem[{{Fallscheer} {et~al.}(2011){Fallscheer}, {Beuther}, {Sauter}, {Wolf},
  \& {Zhang}}]{Fallscheer2011}
{Fallscheer}, C., {Beuther}, H., {Sauter}, J., {Wolf}, S., \& {Zhang}, Q. 2011,
  \apj, 729, 66

\bibitem[{{Fallscheer} {et~al.}(2009){Fallscheer}, {Beuther}, {Zhang}, {Keto},
  \& {Sridharan}}]{Fallscheer2009}
{Fallscheer}, C., {Beuther}, H., {Zhang}, Q., {Keto}, E., \& {Sridharan}, T.~K.
  2009, \aap, 504, 127

\bibitem[{{Fa{\'u}ndez} {et~al.}(2004){Fa{\'u}ndez}, {Bronfman}, {Garay},
  {Chini}, {Nyman}, \& {May}}]{Faundez2004}
{Fa{\'u}ndez}, S., {Bronfman}, L., {Garay}, G., {et~al.} 2004, \aap, 426, 97

\bibitem[{{Goldsmith}(2001)}]{Goldsmith2001}
{Goldsmith}, P.~F. 2001, \apj, 557, 736

\bibitem[{{G{\'o}mez} {et~al.}(2011){G{\'o}mez}, {Wyrowski}, {Pillai},
  {Leurini}, \& {Menten}}]{Gomez2011}
{G{\'o}mez}, L., {Wyrowski}, F., {Pillai}, T., {Leurini}, S., \& {Menten},
  K.~M. 2011, \aap, 529, A161

\bibitem[{{Griffin} {et~al.}(2010){Griffin}, {Abergel}, {Abreu}, {Ade},
  {Andr{\'e}}, {Augueres}, {Babbedge}, {Bae}, {Baillie}, {Baluteau}, {Barlow},
  {Bendo}, {Benielli}, {Bock}, {Bonhomme}, {Brisbin}, {Brockley-Blatt},
  {Caldwell}, {Cara}, {Castro-Rodriguez}, {Cerulli}, {Chanial}, {Chen},
  {Clark}, {Clements}, {Clerc}, {Coker}, {Communal}, {Conversi}, {Cox},
  {Crumb}, {Cunningham}, {Daly}, {Davis}, {de Antoni}, {Delderfield}, {Devin},
  {di Giorgio}, {Didschuns}, {Dohlen}, {Donati}, {Dowell}, {Dowell}, {Duband},
  {Dumaye}, {Emery}, {Ferlet}, {Ferrand}, {Fontignie}, {Fox}, {Franceschini},
  {Frerking}, {Fulton}, {Garcia}, {Gastaud}, {Gear}, {Glenn}, {Goizel},
  {Griffin}, {Grundy}, {Guest}, {Guillemet}, {Hargrave}, {Harwit}, {Hastings},
  {Hatziminaoglou}, {Herman}, {Hinde}, {Hristov}, {Huang}, {Imhof}, {Isaak},
  {Israelsson}, {Ivison}, {Jennings}, {Kiernan}, {King}, {Lange}, {Latter},
  {Laurent}, {Laurent}, {Leeks}, {Lellouch}, {Levenson}, {Li}, {Li},
  {Lilienthal}, {Lim}, {Liu}, {Lu}, {Madden}, {Mainetti}, {Marliani}, {McKay},
  {Mercier}, {Molinari}, {Morris}, {Moseley}, {Mulder}, {Mur}, {Naylor},
  {Nguyen}, {O'Halloran}, {Oliver}, {Olofsson}, {Olofsson}, {Orfei}, {Page},
  {Pain}, {Panuzzo}, {Papageorgiou}, {Parks}, {Parr-Burman}, {Pearce},
  {Pearson}, {P{\'e}rez-Fournon}, {Pinsard}, {Pisano}, {Podosek}, {Pohlen},
  {Polehampton}, {Pouliquen}, {Rigopoulou}, {Rizzo}, {Roseboom}, {Roussel},
  {Rowan-Robinson}, {Rownd}, {Saraceno}, {Sauvage}, {Savage}, {Savini},
  {Sawyer}, {Scharmberg}, {Schmitt}, {Schneider}, {Schulz}, {Schwartz},
  {Shafer}, {Shupe}, {Sibthorpe}, {Sidher}, {Smith}, {Smith}, {Smith},
  {Spencer}, {Stobie}, {Sudiwala}, {Sukhatme}, {Surace}, {Stevens}, {Swinyard},
  {Trichas}, {Tourette}, {Triou}, {Tseng}, {Tucker}, {Turner}, {Vaccari},
  {Valtchanov}, {Vigroux}, {Virique}, {Voellmer}, {Walker}, {Ward}, {Waskett},
  {Weilert}, {Wesson}, {White}, {Whitehouse}, {Wilson}, {Winter}, {Woodcraft},
  {Wright}, {Xu}, {Zavagno}, {Zemcov}, {Zhang}, \& {Zonca}}]{Griffin2010}
{Griffin}, M.~J., {Abergel}, A., {Abreu}, A., {et~al.} 2010, \aap, 518, L3

\bibitem[{{Gritschneder} {et~al.}(2012){Gritschneder}, {Lin}, {Murray}, {Yin},
  \& {Gong}}]{Gritschneder2012}
{Gritschneder}, M., {Lin}, D.~N.~C., {Murray}, S.~D., {Yin}, Q.-Z., \& {Gong},
  M.-N. 2012, \apj, 745, 22

\bibitem[{{Hatchell} \& {van der Tak}(2003)}]{Hatchell2003}
{Hatchell}, J. \& {van der Tak}, F.~F.~S. 2003, \aap, 409, 589

\bibitem[{{Heitsch} \& {Hartmann}(2008)}]{Heitsch2008}
{Heitsch}, F. \& {Hartmann}, L. 2008, \apj, 689, 290

\bibitem[{{Hennemann} {et~al.}(2008){Hennemann}, {Birkmann}, {Krause}, \&
  {Lemke}}]{Hennemann2008}
{Hennemann}, M., {Birkmann}, S.~M., {Krause}, O., \& {Lemke}, D. 2008, \aap,
  485, 753

\bibitem[{{Hennemann} {et~al.}(2012){Hennemann}, {Motte}, {Schneider},
  {Didelon}, {Hill}, {Arzoumanian}, {Bontemps}, {Csengeri}, {Andr{\'e}},
  {Konyves}, {Louvet}, {Marston}, {Men'shchikov}, {Minier}, {Nguyen Luong},
  {Palmeirim}, {Peretto}, {Sauvage}, {Zavagno}, {Anderson}, {Bernard}, {Di
  Francesco}, {Elia}, {Li}, {Martin}, {Molinari}, {Pezzuto}, {Russeil}, {Rygl},
  {Schisano}, {Spinoglio}, {Sousbie}, {Ward-Thompson}, \&
  {White}}]{Hennemann2012}
{Hennemann}, M., {Motte}, F., {Schneider}, N., {et~al.} 2012, \aap, 543, L3

\bibitem[{{Hildebrand}(1983)}]{Hildebrand1983}
{Hildebrand}, R.~H. 1983, \qjras, 24, 267

\bibitem[{{Hill} {et~al.}(2011){Hill}, {Motte}, {Didelon}, {Bontemps},
  {Minier}, {Hennemann}, {Schneider}, {Andr{\'e}}, {Men'shchikov}, {Anderson},
  {Arzoumanian}, {Bernard}, {di Francesco}, {Elia}, {Giannini}, {Griffin},
  {K{\"o}nyves}, {Kirk}, {Marston}, {Martin}, {Molinari}, {Nguyen Luong},
  {Peretto}, {Pezzuto}, {Roussel}, {Sauvage}, {Sousbie}, {Testi},
  {Ward-Thompson}, {White}, {Wilson}, \& {Zavagno}}]{Hill2011}
{Hill}, T., {Motte}, F., {Didelon}, P., {et~al.} 2011, \aap, 533, A94

\bibitem[{{Johnstone} {et~al.}(2003){Johnstone}, {Fiege}, {Redman}, {Feldman},
  \& {Carey}}]{Johnstone2003}
{Johnstone}, D., {Fiege}, J.~D., {Redman}, R.~O., {Feldman}, P.~A., \& {Carey},
  S.~J. 2003, \apjl, 588, L37

\bibitem[{{J{\o}rgensen} {et~al.}(2004){J{\o}rgensen}, {Sch{\"o}ier}, \& {van
  Dishoeck}}]{Jorgensen2004}
{J{\o}rgensen}, J.~K., {Sch{\"o}ier}, F.~L., \& {van Dishoeck}, E.~F. 2004,
  \aap, 416, 603

\bibitem[{{Kessler} {et~al.}(1996){Kessler}, {Steinz}, {Anderegg}, {Clavel},
  {Drechsel}, {Estaria}, {Faelker}, {Riedinger}, {Robson}, {Taylor}, \&
  {Xim{\'e}nez de Ferr{\'a}n}}]{Kessler1996}
{Kessler}, M.~F., {Steinz}, J.~A., {Anderegg}, M.~E., {et~al.} 1996, \aap, 315,
  L27

\bibitem[{{Klessen} {et~al.}(2005){Klessen}, {Ballesteros-Paredes},
  {V{\'a}zquez-Semadeni}, \& {Dur{\'a}n-Rojas}}]{Klessen2005}
{Klessen}, R.~S., {Ballesteros-Paredes}, J., {V{\'a}zquez-Semadeni}, E., \&
  {Dur{\'a}n-Rojas}, C. 2005, \apj, 620, 786

\bibitem[{{Ladd} {et~al.}(2005){Ladd}, {Purcell}, {Wong}, \&
  {Robertson}}]{Ladd2005}
{Ladd}, N., {Purcell}, C., {Wong}, T., \& {Robertson}, S. 2005, \pasa, 22, 62

\bibitem[{{Launhardt} {et~al.}(2013){Launhardt}, {Stutz}, {Schmiedeke},
  {Henning}, {Krause}, {Balog}, {Beuther}, {Birkmann}, {Hennemann},
  {Kainulainen}, {Khanzadyan}, {Linz}, {Lippok}, {Nielbock}, {Pitann}, {Ragan},
  {Risacher}, {Schmalzl}, {Shirley}, {Stecklum}, {Steinacker}, \&
  {Tackenberg}}]{Launhardt2013}
{Launhardt}, R., {Stutz}, A.~M., {Schmiedeke}, A., {et~al.} 2013, ArXiv
  e-prints

\bibitem[{{L{\'o}pez-Sepulcre} {et~al.}(2011){L{\'o}pez-Sepulcre}, {Walmsley},
  {Cesaroni}, {Codella}, {Schuller}, {Bronfman}, {Carey}, {Menten}, {Molinari},
  \& {Noriega-Crespo}}]{Lopez-Sepulcre2011}
{L{\'o}pez-Sepulcre}, A., {Walmsley}, C.~M., {Cesaroni}, R., {et~al.} 2011,
  \aap, 526, L2

\bibitem[{{Mac Low} \& {Klessen}(2004)}]{MacLow2004}
{Mac Low}, M.-M. \& {Klessen}, R.~S. 2004, Reviews of Modern Physics, 76, 125

\bibitem[{{MacLaren} {et~al.}(1988){MacLaren}, {Richardson}, \&
  {Wolfendale}}]{MacLaren1988}
{MacLaren}, I., {Richardson}, K.~M., \& {Wolfendale}, A.~W. 1988, \apj, 333,
  821

\bibitem[{{McKee} \& {Tan}(2003)}]{McKee2003}
{McKee}, C.~F. \& {Tan}, J.~C. 2003, \apj, 585, 850

\bibitem[{{Men'shchikov} {et~al.}(2010){Men'shchikov}, {Andr{\'e}}, {Didelon},
  {K{\"o}nyves}, {Schneider}, {Motte}, {Bontemps}, {Arzoumanian}, {Attard},
  {Abergel}, {Baluteau}, {Bernard}, {Cambr{\'e}sy}, {Cox}, {di Francesco}, {di
  Giorgio}, {Griffin}, {Hargrave}, {Huang}, {Kirk}, {Li}, {Martin}, {Minier},
  {Miville-Desch{\^e}nes}, {Molinari}, {Olofsson}, {Pezzuto}, {Roussel},
  {Russeil}, {Saraceno}, {Sauvage}, {Sibthorpe}, {Spinoglio}, {Testi},
  {Ward-Thompson}, {White}, {Wilson}, {Woodcraft}, \&
  {Zavagno}}]{Menshchikov2010}
{Men'shchikov}, A., {Andr{\'e}}, P., {Didelon}, P., {et~al.} 2010, \aap, 518,
  L103

\bibitem[{{Molinari} {et~al.}(2010){Molinari}, {Swinyard}, {Bally}, {Barlow},
  {Bernard}, {Martin}, {Moore}, {Noriega-Crespo}, {Plume}, {Testi}, {Zavagno},
  {Abergel}, {Ali}, {Andr{\'e}}, {Baluteau}, {Benedettini}, {Bern{\'e}},
  {Billot}, {Blommaert}, {Bontemps}, {Boulanger}, {Brand}, {Brunt}, {Burton},
  {Campeggio}, {Carey}, {Caselli}, {Cesaroni}, {Cernicharo}, {Chakrabarti},
  {Chrysostomou}, {Codella}, {Cohen}, {Compiegne}, {Davis}, {de Bernardis}, {de
  Gasperis}, {Di Francesco}, {di Giorgio}, {Elia}, {Faustini}, {Fischera},
  {Fukui}, {Fuller}, {Ganga}, {Garcia-Lario}, {Giard}, {Giardino}, {Glenn},
  {Goldsmith}, {Griffin}, {Hoare}, {Huang}, {Jiang}, {Joblin}, {Joncas},
  {Juvela}, {Kirk}, {Lagache}, {Li}, {Lim}, {Lord}, {Lucas}, {Maiolo},
  {Marengo}, {Marshall}, {Masi}, {Massi}, {Matsuura}, {Meny}, {Minier},
  {Miville-Desch{\^e}nes}, {Montier}, {Motte}, {M{\"u}ller}, {Natoli}, {Neves},
  {Olmi}, {Paladini}, {Paradis}, {Pestalozzi}, {Pezzuto}, {Piacentini},
  {Pomar{\`e}s}, {Popescu}, {Reach}, {Richer}, {Ristorcelli}, {Roy}, {Royer},
  {Russeil}, {Saraceno}, {Sauvage}, {Schilke}, {Schneider-Bontemps},
  {Schuller}, {Schultz}, {Shepherd}, {Sibthorpe}, {Smith}, {Smith},
  {Spinoglio}, {Stamatellos}, {Strafella}, {Stringfellow}, {Sturm}, {Taylor},
  {Thompson}, {Tuffs}, {Umana}, {Valenziano}, {Vavrek}, {Viti}, {Waelkens},
  {Ward-Thompson}, {White}, {Wyrowski}, {Yorke}, \& {Zhang}}]{Molinari2010}
{Molinari}, S., {Swinyard}, B., {Bally}, J., {et~al.} 2010, \pasp, 122, 314

\bibitem[{{Motte} {et~al.}(2003){Motte}, {Schilke}, \& {Lis}}]{Motte2003}
{Motte}, F., {Schilke}, P., \& {Lis}, D.~C. 2003, \apj, 582, 277

\bibitem[{{M{\"u}ller} {et~al.}(2005){M{\"u}ller}, {Schl{\"o}der}, {Stutzki},
  \& {Winnewisser}}]{Muller2005}
{M{\"u}ller}, H.~S.~P., {Schl{\"o}der}, F., {Stutzki}, J., \& {Winnewisser}, G.
  2005, Journal of Molecular Structure, 742, 215

\bibitem[{{Nguyen Luong} {et~al.}(2011){Nguyen Luong}, {Motte}, {Schuller},
  {Schneider}, {Bontemps}, {Schilke}, {Menten}, {Heitsch}, {Wyrowski},
  {Carlhoff}, {Bronfman}, \& {Henning}}]{NguyenLuong2011}
{Nguyen Luong}, Q., {Motte}, F., {Schuller}, F., {et~al.} 2011, \aap, 529, A41

\bibitem[{{Ossenkopf} \& {Henning}(1994)}]{Ossenkopf1994}
{Ossenkopf}, V. \& {Henning}, T. 1994, \aap, 291, 943

\bibitem[{{Ott}(2010)}]{Ott2010}
{Ott}, S. 2010, in Astronomical Society of the Pacific Conference Series, Vol.
  434, Astronomical Data Analysis Software and Systems XIX, ed. Y.~{Mizumoto},
  K.-I. {Morita}, \& M.~{Ohishi}, 139

\bibitem[{{Perault} {et~al.}(1996){Perault}, {Omont}, {Simon}, {Seguin},
  {Ojha}, {Blommaert}, {Felli}, {Gilmore}, {Guglielmo}, {Habing}, {Price},
  {Robin}, {de Batz}, {Cesarsky}, {Elbaz}, {Epchtein}, {Fouque}, {Guest},
  {Levine}, {Pollock}, {Prusti}, {Siebenmorgen}, {Testi}, \&
  {Tiphene}}]{Perault1996}
{Perault}, M., {Omont}, A., {Simon}, G., {et~al.} 1996, \aap, 315, L165

\bibitem[{{Peretto} {et~al.}(2006){Peretto}, {Andr{\'e}}, \&
  {Belloche}}]{Peretto2006}
{Peretto}, N., {Andr{\'e}}, P., \& {Belloche}, A. 2006, \aap, 445, 979

\bibitem[{{Peretto} {et~al.}(2012){Peretto}, {Andr{\'e}}, {K{\"o}nyves},
  {Schneider}, {Arzoumanian}, {Palmeirim}, {Didelon}, {Attard}, {Bernard}, {Di
  Francesco}, {Elia}, {Hennemann}, {Hill}, {Kirk}, {Men'shchikov}, {Motte},
  {Nguyen Luong}, {Roussel}, {Sousbie}, {Testi}, {Ward-Thompson}, {White}, \&
  {Zavagno}}]{Peretto2012}
{Peretto}, N., {Andr{\'e}}, P., {K{\"o}nyves}, V., {et~al.} 2012, \aap, 541,
  A63

\bibitem[{{Peretto} \& {Fuller}(2009)}]{Peretto2009}
{Peretto}, N. \& {Fuller}, G.~A. 2009, \aap, 505, 405

\bibitem[{{Peretto} \& {Fuller}(2010)}]{Peretto2010}
{Peretto}, N. \& {Fuller}, G.~A. 2010, \apj, 723, 555

\bibitem[{{Peters} {et~al.}(2010){Peters}, {Banerjee}, {Klessen}, {Mac Low},
  {Galv{\'a}n-Madrid}, \& {Keto}}]{Peters2010}
{Peters}, T., {Banerjee}, R., {Klessen}, R.~S., {et~al.} 2010, \apj, 711, 1017

\bibitem[{{Pilbratt} {et~al.}(2010){Pilbratt}, {Riedinger}, {Passvogel},
  {Crone}, {Doyle}, {Gageur}, {Heras}, {Jewell}, {Metcalfe}, {Ott}, \&
  {Schmidt}}]{Pilbratt2010}
{Pilbratt}, G.~L., {Riedinger}, J.~R., {Passvogel}, T., {et~al.} 2010, \aap,
  518, L1

\bibitem[{{Pillai} {et~al.}(2006){Pillai}, {Wyrowski}, {Carey}, \&
  {Menten}}]{Pillai2006}
{Pillai}, T., {Wyrowski}, F., {Carey}, S.~J., \& {Menten}, K.~M. 2006, \aap,
  450, 569

\bibitem[{{Poglitsch} {et~al.}(2010){Poglitsch}, {Waelkens}, {Geis},
  {Feuchtgruber}, {Vandenbussche}, {Rodriguez}, {Krause}, {Renotte}, {van
  Hoof}, {Saraceno}, {Cepa}, {Kerschbaum}, {Agn{\`e}se}, {Ali}, {Altieri},
  {Andreani}, {Augueres}, {Balog}, {Barl}, {Bauer}, {Belbachir}, {Benedettini},
  {Billot}, {Boulade}, {Bischof}, {Blommaert}, {Callut}, {Cara}, {Cerulli},
  {Cesarsky}, {Contursi}, {Creten}, {De Meester}, {Doublier}, {Doumayrou},
  {Duband}, {Exter}, {Genzel}, {Gillis}, {Gr{\"o}zinger}, {Henning},
  {Herreros}, {Huygen}, {Inguscio}, {Jakob}, {Jamar}, {Jean}, {de Jong},
  {Katterloher}, {Kiss}, {Klaas}, {Lemke}, {Lutz}, {Madden}, {Marquet},
  {Martignac}, {Mazy}, {Merken}, {Montfort}, {Morbidelli}, {M{\"u}ller},
  {Nielbock}, {Okumura}, {Orfei}, {Ottensamer}, {Pezzuto}, {Popesso},
  {Putzeys}, {Regibo}, {Reveret}, {Royer}, {Sauvage}, {Schreiber}, {Stegmaier},
  {Schmitt}, {Schubert}, {Sturm}, {Thiel}, {Tofani}, {Vavrek}, {Wetzstein},
  {Wieprecht}, \& {Wiezorrek}}]{Poglitsch2010}
{Poglitsch}, A., {Waelkens}, C., {Geis}, N., {et~al.} 2010, \aap, 518, L2

\bibitem[{{Ragan} {et~al.}(2012{\natexlab{a}}){Ragan}, {Henning}, {Krause},
  {Pitann}, {Beuther}, {Linz}, {Tackenberg}, {Balog}, {Hennemann}, {Launhardt},
  {Lippok}, {Nielbock}, {Schmiedeke}, {Schuller}, {Steinacker}, {Stutz}, \&
  {Vasyunina}}]{Ragan2012}
{Ragan}, S., {Henning}, T., {Krause}, O., {et~al.} 2012{\natexlab{a}}, \aap,
  547, A49

\bibitem[{{Ragan} {et~al.}(2012{\natexlab{b}}){Ragan}, {Heitsch}, {Bergin}, \&
  {Wilner}}]{Ragan2012a}
{Ragan}, S.~E., {Heitsch}, F., {Bergin}, E.~A., \& {Wilner}, D.
  2012{\natexlab{b}}, \apj, 746, 174

\bibitem[{{Roussel}(2012)}]{Roussel2012}
{Roussel}, H. 2012, ArXiv e-prints

\bibitem[{{Sakai} {et~al.}(2010){Sakai}, {Sakai}, {Hirota}, \&
  {Yamamoto}}]{Sakai2010}
{Sakai}, T., {Sakai}, N., {Hirota}, T., \& {Yamamoto}, S. 2010, \apj, 714, 1658

\bibitem[{{Sawada} {et~al.}(2008){Sawada}, {Ikeda}, {Sunada}, {Kuno},
  {Kamazaki}, {Morita}, {Kurono}, {Koura}, {Abe}, {Kawase}, {Maekawa},
  {Horigome}, \& {Yanagisawa}}]{Sawada2008}
{Sawada}, T., {Ikeda}, N., {Sunada}, K., {et~al.} 2008, \pasj, 60, 445

\bibitem[{{Schlingman} {et~al.}(2011){Schlingman}, {Shirley}, {Schenk},
  {Rosolowsky}, {Bally}, {Battersby}, {Dunham}, {Ellsworth-Bowers}, {Evans},
  {Ginsburg}, \& {Stringfellow}}]{Schlingman2011}
{Schlingman}, W.~M., {Shirley}, Y.~L., {Schenk}, D.~E., {et~al.} 2011, \apjs,
  195, 14

\bibitem[{{Schneider} {et~al.}(2010){Schneider}, {Csengeri}, {Bontemps},
  {Motte}, {Simon}, {Hennebelle}, {Federrath}, \& {Klessen}}]{Schneider2010}
{Schneider}, N., {Csengeri}, T., {Bontemps}, S., {et~al.} 2010, \aap, 520, A49

\bibitem[{{Sch{\"o}ier} {et~al.}(2005){Sch{\"o}ier}, {van der Tak}, {van
  Dishoeck}, \& {Black}}]{Schoier2005}
{Sch{\"o}ier}, F.~L., {van der Tak}, F.~F.~S., {van Dishoeck}, E.~F., \&
  {Black}, J.~H. 2005, \aap, 432, 369

\bibitem[{{Schuller} {et~al.}(2009){Schuller}, {Menten}, {Contreras},
  {Wyrowski}, {Schilke}, {Bronfman}, {Henning}, {Walmsley}, {Beuther},
  {Bontemps}, {Cesaroni}, {Deharveng}, {Garay}, {Herpin}, {Lefloch}, {Linz},
  {Mardones}, {Minier}, {Molinari}, {Motte}, {Nyman}, {Reveret}, {Risacher},
  {Russeil}, {Schneider}, {Testi}, {Troost}, {Vasyunina}, {Wienen}, {Zavagno},
  {Kovacs}, {Kreysa}, {Siringo}, \& {Wei{\ss}}}]{Schuller2009}
{Schuller}, F., {Menten}, K.~M., {Contreras}, Y., {et~al.} 2009, \aap, 504, 415

\bibitem[{{Smith} {et~al.}(2009){Smith}, {Longmore}, \& {Bonnell}}]{Smith2009}
{Smith}, R.~J., {Longmore}, S., \& {Bonnell}, I. 2009, \mnras, 400, 1775

\bibitem[{{Smith} {et~al.}(2013){Smith}, {Shetty}, {Beuther}, {Klessen}, \&
  {Bonnell}}]{Smith2013}
{Smith}, R.~J., {Shetty}, R., {Beuther}, H., {Klessen}, R.~S., \& {Bonnell},
  I.~A. 2013, \apj, 771, 24

\bibitem[{{Sridharan} {et~al.}(2005){Sridharan}, {Beuther}, {Saito},
  {Wyrowski}, \& {Schilke}}]{Sridharan2005}
{Sridharan}, T.~K., {Beuther}, H., {Saito}, M., {Wyrowski}, F., \& {Schilke},
  P. 2005, \apjl, 634, L57

\bibitem[{{Sridharan} {et~al.}(2002){Sridharan}, {Beuther}, {Schilke},
  {Menten}, \& {Wyrowski}}]{Sridharan2002}
{Sridharan}, T.~K., {Beuther}, H., {Schilke}, P., {Menten}, K.~M., \&
  {Wyrowski}, F. 2002, \apj, 566, 931

\bibitem[{{Sunada} {et~al.}(2000){Sunada}, {Yamaguchi}, {Kuno}, {Okumura},
  {Nakai}, \& {Ukita}}]{Sunada2000}
{Sunada}, K., {Yamaguchi}, C., {Kuno}, N., {et~al.} 2000, in Astronomical
  Society of the Pacific Conference Series, Vol. 217, Imaging at Radio through
  Submillimeter Wavelengths, ed. J.~G. {Mangum} \& S.~J.~E. {Radford}, 19

\bibitem[{{Tackenberg} {et~al.}(2013){Tackenberg}, {Beuther}, {Henning},
  {Linz}, {Sakai}, {Ragan}, {Stutz}, {Krause}, {Nielbock}, {Hennemann},
  {Pitann}, \& {Schmiedecke}}]{Tackenberg2013a}
{Tackenberg}, J., {Beuther}, H., {Henning}, T., {et~al.} 2013, \aap, submitted

\bibitem[{{Tielens}(2005)}]{Tielens2005}
{Tielens}, A.~G.~G.~M. 2005, {The Physics and Chemistry of the Interstellar
  Medium}

\bibitem[{{Tobin} {et~al.}(2012){Tobin}, {Hartmann}, {Bergin}, {Chiang},
  {Looney}, {Chandler}, {Maret}, \& {Heitsch}}]{Tobin2012}
{Tobin}, J.~J., {Hartmann}, L., {Bergin}, E., {et~al.} 2012, \apj, 748, 16

\bibitem[{{Vasyunina} {et~al.}(2009){Vasyunina}, {Linz}, {Henning}, {Stecklum},
  {Klose}, \& {Nyman}}]{Vasyunina2009}
{Vasyunina}, T., {Linz}, H., {Henning}, T., {et~al.} 2009, \aap, 499, 149

\bibitem[{{Vasyunina} {et~al.}(2011){Vasyunina}, {Linz}, {Henning},
  {Zinchenko}, {Beuther}, \& {Voronkov}}]{Vasyunina2011}
{Vasyunina}, T., {Linz}, H., {Henning}, T., {et~al.} 2011, \aap, 527, A88

\bibitem[{{Wang} {et~al.}(2008){Wang}, {Zhang}, {Pillai}, {Wyrowski}, \&
  {Wu}}]{Wang2008}
{Wang}, Y., {Zhang}, Q., {Pillai}, T., {Wyrowski}, F., \& {Wu}, Y. 2008, \apjl,
  672, L33

\bibitem[{{Wienen} {et~al.}(2012){Wienen}, {Wyrowski}, {Schuller}, {Menten},
  {Walmsley}, {Bronfman}, \& {Motte}}]{Wienen2012}
{Wienen}, M., {Wyrowski}, F., {Schuller}, F., {et~al.} 2012, \aap, 544, A146

\bibitem[{{Wilcock} {et~al.}(2011){Wilcock}, {Kirk}, {Stamatellos},
  {Ward-Thompson}, {Whitworth}, {Battersby}, {Brunt}, {Fuller}, {Griffin},
  {Molinari}, {Martin}, {Mottram}, {Peretto}, {Plume}, {Smith}, \&
  {Thompson}}]{Wilcock2011}
{Wilcock}, L.~A., {Kirk}, J.~M., {Stamatellos}, D., {et~al.} 2011, \aap, 526,
  A159+

\bibitem[{{Wilcock} {et~al.}(2012){Wilcock}, {Ward-Thompson}, {Kirk},
  {Stamatellos}, {Whitworth}, {Elia}, {Fuller}, {DiGiorgio}, {Griffin},
  {Molinari}, {Martin}, {Mottram}, {Peretto}, {Pestalozzi}, {Schisano},
  {Plume}, {Smith}, \& {Thompson}}]{Wilcock2012}
{Wilcock}, L.~A., {Ward-Thompson}, D., {Kirk}, J.~M., {et~al.} 2012, \mnras,
  422, 1071

\bibitem[{{Williams} {et~al.}(1994){Williams}, {de Geus}, \&
  {Blitz}}]{Williams1994}
{Williams}, J.~P., {de Geus}, E.~J., \& {Blitz}, L. 1994, \apj, 428, 693

\bibitem[{{Zinnecker} \& {Yorke}(2007)}]{Zinnecker2007}
{Zinnecker}, H. \& {Yorke}, H.~W. 2007, \araa, 45, 481

\end{thebibliography}
